\documentclass[a4paper,11pt]{article}
\pdfoutput=1 

\usepackage{jinstpub} 
\usepackage{lineno}
\usepackage{textcomp}
\usepackage{xcolor}
\usepackage{graphicx}
\usepackage{mathtools}
\usepackage{subcaption}
\usepackage{upgreek}
\usepackage[LGRgreek]{mathastext}

\title{\boldmath Construction and commissioning of CMS CE prototype silicon modules}

\author{B.~Acar,}
\author{G.~Adamov,}
\author{C.~Adloff,}
\author{S.~Afanasiev ,}
\author{N.~Akchurin,}
\author{B.~Akg\"{u}n,}
\author{M.~Alhusseini,}
\author{J.~Alison,}
\author{G.~Altopp,}
\author{M.~Alyari,}
% ...

\emailAdd{arnaud.steen@cern.ch, bora.akgun@cern.ch}

\abstract{As part of its HL-LHC upgrade program, the CMS Collaboration is developing a High Granularity Calorimeter (CE) to replace the existing endcap calorimeters. The CE is a sampling calorimeter with unprecedented transverse and longitudinal readout for both electromagnetic (CE-E) and hadronic (CE-H) compartments. The calorimeter will be built with $\sim$30,000 hexagonal silicon modules. Prototype modules have been constructed with 6-inch hexagonal silicon sensors with cell areas of 1.1~$cm^2$, and the SKIROC2-CMS readout ASIC. Beam tests of different sampling configurations were conducted with the prototype modules at DESY and CERN in 2017 and 2018. This paper describes the construction and commissioning of the CE calorimeter prototype, the silicon modules used in the construction, their basic performance, and the methods used for their calibration.} 

\graphicspath{{figs/}}

\begin{document}
\maketitle
\flushbottom


\section{Introduction}
\label{sec:intro}
The CERN High-Luminosity LHC (HL-LHC) will operate with a higher instantaneous luminosity than the CERN LHC and is expected to record ten times more data. The increase in the instantaneous luminosity is a challenge for detector design, due to the needs for increased radiation tolerance and for the mitigation of effects due to overlapping events (pile-up), which is expected to be as high as 200 collisions per bunch crossing. To cope with these conditions, the CMS Collaboration has undertaken an extensive R$\&$D program to upgrade many parts of the detector, including the replacement of the calorimeter endcaps~\cite{bib.cms}. There are two key requirements that the new endcap calorimeters must meet.  Firstly, they should maintain acceptable performance after the delivery of the expected HL-LHC integrated luminosity (3000 fb$^{-1}$), when the total neutron fluence in the innermost region will be $10^{16}~n_{eq}/cm^2$ and the total ionizing dose will be 2 MGy. Secondly, the detector needs to have $\sim$50~ps timing resolution to mitigate the pile-up. 

\begin{figure}[htbp]
  \centering
  \includegraphics[width=.9\linewidth]{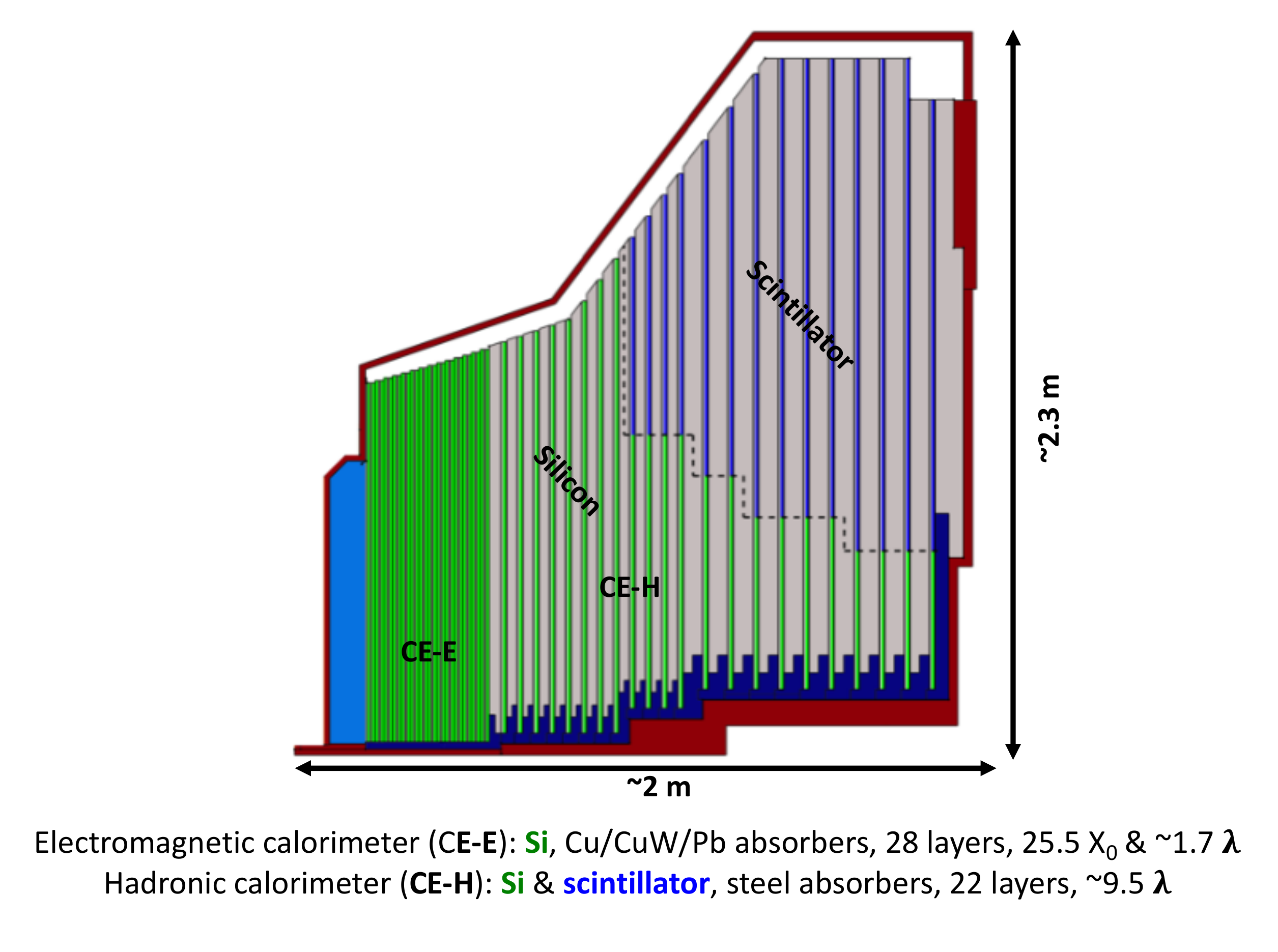}
  \caption{\label{fig:cms-hgcal} Schematic view and key parameters of the CMS High Granularity Calorimeter Endcap.} %
\end{figure}

The CE~\cite{bib.cms-tdr}, shown schematically in Figure~\ref{fig:cms-hgcal}, is a high granularity sampling calorimeter with 50 active layers and more than 6 million channels. Silicon modules with a hexagonal sensor, an absorber plate, and readout electronics will be the building blocks of the calorimeter. In the electromagnetic section (CE-E) of the calorimeter, silicon modules will be interleaved with lead and copper absorbers. The silicon sensors will be segmented into hexagonal cells with an area of approximately $1.1~cm^2$. In the innermost region the segmentation will instead result in cells with an area of $0.5~cm^2$, where the fluence will be highest. The hadronic section (CE-H) will also use silicon sensors in the region where the radiation is high, and plastic scintillator tiles readout by on-tile silicon photomultipliers (SiPM) where it is low. The main absorber of the hadronic calorimeter will be steel. The full calorimeter will be inside a cold volume kept at -30$^{\circ}$C to reduce the dark currents in the silicon sensors and the SiPMs. This highly-segmented calorimeter will provide transverse, longitudinal and precision timing information on showers that will be essential for pile-up mitigation, event reconstruction, and analysis.

Several  tests of calorimeters built with prototype silicon modules have taken place in beams at CERN, Fermilab and DESY. The goals for these tests were to validate the basic design of the CE, to study the calorimetric performance of a silicon-based calorimeter, and to compare the Geant4 simulation~\cite{bib.geant4} of the calorimeter with experimental data. The first prototypes of hexagonal silicon modules were tested in beams at CERN and Fermilab in 2016, with up to 16 silicon modules, equipped with the SKIROC2 ASIC~\cite{bib.Skiroc2}. Despite the limited number of silicon modules, encouraging results were achieved in terms of energy resolution, and there was good agreement with a Geant4 simulation of the detector~\cite{bib.hgcal-paper}. In addition, during a separate beam test in 2016, the timing performance of sets of non-irradiated and irradiated silicon diodes were evaluated. Their measured timing resolution was about tens of picoseconds~\cite{bib.hgcal-paper}.

In October 2018, a two-week beam test, at the H2 beam line of the CERN Super Proton Synchrotron (SPS), was conducted with a calorimeter built with 94 prototype silicon modules, that were equipped with a new version of the readout ASIC, the SKIROC2-CMS~\cite{bib.Skiroc2cms}. The response of the calorimeter was measured with beams of charged hadrons, electrons and muons that had momenta from 20 to 300 GeV/c.

This paper describes the construction and commissioning of CMS CE prototype silicon modules and their assembly into the prototype calorimeter. Section~\ref{sec:modules} describes the silicon module components, and their assembly and testing. Section~\ref{sec:setup} describes the setup used during the 2018 beam test. Section~\ref{sec:performances} shows the performance of the prototype modules, and in section~\ref{sec:calibrations} the calibration procedures used, including channel-to-channel response equalization, and the gain linearization are presented.

\section{Module components, construction and testing}
\label{sec:modules}

The building block of the CE is the silicon module. It consists of a baseplate for mechanical support, a silicon sensor, and a printed circuit board (PCB) with embedded electronics. The construction procedure for prototype modules used in the 2016 beam tests is documented in~\cite{bib.hgcal-paper}. In 2018, the semi-automated module assemlby process was demonstrated with the construction of 94 modules equipped with 6-inch silicon sensor, reaching the targeted production rate of 6 modules per day. The limiting factor was the time for glue curing at room temperature. The assembly and testing procedures, established at the University of California, Santa Barabara (UCSB) pilot centre, were tested and improved during the prototype module production. 

The CE detector will have approximately 30,000 silicon modules built with 8-inch silicon sensors. For their assembly, it is planned that there will be up to six module assembly centres (MACs). The planned production rate is 24 modules per day at each MAC during the construction phase. The assembly and testing procedures developed for the 6-inch modules are now being applied to the 8-inch modules, which is the baseline design of the CE detector~\cite{bib.cms-tdr}.

\subsection{Module construction}
\label{subsec:modules}

\begin{figure}[htbp]
 \centering
  \includegraphics[width=.7\linewidth]{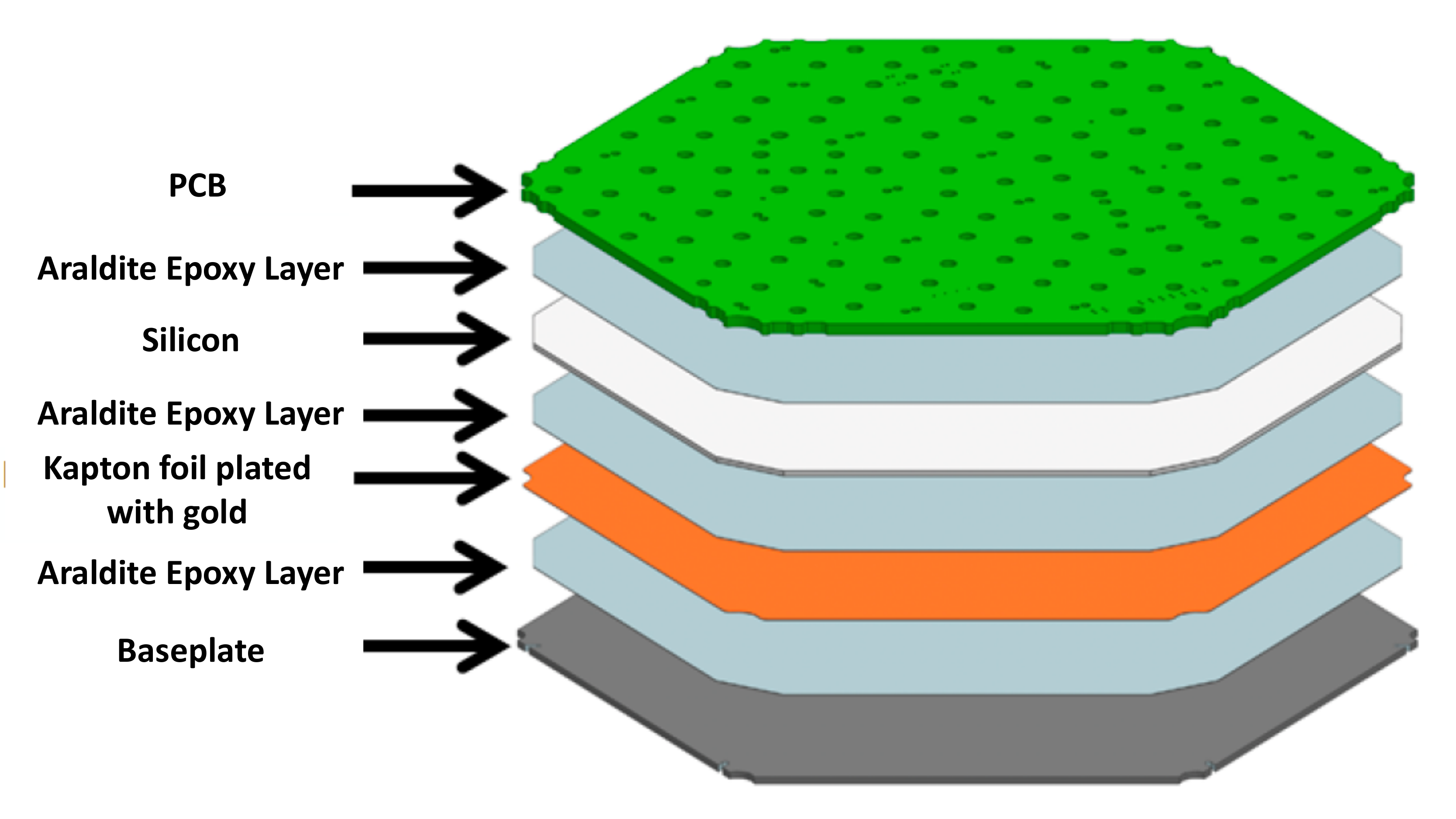}
  \caption{\label{fig:modulestack} The module components are epoxied with Araldite\textregistered 2011 to form a stack: a baseplate at the bottom, a Kapton\texttrademark~foil with gold layer on top, a silicon sensor, and a PCB.}
\end{figure}

\begin{figure}[htbp]
  \centering
  \includegraphics[width=.7\linewidth]{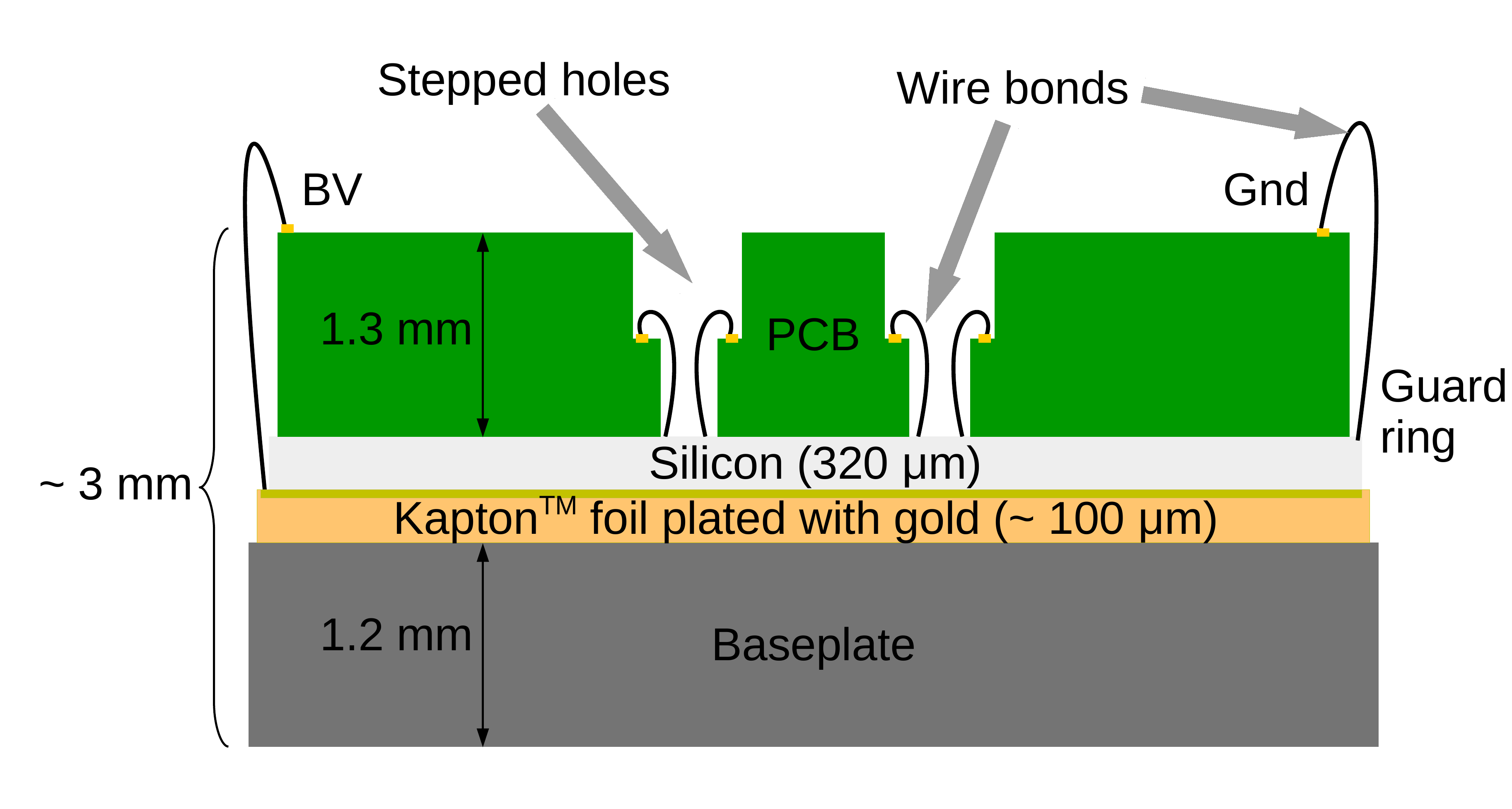}
  \caption{\label{fig:modulescheme} The side-view schematic of a module showing stack layers and wire bonds.}
\end{figure}

The construction of a silicon module is shown schematicaly in Figure~\ref{fig:modulestack}. They consist of: a baseplate in copper or copper-tungsten, a 100~$\upmu m$ thick gold-plated Kapton\texttrademark~sheet, a hexagonal silicon sensor, and a printed circuit board, called `hexaboard', holding four readout ASICs. 
\begin{figure}[htbp]
  \centering
  \includegraphics[width=.7\linewidth]{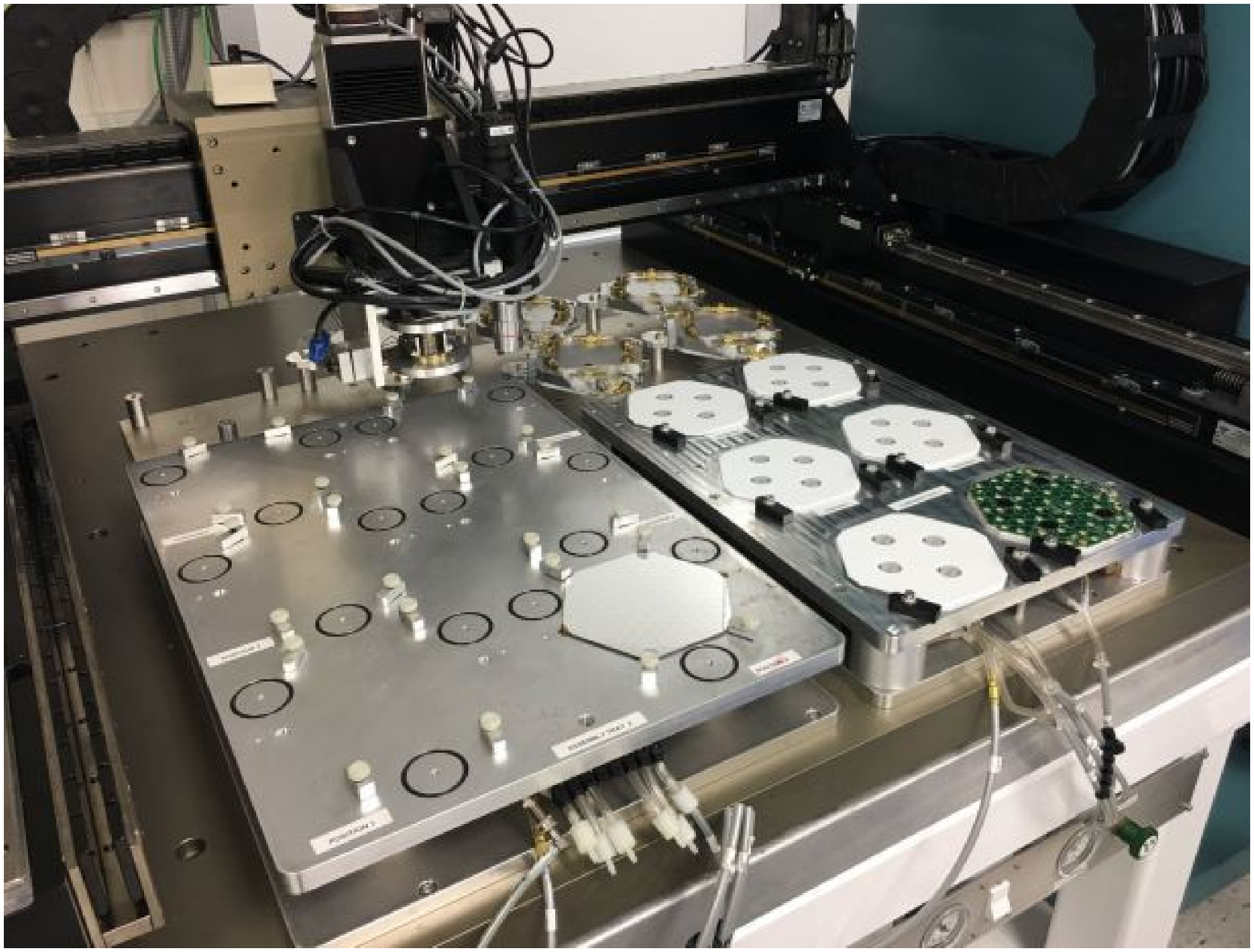}
  \caption{\label{fig:gantry} The UCSB gantry workspace for 6-inch module assembly capable of assembling six 6-inch modules at a time.}
\end{figure}
The baseplate provides mechanical support and thermal conductivity between the module and the cooling layer, as described in Section~\ref{sec:setup}. The baseplate flatness is within 100 $\upmu$m, and the thickness tolerance is less than 30 $\upmu$m. Two different baseplate materials are used: copper for CE-H and copper-tungsten (25$\%$ Cu and 75$\%$ W) for CE-E prototypes. In the CE-E, the denser baseplate is used to increase compactness of the calorimeter. The gold plated Kapton\texttrademark~sheet, epoxied onto the baseplate, serves two functions: it insulates the silicon sensor from the baseplate, and provides a bias connection to the back side of the silicon sensor via the gold plating. The silicon sensor is glued to the gold layer of the Kapton\texttrademark~sheet with a silver epoxy to provide the electrical connection. The PCB glued to the silicon is 1.3 mm thick. It holds four SKIROC2-CMS ASICs~\cite{bib.Skiroc2cms} and contains stepped holes where wire-bonds are attached. The wire bonds provide the electrical connection between the silicon sensor cells and the PCB. They are also used to connect the ground pads (Gnd) on the PCB to the silicon sensor guard ring, and the sensor bias voltage (BV) on the PCB to the gold layer of the Kapton sheet, as shown in Figure~\ref{fig:modulescheme}.

A robotic gantry (shown in Figure~\ref{fig:gantry}) equipped with custom tooling for precision pick-and-place and epoxy dispensing tasks was used in the assembly and a placement precision of $\pm$30~$\upmu$m was achieved.

\begin{figure}[htbp]
  \includegraphics[width=.5\linewidth]{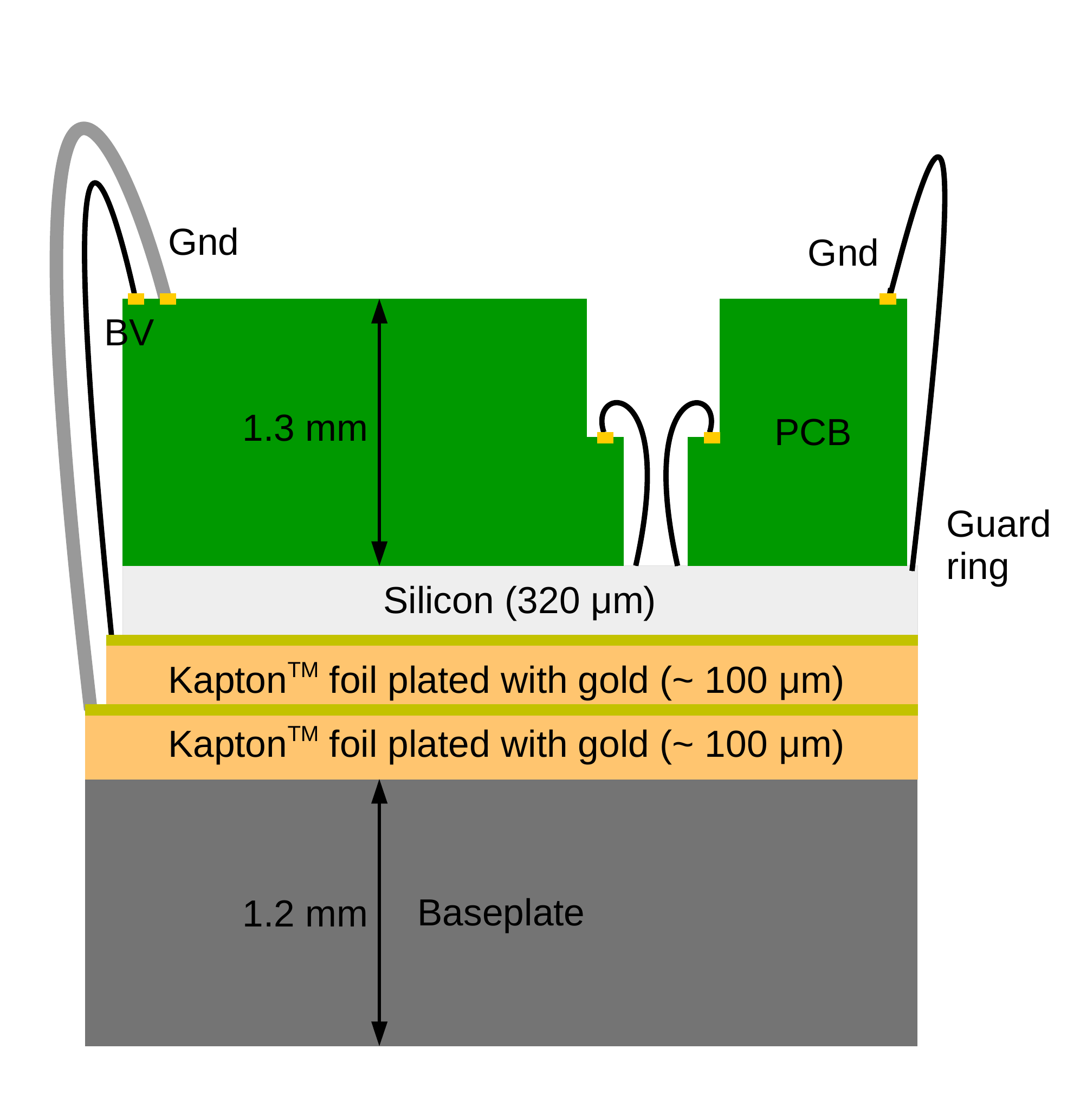}
  \includegraphics[width=.5\linewidth]{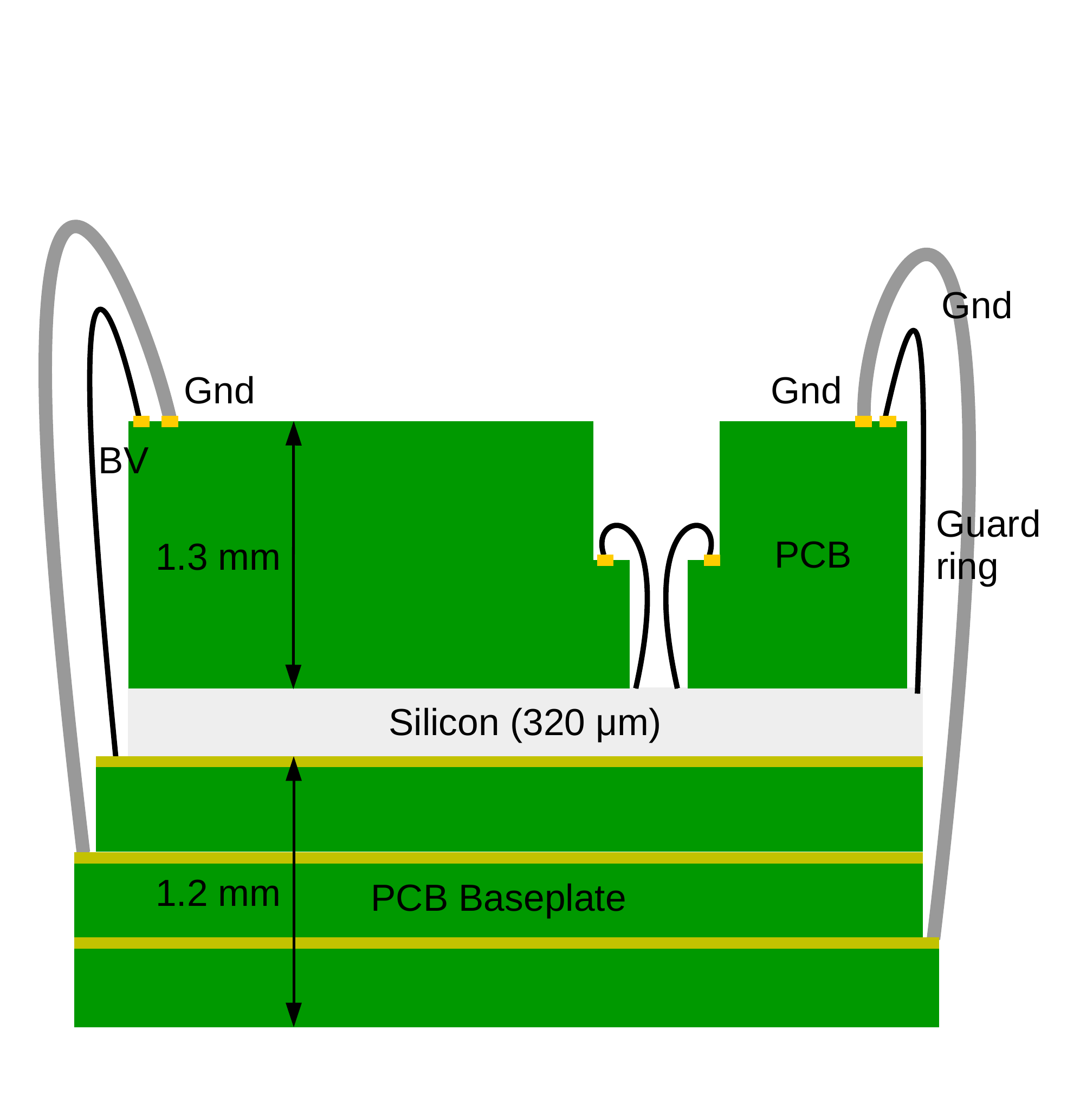}
  \caption{\label{fig:dk_pcb} The side-view schematic of a double-Kapton\texttrademark~foil module showing stack layers and wire bonds (left). The side-view schematic of a PCB baseplate module showing stack layers and wire bonds (right).}
\end{figure}

Several modules of the CE-H were constructed differently to explore two different grounding schemes. The first, called `double Kapton\texttrademark'~is shown in Figure~\ref{fig:dk_pcb} (left). It contains a second gold plated Kapton\texttrademark~sheet epoxied to the first one. 
The top sheet is connected by wire bonds to the bias voltage pads on the PCB, while the bottom is connected to a ground pad on the PCB with a soldered wire. The second grounding scheme, called `PCB baseplate' is shown in Figure~\ref{fig:dk_pcb} (right). For this configuration, the baseplate was replaced by a PCB with two ground planes. It was found that the optimal performance in terms of noise is achieved by connecting the middle layer of the PCB baseplate to a ground pad of the hexaboard. In this configuration, the bottom layer of the PCB baseplate is also connected for redundancy as shown in Figure~\ref{fig:dk_pcb}. The noise performance with different grounding schemes is discussed in Section~\ref{sec:pedestal-noise}.

\subsection{Silicon sensors}
\label{subsec:sensor}

\begin{figure}[htbp]
  \centering
  \includegraphics[width=.7\linewidth]{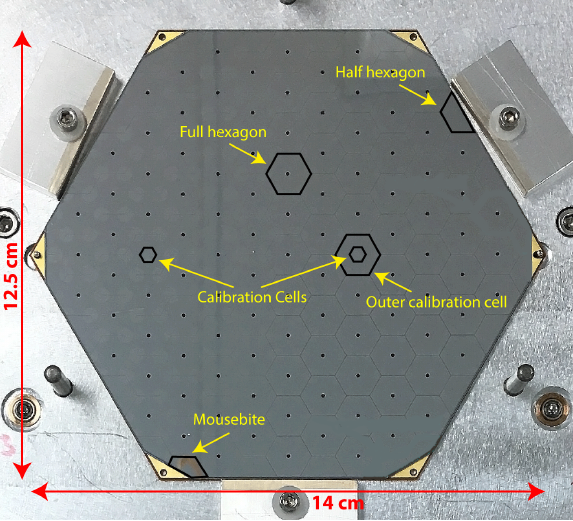}
  \caption{\label{fig:sensorpic} Picture of one 6-inch silicon sensor where various cell types are highlighted. }
\end{figure}

Silicon sensors with 135 cells on a 6-inch wafer were produced by HPK\footnote{Hamamatsu Photonics K.K.}. A picture of one silicon sensor is shown in Figure~\ref{fig:sensorpic}.
All sensors were made with float-zone p-on-n silicon wafers. Ninety of the sensors were made with a 300~$\upmu$m thick depletion zone and for four sensors it was 200~$\upmu$m. The physical thickness of all the silicon sensors is 320~$\upmu$m. The majority of the cells (107/135) on a sensor are hexagonal with an area of 1.1~$\text{cm}^2$. Two hexagonal cells are divided into two parts: an `inner calibration' cell, having an area of about 1/9th of the area of the full hexagonal cell, and the surrounding `outer calibration' cell. The former facilitates calibration with single minimum ionizing particles (MIPs) after irradiation, when the signal over noise ratio from a full cell might be too small to detect single MIPs efficiently. The smaller cell has a smaller capacitance, which reduces the intrinsic noise making the MIP signal easier to detect. A small increase in signal size was also observed for these cells, as discussed in Section~\ref{subsec:mip}. The sensors also have half-hexagonal cells at their edges and odd-shaped or `mousebitten' cells at the corners.

\begin{figure}[!ht]
	\centering
	\includegraphics[width=0.48\textwidth]{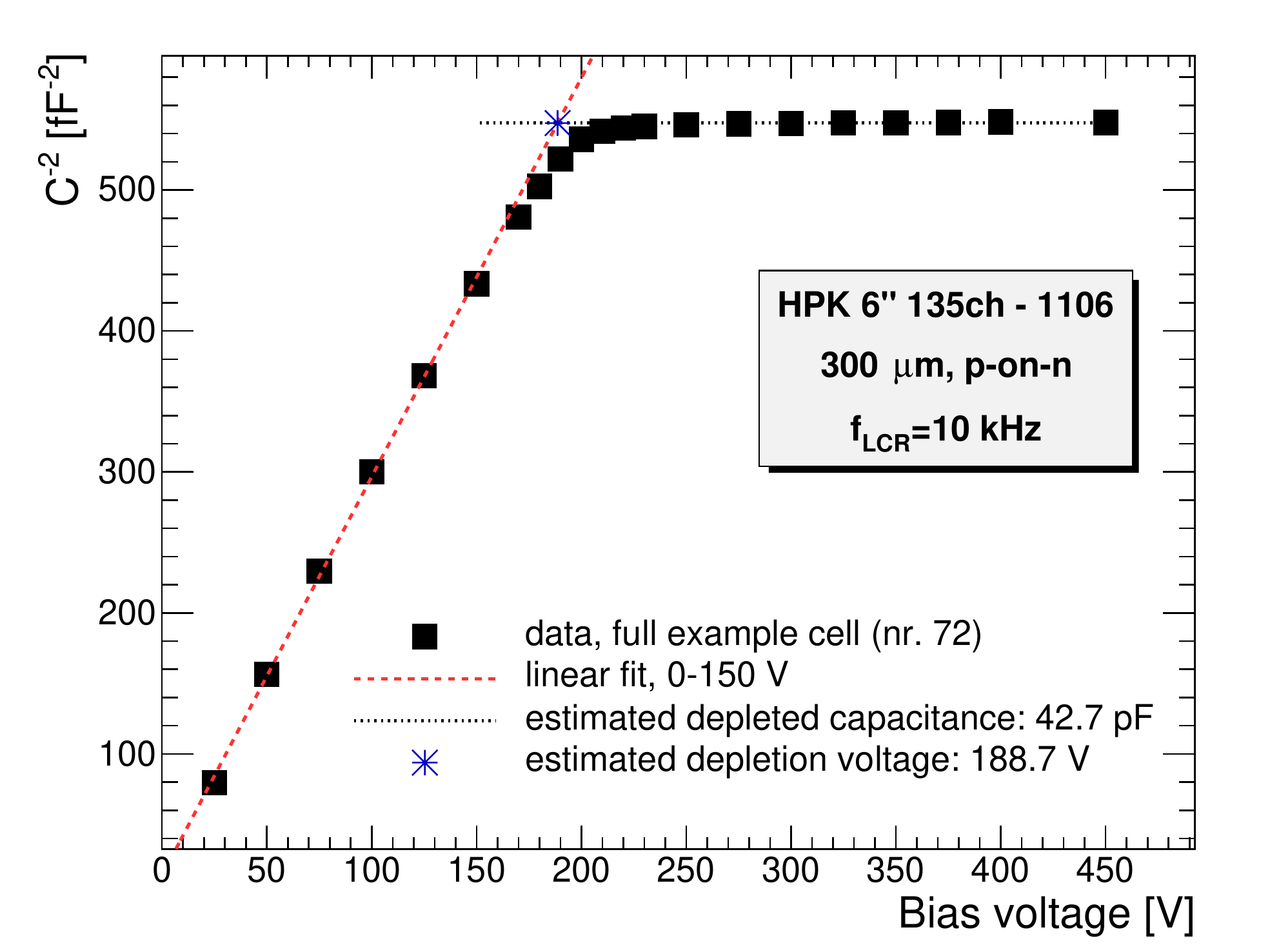}
	\includegraphics[width=0.48\textwidth]{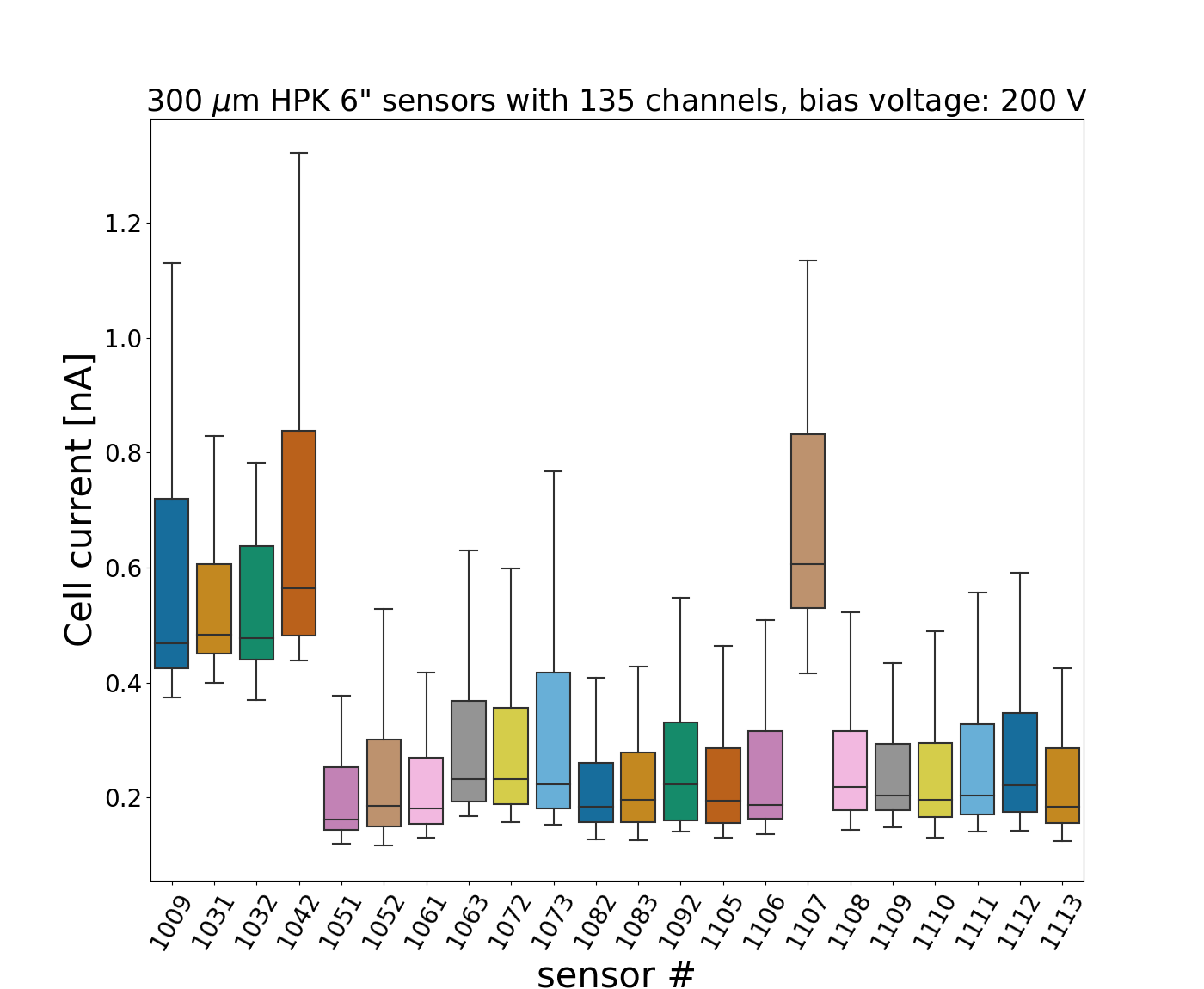}
	\caption{The depletion-voltage estimation for a cell on an example 300~$\upmu$m thick  sensor (left). The cell leakage current at a bias voltage of 200~V for a subset of the silicon prototype sensors characterized for October 2018 beam test (right). The line inside the boxes indicates the leakage current median, the box indicates the interquartile range, and the whiskers indicate the 5th and the 95th percentiles. Only full hexagonal cells were included.}
	\label{fig:cv_c2v}
\end{figure}

A probe-card-based system~\cite{bib.cv-paper} was used to measure the leakage-current and capacitance at biases up to 1000~V before the sensors were assembled into modules. Figure~\ref{fig:cv_c2v} (left) displays the capacitance measurement for a single cell, where the full depletion was reached at 189 V, and the right plots shows the leakage current measurements at 200 V for a selection of the silicon sensors. The average full cell leakage current was less than 1 nA for nearly all the sensors.

\subsection{Front-end electronics}
\label{subsec:felectronic}

The readout PCB is epoxied onto the silicon sensor and connected to the silicon cells with wire bonds through stepped holes in the PCB. The hexaboard holds four SKIROC2-CMS ASICs and a MAX\textregistered10 FPGA. The SKIROC2-CMS ASIC is derived from the SKIROC2~\cite{bib.Skiroc2}  ASIC, developed for the CALICE Si-W ECAL~\cite{bib.siwecal}. The FPGA receives the clock, trigger signals and fast commands from the back-end data acquisition (DAQ) boards, and aggregates and transmits the data from the SKIROC2-CMS ASICs to the DAQ boards. The details of the DAQ system developed for the beam tests are described in~\cite{bib.h1-hgc}. 

The SKIROC2-CMS ASIC was designed to measure signals ranging from a few fC to 10 pC with a low power consumption, 10 mW per channel.
\begin{figure}[htp]
  \centering
  \includegraphics[width=.9\linewidth]{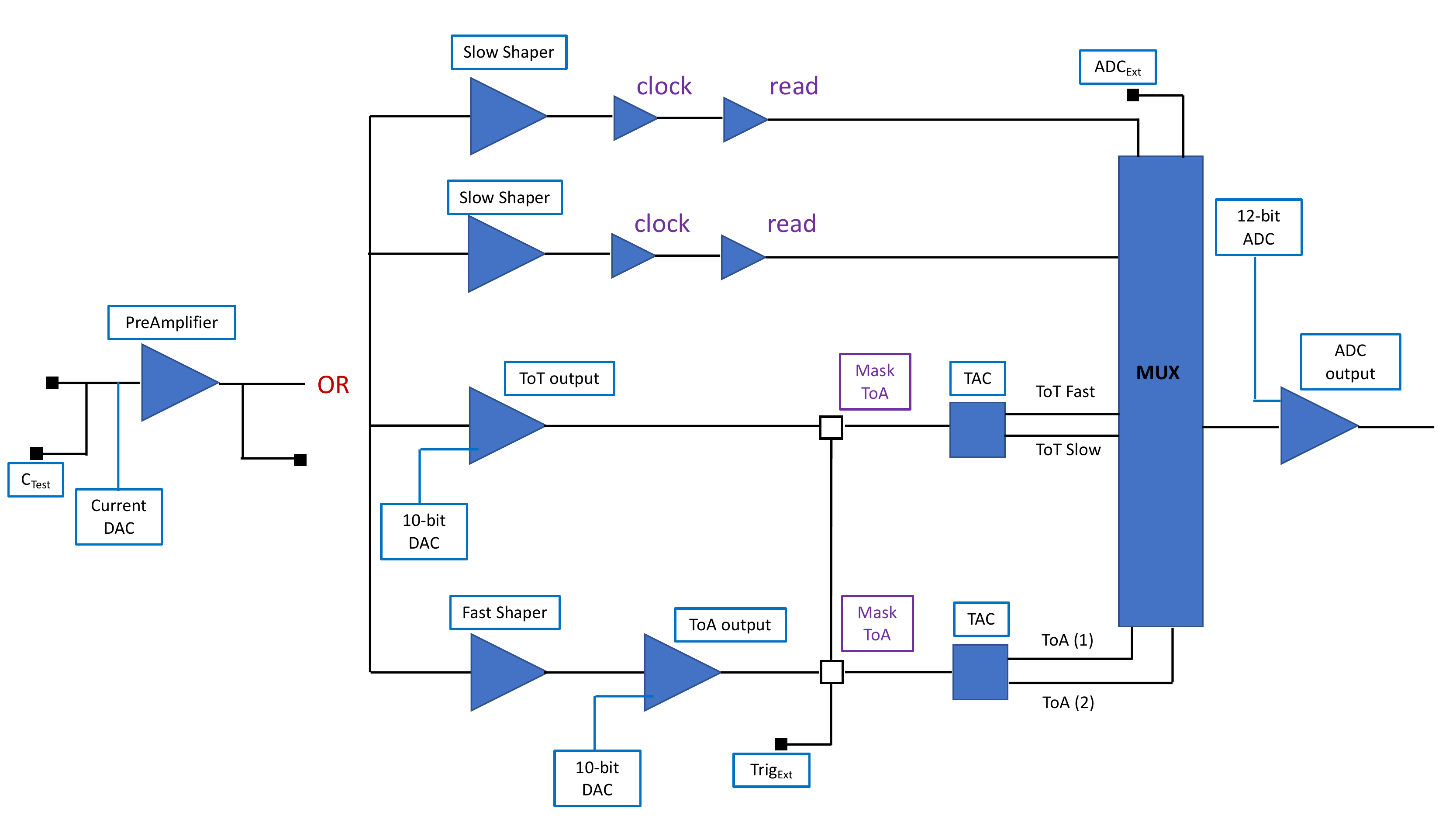}
  \caption{\label{fig:skiroc2cms} The simplified schematic diagram of the SKIROC2-CMS ASIC. Each preamplifier is followed by two slow shapers for the charge measurement and a ToT circuit, and by a fast shaper for timing measurement. The slow shapers consists of a low and a high-gain components to handle the large dynamic range. The shaping time can be tuned from 10 to 70~ns with a 4-bit slow-control parameter. The shaping time of the fast shaper can be adjusted between 2 to 5 ns.}
\end{figure}
Figure~\ref{fig:skiroc2cms} shows a simplified schematic diagram of the ASIC. It has 64 channels each  with a low-noise pre-amplifier followed by high- and low-gain amplifiers, and slow shapers with a shaping time of 40 ns. In order to minimize signal path lengths from the silicon sensor to the ASIC, the hexaboard has four ASICs, each with only 32 active channels used to readout, in total, 128 silicon cells out of the 135 (one cell in each corner of the sensor and one calibration cell are not connected).

The signals from each silicon cell are sampled every 25 ns and stored in a 13-deep analog memory. When a trigger signal is received, the 13 samples are digitized with 12-bit analog-to-digital converter (ADC) and a bipolar waveform is obtained. For signals above 600 fC, corresponding to about 200 MIPs crossing a 300 $\upmu m $ thick silicon cell, a time-over-threshold measurement is used to estimate the signal's amplitude. The full dynamic range is achieved by combining the two outputs of the ADCs and the ToT.

The time-of-arrival is obtained from a fast shaper placed after the pre-amplifier and followed by a discriminator and an internal time-to-digital converter with a 25~ps time bin. The expected time resolution for signals above a few hundred MIPs is about 50~ps. Details of the SKIROC2-CMS ASIC can be found in~\cite{bib.Skiroc2cms}.

\subsection{Module testing}
\label{subsec:module-test}

Two quality control tests were performed during the module construction: the current as a function of the applied voltage (IV) was measured to assure proper module biasing, and the front-end electronics was tested. These tests are described in this section.

\subsubsection{IV tests}
The leakage current as a function of the applied voltage was measured in the prototype modules at the module assembly centre in UCSB and upon reception at CERN. For these 94 modules, a Keithley 2410 source meter was used to bias the silicon sensor and record its total leakage current. The bias voltage was scanned between 0 and 300~V in steps of 10 V. 
\begin{figure}[htp]
  \centering
  \includegraphics[width=1.0\linewidth]{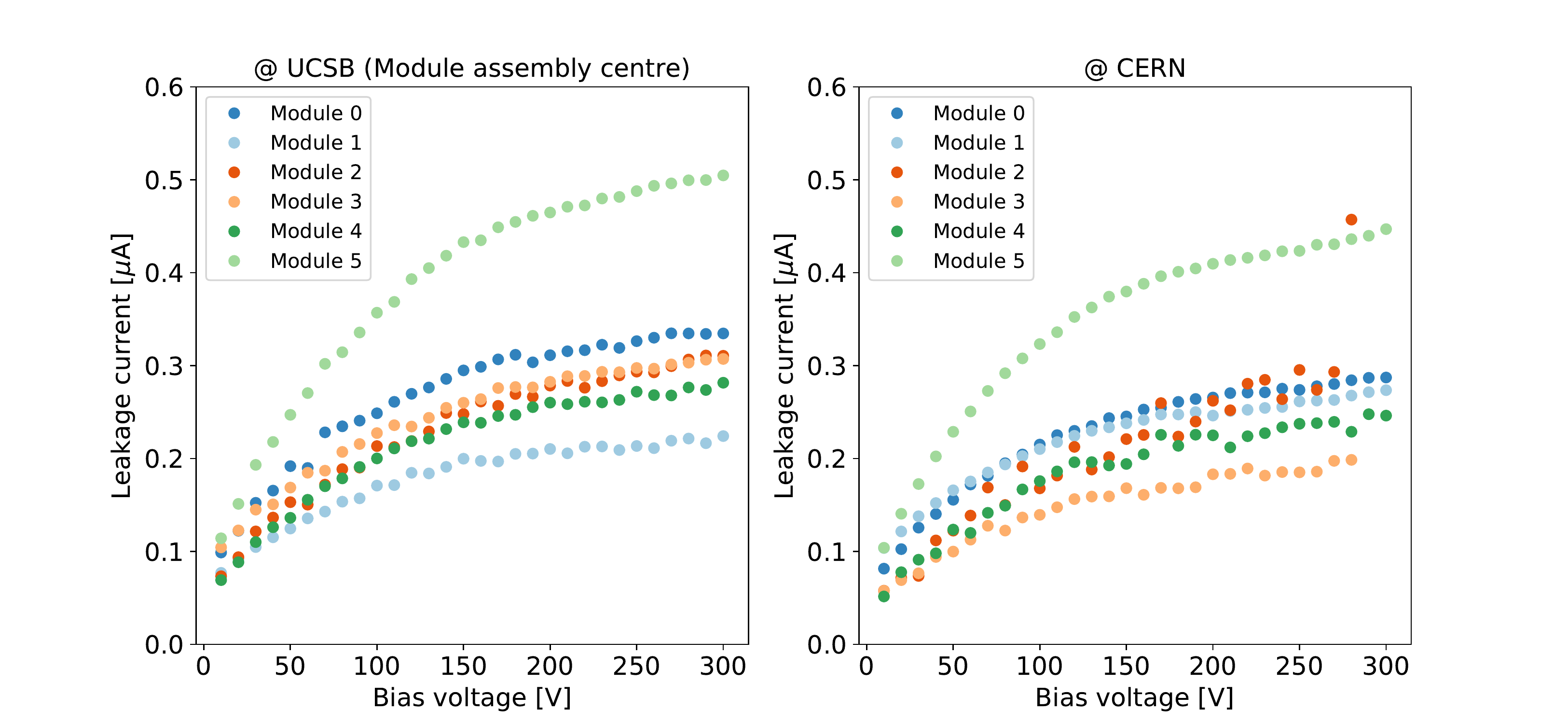}
  \caption{\label{fig:iv-curves} The IV curves for a set of 6 modules tested at UCSB after assembly (left) and at CERN after reception (right).}
\end{figure}

\begin{figure}[htp]
  \centering
  \includegraphics[width=.7\linewidth]{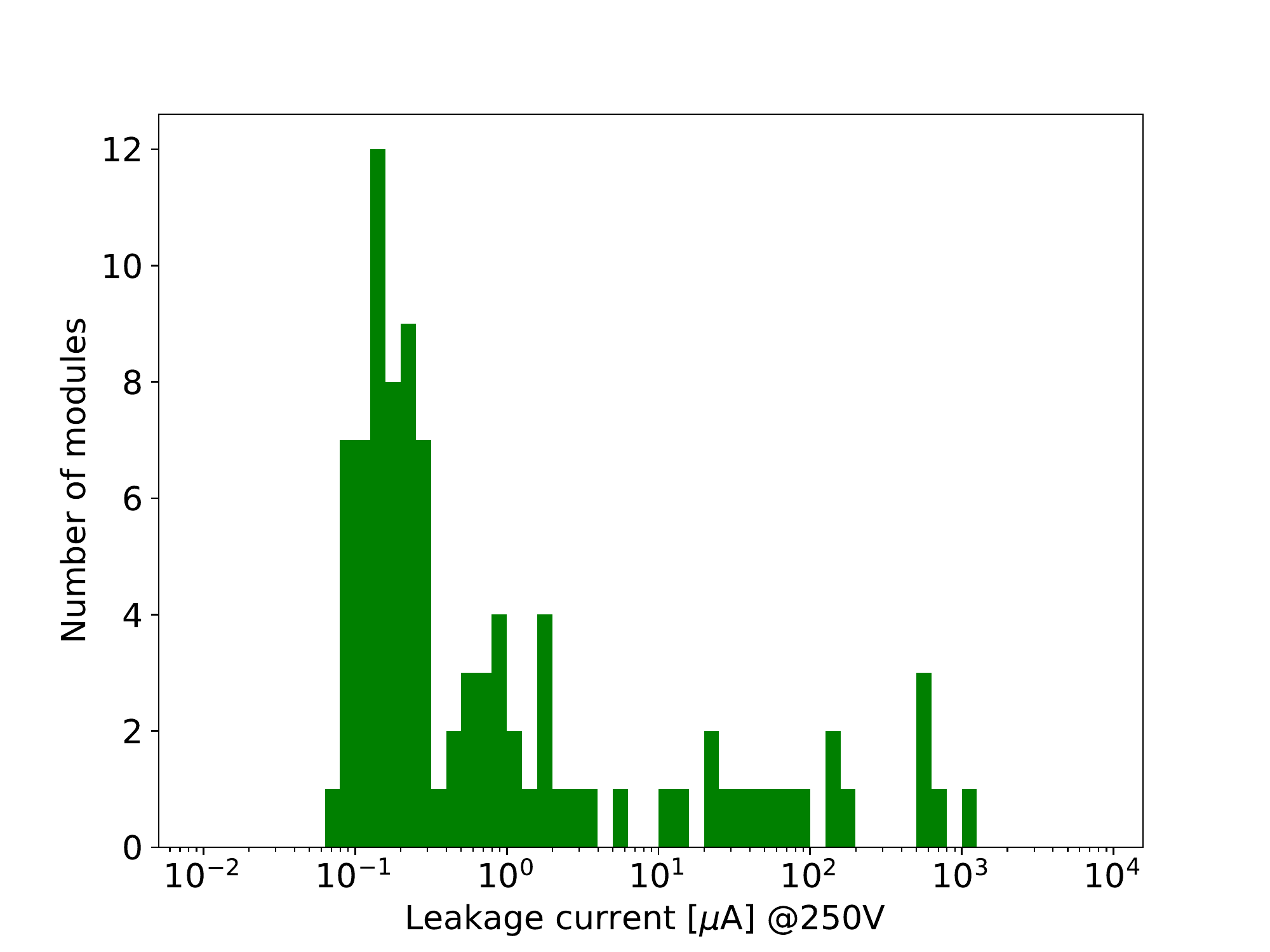}
  \caption{\label{fig:I250} The total leakage current distribution of the 94 prototype modules at 250 V.}
\end{figure}

Figure~\ref{fig:iv-curves} shows the IV measurement for the same 6 modules at UCSB and CERN. The differences between them were small and could be attributed to differences in the environments of the testing stations. Module 2 showed breakdown at a lower voltage which may indicate potential sensor damage during shipment from UCSB to CERN. Figure~\ref{fig:I250} shows the leakage current distribution at 250~V for each module. About 75$\%$ of the modules had leakage currents below 1~$\upmu$A. However, eight modules had leakage currents in excess of 100~$\upmu$A, which may indicate that the sensors were damaged during module construction, shipment or handling. Among these eight modules, six of them were from an early production after which the module assembly, shipment, packaging and the module handling procedures were improved and standardized. 

\subsubsection{Tests on the front-end electronics}
\label{subsubsec:module-test-electronic}

\begin{figure}[htp]
  \centering
  \includegraphics[width=.8\linewidth]{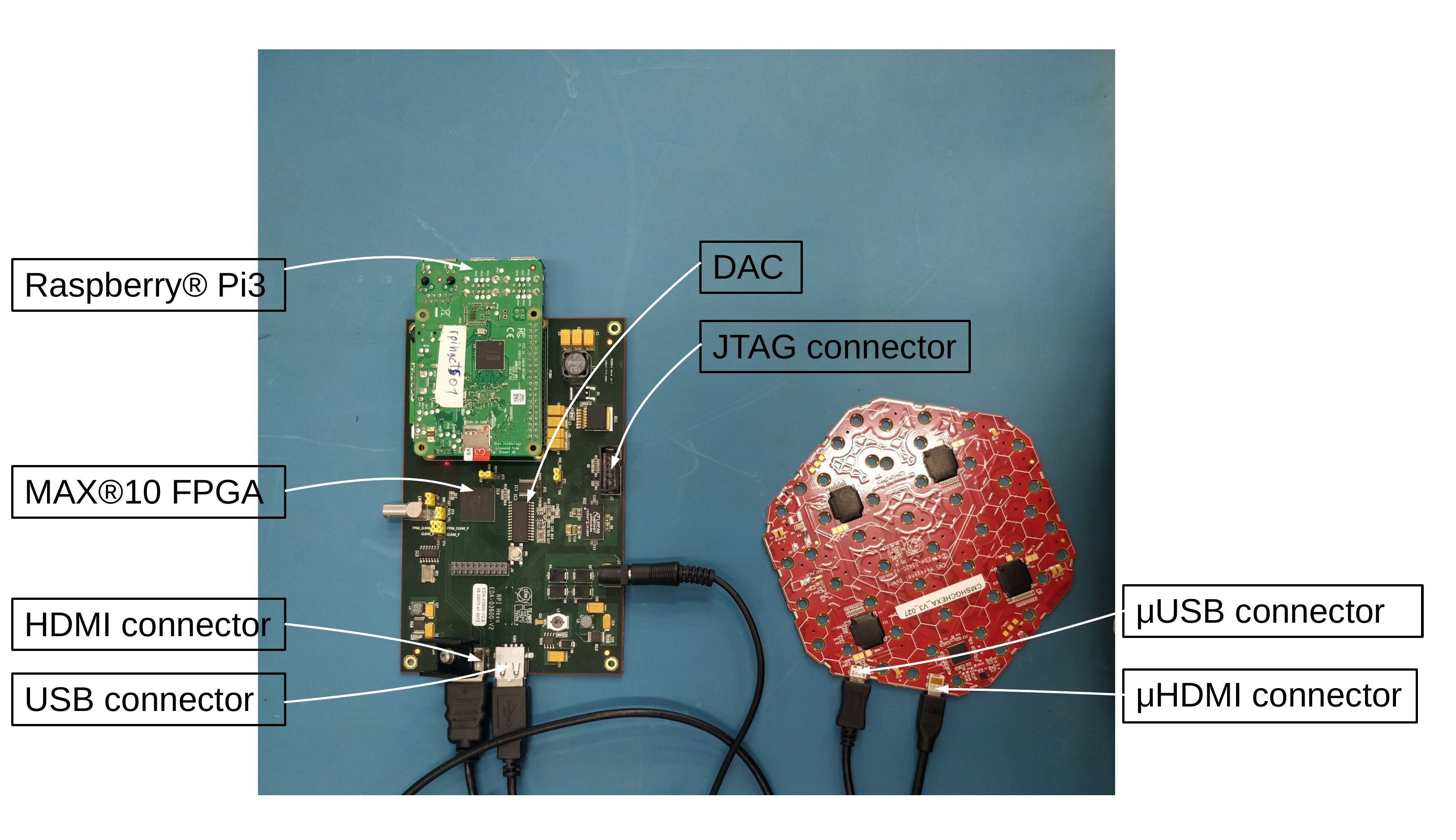}
  \caption{\label{fig:rpi-teststand} A photograph of the test-board connected to a hexaboard. The USB and HDMI connectors are connected respectively to the $\upmu$USB and the $\upmu$HDMI connectors of the hexaboard.}
\end{figure}

The ASICs and hexaboards were also tested at different stages of the module construction. A custom test-board, equipped with Raspberry\textregistered Pi3, shown in Figure~\ref{fig:rpi-teststand}, was used to perform these tests. The functions of the board were to program the MAX\textregistered10 FPGA of the hexaboards, to provide the low voltage, clock and trigger signal to the hexaboard, to configure the front-end ASICs, and to acquire the data.

The core of the test-board is a MAX\textregistered 10 FPGA, which communicates with an on-board Raspberry\textregistered Pi3 and the MAX\textregistered 10 of the hexaboard. The interface with the Raspberry\textregistered Pi3 is made via the general purpose input/output (GPIO) pins. In addition, the test-board contains a 12-bit Digital-to-Analog converter (DAC) to generate a calibration-pulse, and a 40 MHz quartz oscillator to generate the clocks. The communication between the test-board and the hexaboard is done through USB and HDMI connectors. The USB connection carries the JTAG signals to program the hexaboard MAX\textregistered 10, while the HDMI carries the low voltage to the hexaboard, the clocks (160 and 40~MHz), the configuration and fast commands, and the calibration pulses from the DAC. The HDMI is also used to transmit the readout data.

The test-board was used to test the hexaboards (without any silicon sensor) and the silicon modules. After visual inspection, the firmware was loaded onto the MAX\textregistered10 FPGA via the test-board. Pedestal data were acquired and checked following the configuration of the SKIROC2-CMS. About 20$\%$ of the hexaboards failed these tests and were discarded. The main two reasons for these failures were a failure in programming the MAX\textregistered10 FPGA, or a failure during the configuration of at least one of the four ASICs\footnote{Note that the ASICs were not tested prior to assembly on the hexaboards.}. The pedestal values were between 200 and 250 ADC counts, as expected, for the remaining 80$\%$ of the hexaboards. For most of the channels, the total noise was below 5 ADC counts for the high gain\footnote{A typical MIP signal is about 40 ADC counts in high gain.} and below 2 ADC counts for the low gain.

\begin{figure}[htp]
  \centering
  \includegraphics[width=.49\linewidth]{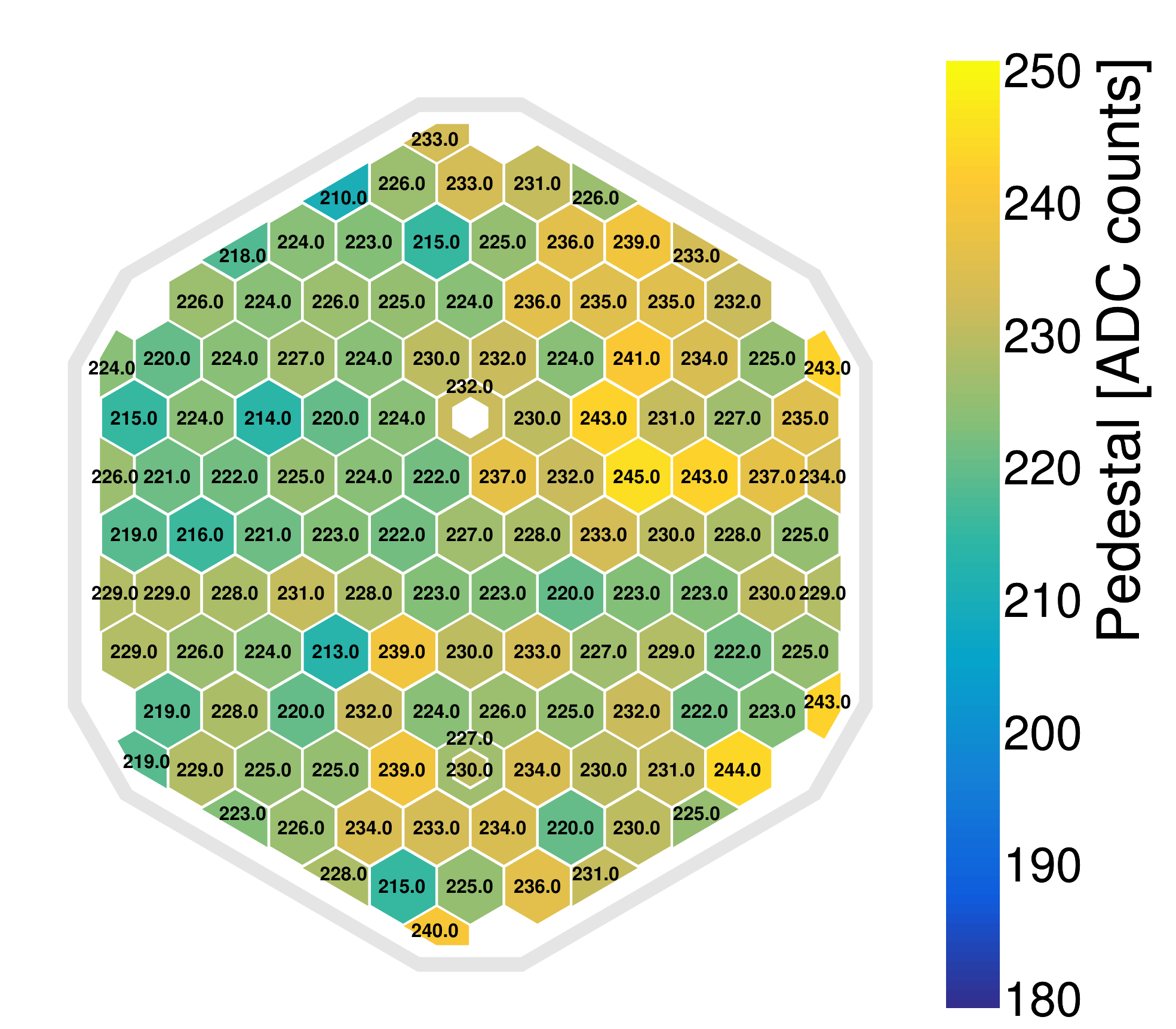}
  \includegraphics[width=.49\linewidth]{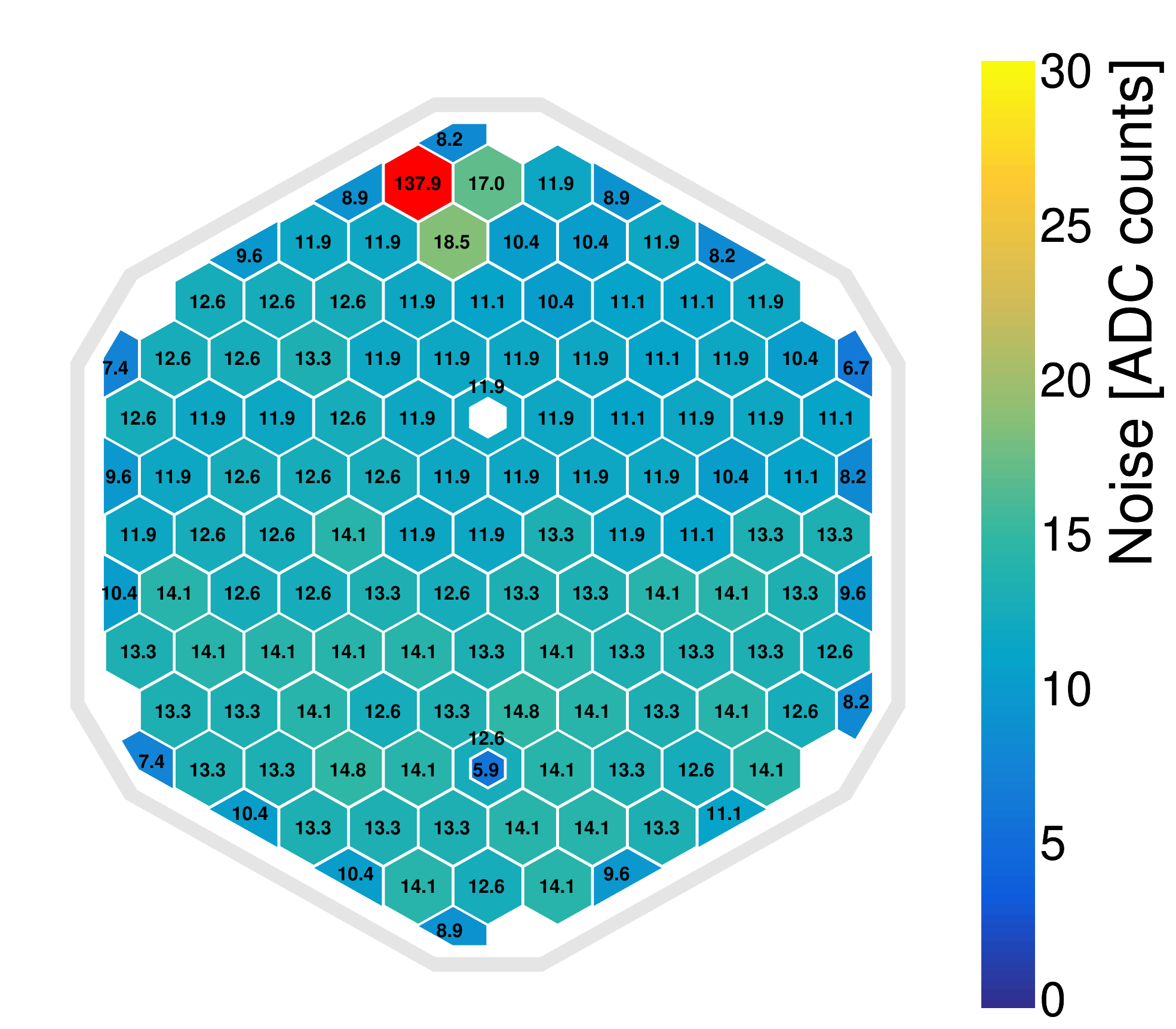}
  \caption{\label{fig:pedestal-noise} Example maps of pedestal (left) and noise (right) values for the high-gain shaper of a prototype module. One of the two calibration cells and one of the two mousebitten silicon cells, in each corner, are not connected to a front-end ASIC channel. These channels are shown in white.}
\end{figure}

Pedestal data were also acquired after modules were assembled. Figure~\ref{fig:pedestal-noise} shows maps of pedestal (left) and noise (right) values for the high-gain ADC of a prototype module. The typical values of the noise were found to be between 8 to 20 ADC counts in high gain, larger than the noise of bare hexaboards, due to the silicon capacitance and, possibly, external pickup. The details of the noise studies are discussed in Section~\ref{sec:pedestal-noise}.

One channel, shown in red on the right plot of Figure~\ref{fig:pedestal-noise}, was found to be extremely noisy (>100 ADC counts for the high gain) in all modules. This was the channel connected to the silicon cell located below the $\upmu$HDMI connector. This large noise was likely caused by the digital signals through nearby vias of the PCB. This channel was masked in all prototype modules during the beam test operation.

\section{Final beam test setup}
\label{sec:setup}
\begin{figure}[!ht]
  \centering
  \includegraphics[width=.8\linewidth]{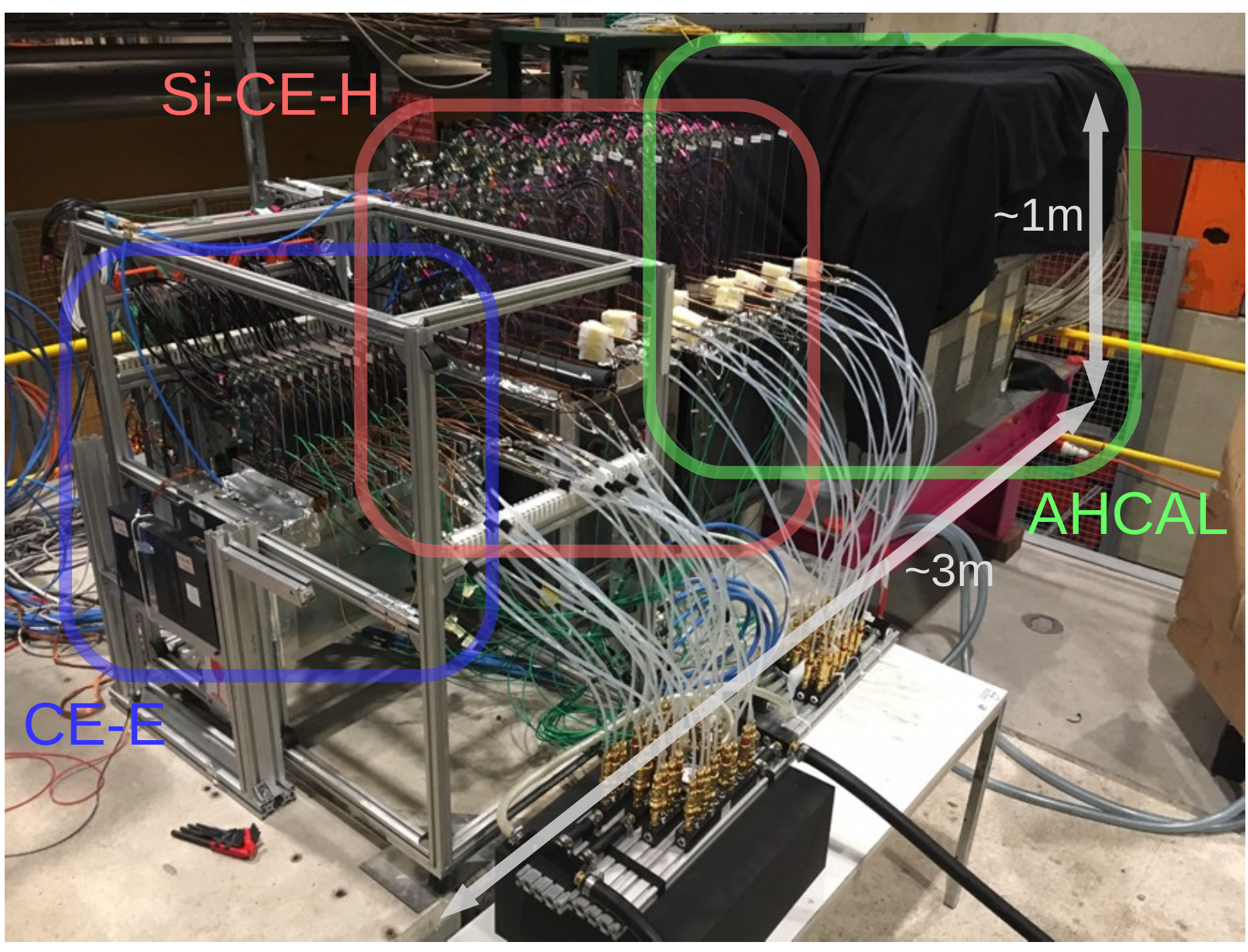}
  \caption{\label{fig:set-up-photo} A picture of the October 2018 beam test setup in the CERN SPS H2 beam zone with the CMS CE prototype followed by the CALICE AHCAL prototype.}
\end{figure}

For the October 2018 beam test, the calorimeter prototype was installed in the CERN SPS H2 beam line on a concrete platform. The prototype comprised a stack of 28 layers of single modules mounted on copper cooling plates for the electromagnetic section. Downstream, there was the CE-H prototype with 9 layers of 7 neighboring modules and 3 layers of one module. The CE-E prototype had about 3500 channels and CE-H prototype had about 8500. The full depth of the silicon-based calorimeter corresponded to about 60 radiation lengths ($\Chi_{0}$) and 5 interaction lengths ($\lambda_I$) with lead, copper and steel as main absorbers. Behind this was the CALICE Analog Hadronic Calorimeter (AHCAL) prototype which is a SiPM-on-tile 4~$\lambda_I$ hadron sampling calorimeter with 39 active layers \cite{bib.ahcal}. A picture of the setup in the CERN SPS H2 beam zone is shown in Figure~\ref{fig:set-up-photo}.

\subsection{CERN-SPS H2 beam line instrumentation and trigger system}
\begin{figure}[!ht]
  \centering  \includegraphics[width=1.0\linewidth]{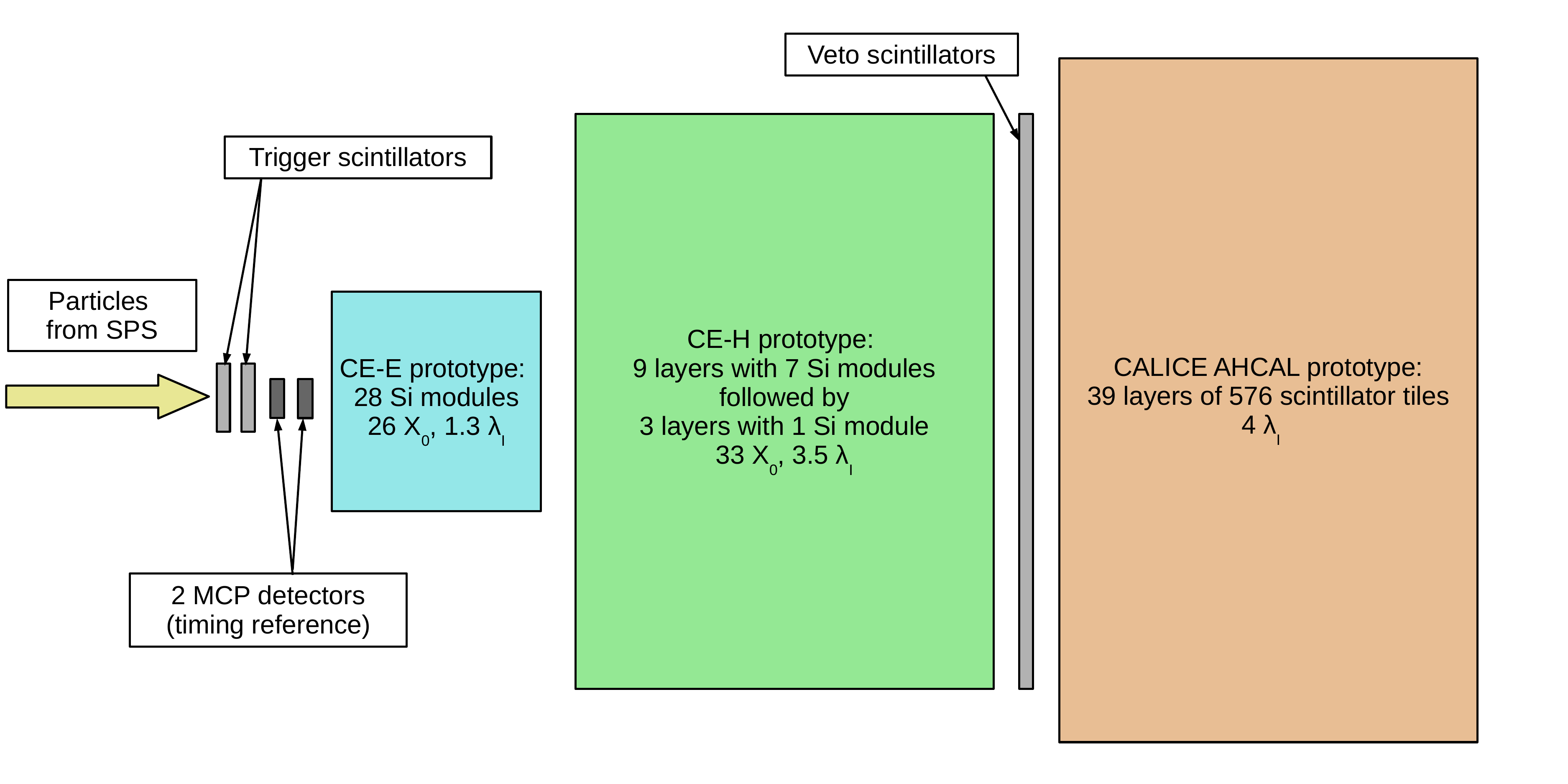}
  \caption{\label{fig:full-set-up} A schematic, not to scale, of the October 2018 beam test setup with the CMS CE prototype followed by the CALICE AHCAL prototype. The four DWCs were upstream of the setup and are not shown in this figure.}
\end{figure}

The H2 beam line has several beam-characterization detectors, located upstream of the CMS CE prototype, to monitor the beam. The detailed description of the beam-characterization detectors can be found in~\cite{bib.h1-hgc}. Four delay wire chambers (DWC)~\cite{bib.dwcs} track the beam particles to determine the transverse particle distribution of the beam and the incident particles position on the prototype. In addition, two plastic scintillator tiles readout by fast photomultipliers were installed in front of the calorimeter to provide a trigger signal. A third plastic scintillator was placed between the silicon CE-H and the AHCAL prototypes and used as a muon and hadronic shower veto during the electron runs. Finally, to provide fast signals for reference timing measurements of the incident particles, two micro-channel plate detectors (MCP) were installed directly upstream of the CE-E. A schematic of the detector arrangement is shown in Figure~\ref{fig:full-set-up}.

\subsection{Silicon electromagnetic calorimeter prototype}
\label{sec:ce-e}
\begin{figure}[!ht]
  \centering
  \includegraphics[width=0.32\linewidth]{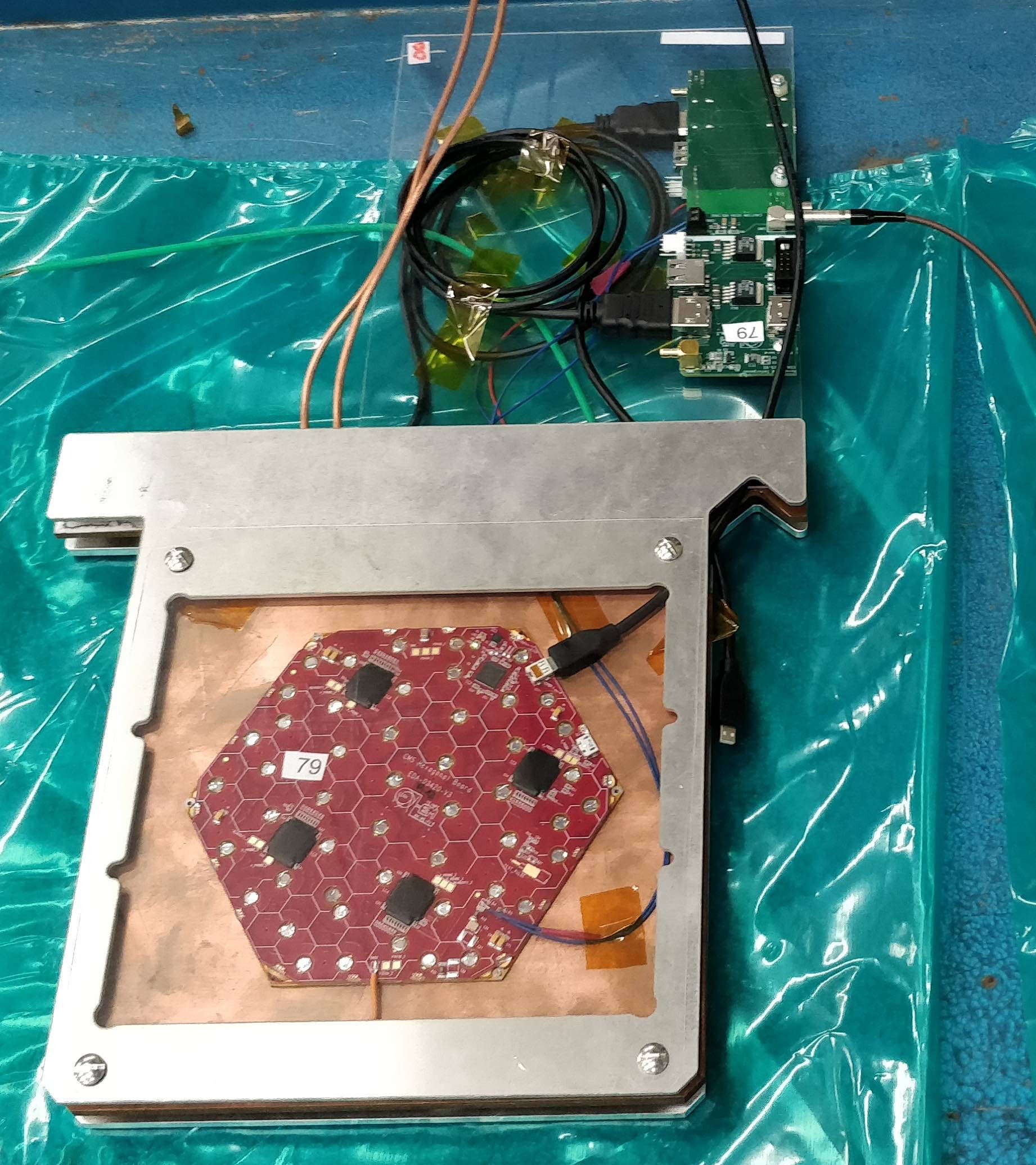}
  \includegraphics[width=0.67\linewidth]{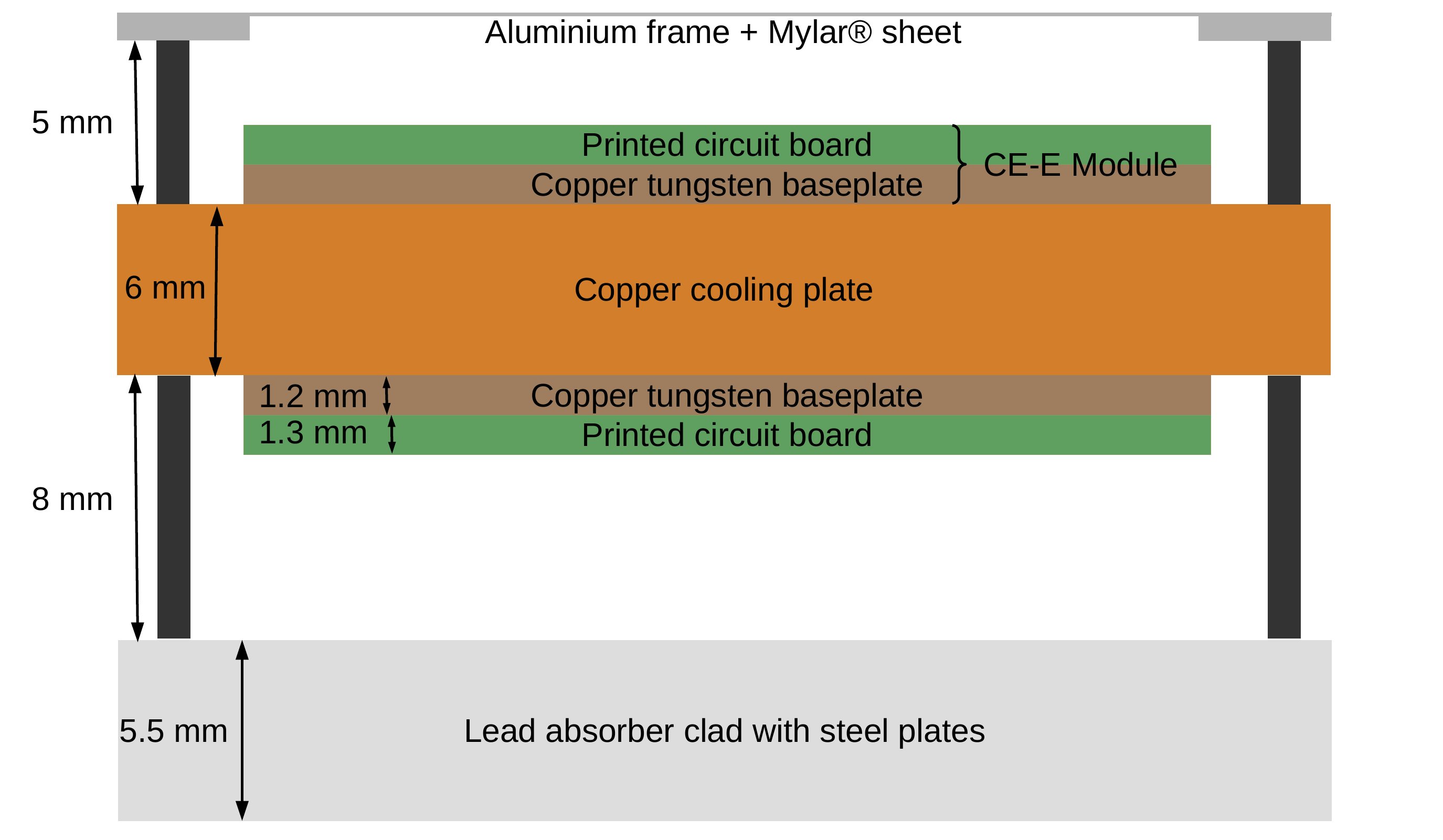}
  \caption{\label{fig:cassettes-layer} Front-view of an example mini-cassette of the CE-E prototype (left). A plate of Makrolon\textregistered~was attached to the mini-cassette to hold an `interposer' board above the cooling plate. Schematic transversal section of a CE-E mini-cassette (right). The aluminium frame was chosen to ensure to have enough space without adding more absorber material and to protect the module. The size of the spacers was driven by the thickness of the $\upmu$HDMI connector and cable on the module.}
\end{figure}

To build the electromagnetic section, pairs of modules were assembled in mini-cassettes. In a mini-cassete, the two modules were mounted on both sides of a 6 mm thick copper cooling plate and connected  to a 4.9 mm thick lead clad with 300~$\upmu m$ thick steel. A photograph of a mini-cassette is shown in Figure~\ref{fig:cassettes-layer}. The mini-cassettes were closed with an aluminium frame and a Mylar\textregistered~sheet. The copper plate had a groove to allow the insertion of a 3~mm-diameter copper pipe through which water at 28$^\circ$C was passed to keep the prototype silicon modules at a constant temperature during the data taking. This temperature was required for the MAX\textregistered10 FPGA on the hexaboard, which could not be operated reliably at a lower temperature. It was also necessary to take data at constant temperature since the ToA and the ToT measurements were temperature sensitive. In characterization studies on the SKIROC2-CMS ASIC, a variation of 5$^\circ$C could lead to an error of 4 to 5$\%$ on the measured charge with the ToT and an error on the reconstructed time with the ToA of between 50~ps to 100~ps.

For each mini-cassette two interposer boards, one per module, were attached with Makrolon\textregistered~plates above the cooling plate as shown in Figure~\ref{fig:cassettes-layer}. The interposer boards regulated the 5~V output from the DAQ boards to the 3.3~V needed by the hexaboards. They also filtered and transmitted the bias voltage coming from the DAQ boards through LEMO\texttrademark~cables to wires soldered to the hexaboards. The mini-cassettes were inserted in a `hanging file' mechanical support system. The total depth of the CE-E prototype corresponded to a total depth of 26~$X_0$, with a physical length of about 50~cm. In comparison, the final CMS CE-E that will have a total depth of 26~$X_0$ for a length of 34~cm.

\begin{figure}[!ht]
  \centering
  \includegraphics[width=0.5\linewidth]{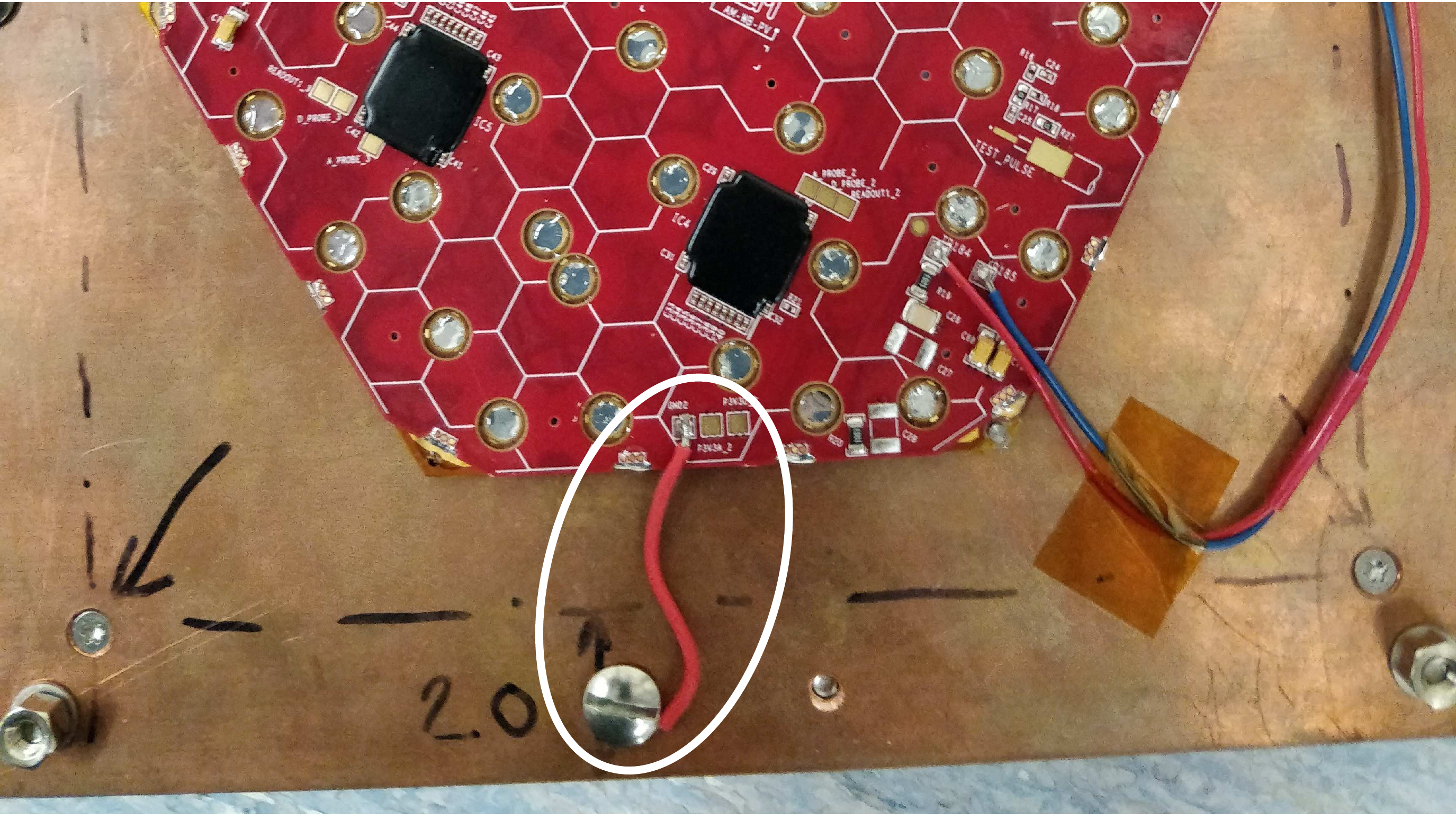}
  \caption{\label{fig:groundEE} An example ground connection used in the CE-E prototype: a short (red) wire was soldered to a ground pad on the hexaboard and screwed to the copper cooling plate.}
\end{figure}

\begin{figure}[!ht]
  \centering
  \includegraphics[width=0.8\linewidth]{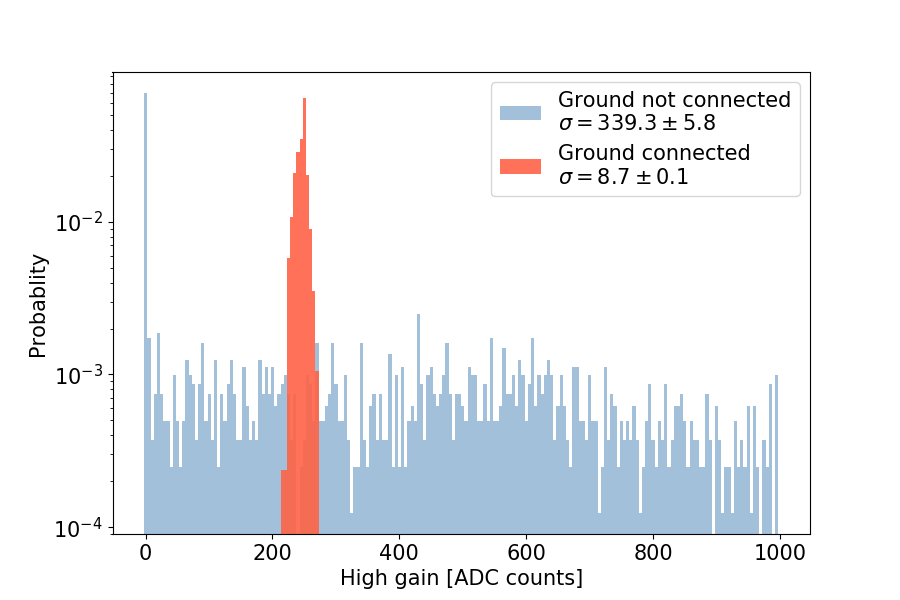}
  \caption{\label{fig:noiseEE} The high-gain pedestal distribution for one channel of a CE-E prototype module with and without a ground connection.}
\end{figure}

All modules exhibited common-mode noise with an amplitude corresponding to a signal of 6-7 MIPs during the initial tests. Several methods were tried to reduce this noise. The optimal method, in terms of noise reduction (and feasibility), was to connect the copper cooling plate with the ground of the hexaboards as shown in Figure~\ref{fig:groundEE}. Figure~\ref{fig:noiseEE} shows the high-gain pedestal distribution of one channel of a CE-E prototype module with and without the copper plate being grounded. Grounding significantly reduced the high gain noise to 1/5 to 1/3 MIPs (8 to 15 ADC counts). After grounding, the low gain noise was about 1/5 MIPs (1-2 ADC counts).

\subsection{Silicon hadronic calorimeter prototype}
\label{sec:ce-h}

\begin{figure}[!ht]
  \centering
  \includegraphics[width=0.8\linewidth]{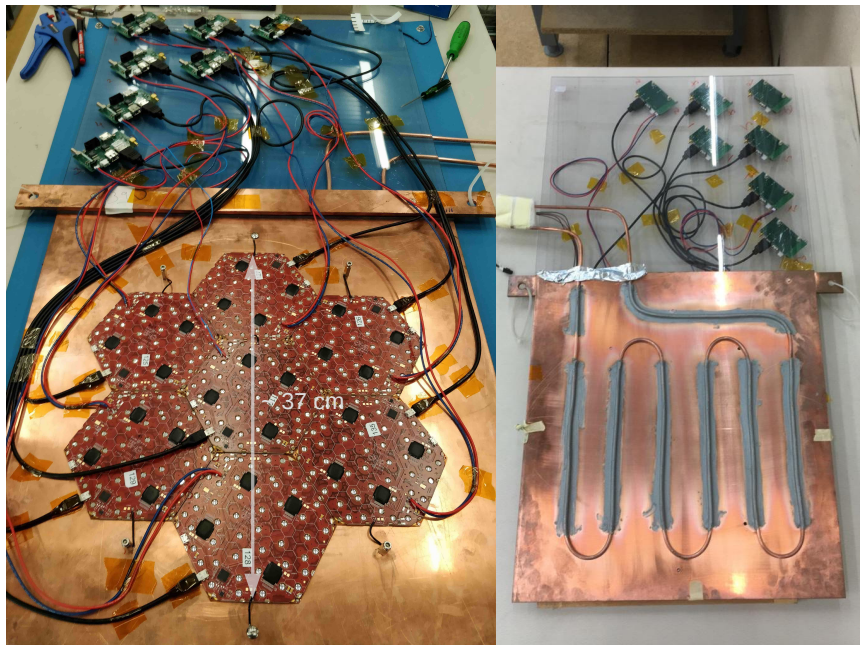}
  \caption{\label{fig:fh-layer} An example layer of the silicon CE-H prototype with 7 prototype silicon modules (left) and the back-side of a CE-H layer with its copper cooling pipe (right).}
\end{figure}

The silicon CE-H prototype was placed downstream of the CE-E prototype and comprised of 12 layers. An example layer is shown in Figure~\ref{fig:fh-layer}. The layers were made of a 6~mm thick copper plate on which up to 7 modules were attached on one side and a copper cooling pipe on the other. Above the copper plate, a sheet of Makrolon\textregistered~was attached to hold the `interposer' boards to connect the modules with the DAQ boards as for the CE-E section. 
The layers were inserted into two mechanical support boxes and were separated by 4~cm-thick steel absorber plates. The total depth of the silicon CE-H prototype corresponded to approximately 3.5~$\lambda_I$.

\begin{figure}[!ht]
  \centering
  \includegraphics[width=0.8\linewidth]{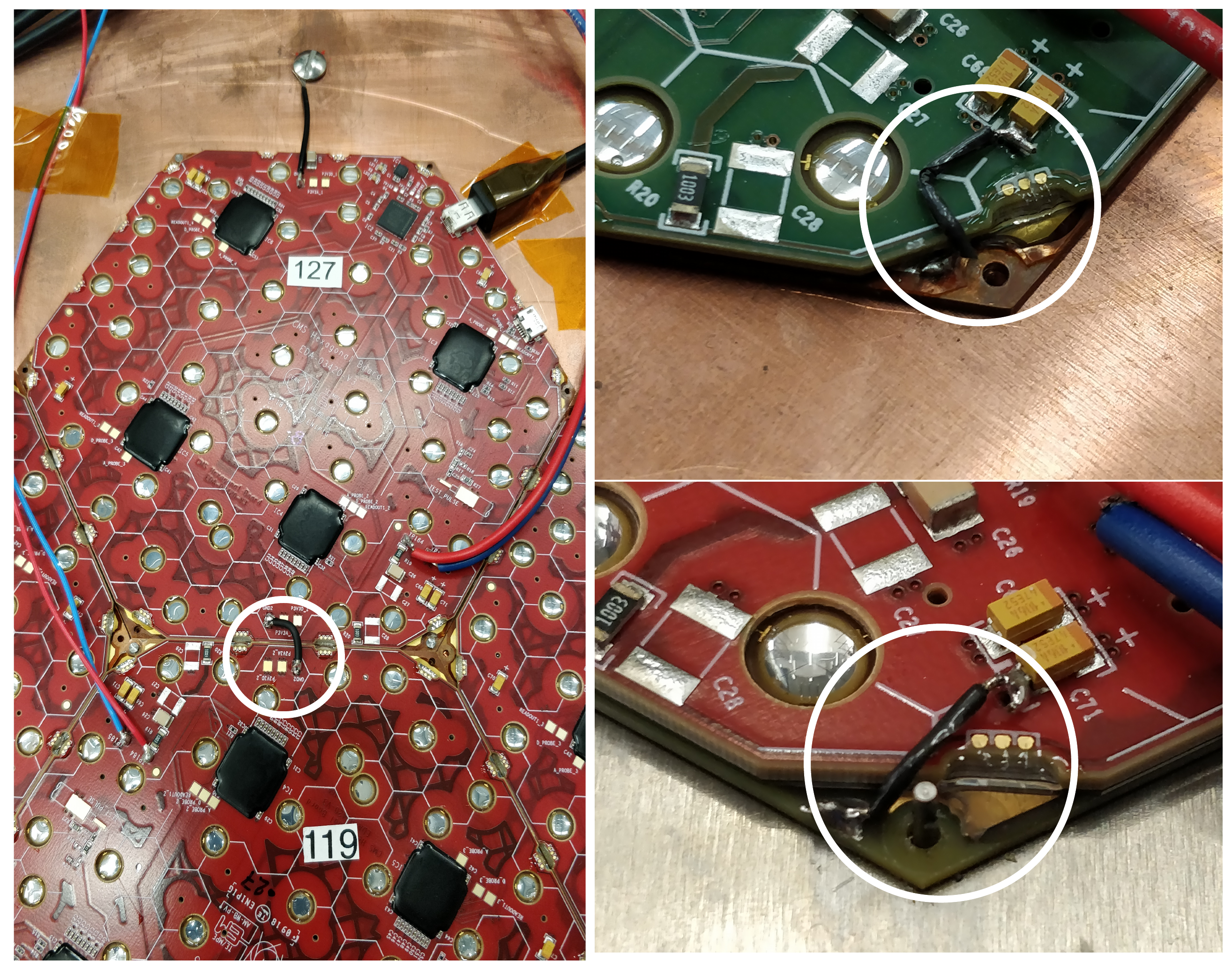}
  \caption{\label{fig:ground_FH} An example layer of the silicon CE-H with the ground connection between its central module and a peripheral module (left). An example of the ground connection for a double-Kapton\texttrademark~ module (top right). An example of the ground connection for a PCB baseplate module (bottom right).}
\end{figure}

The peripheral modules of the silicon CE-H prototype were grounded to the copper cooling plates via short wires. The same scheme was also tried for the central modules but did not perform as well due to the necessity to use longer wires. The grounding for the central modules was established by connecting a ground pad with a ground pad of one of the peripheral modules as shown in Figure~\ref{fig:ground_FH}~(left).

As mentioned in Section~\ref{subsec:modules}, two additional types of module were constructed with different grounding schemes. For the double-Kapton\texttrademark~modules the bottom gold layer was connected to a ground of the hexaboard as shown in Figure~\ref{fig:ground_FH} (top right). Seven double-Kapton\texttrademark~modules were built and used to form one silicon CE-H layer. The gold layers of the PCB baseplate modules were also connected to the ground of the hexaboard as shown in Figure~\ref{fig:ground_FH}~(bottom right). Seven PCB baseplate modules were built and used to construct another silicon CE-H layer. As described in Section~\ref{sec:performances}, the double-Kapton\texttrademark~and the PCB baseplate modules showed good noise performance, and the design of the future hexaboards for the CMS CE project foresees notches and pads dedicated to grounding connections.



\section{Beam test performance}
\label{sec:performances}

\subsection{Event building and analysis procedures}
The data acquisition software was embedded in the EUDAQ~\cite{bib.eudaq} framework. The raw data coming from different readout boards were saved using the `Event' class of the EUDAQ library. The event blocks were synchronized using their local event ID, incremented after each trigger. The synchronization of CMS CE readout boards was checked offline by analyzing the trigger time-stamps. 

A C++ library was developed~\cite{bib.h1-hgc} to convert the raw data from EUDAQ format to ROOT files~\cite{bib.root} that contain the high-gain and low-gain shaper samples, and the ToT values. The first step of the workflow was to calculate the pedestals for all memory cells of the channels of the SKIROC2-CMS ASICs. After subtracting the pedestal values, the common-mode noise was estimated and subtracted on an event-by-event basis, for each module, and every time-sample. Then the signal amplitudes were reconstructed for both the high and low-gain shapers as described in Section~\ref{sec:signal-reco}. Finally the response equalization and gain linearization procedure, discussed in Section~\ref{sec:calibrations}, completed the hit reconstruction.

\subsection{Pedestals and noise: calculation and stability}
\label{sec:pedestal-noise}

\begin{figure}[!ht]
	\centering
	\includegraphics[width=0.6\textwidth]{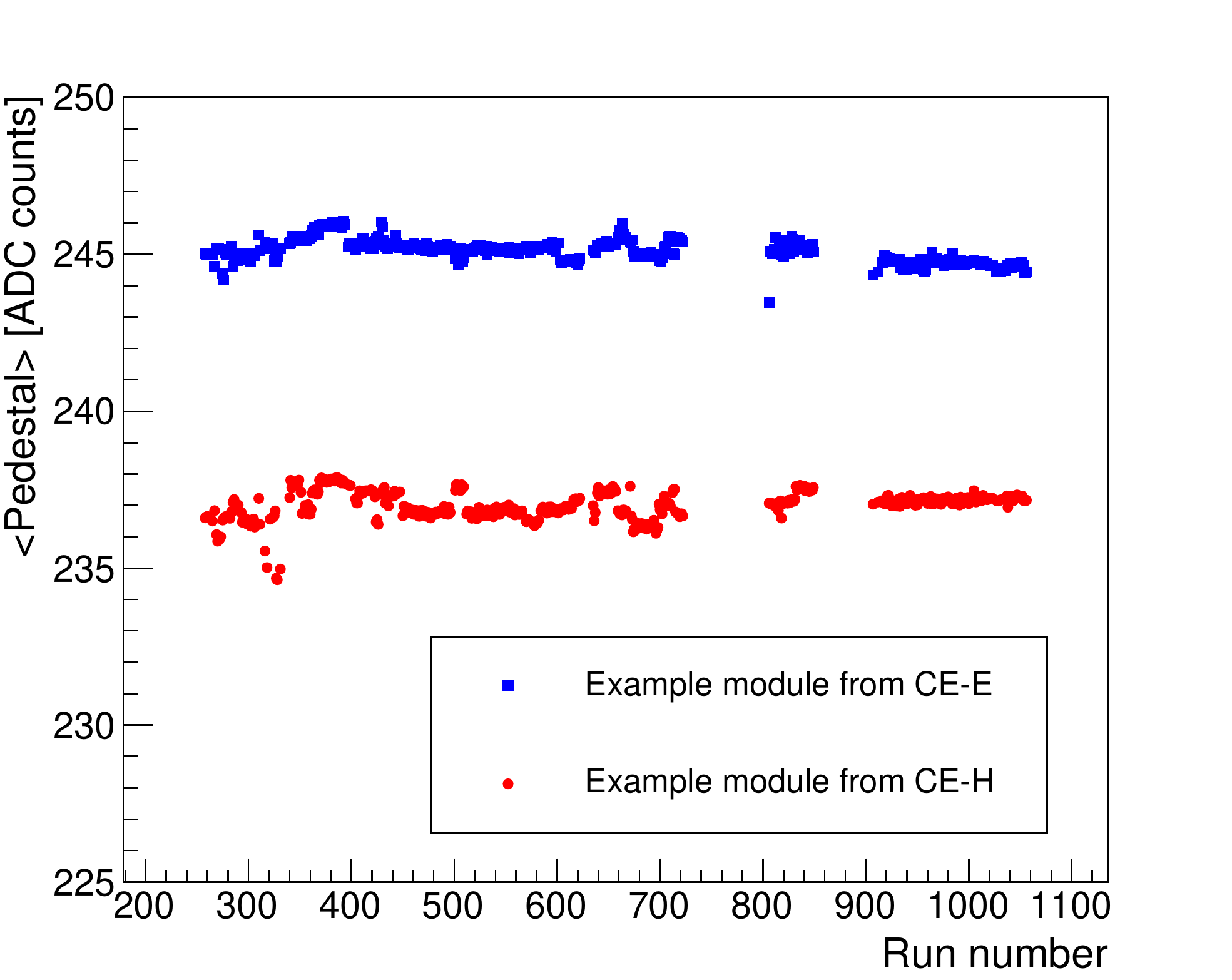}
        \caption{The pedestal values for two example prototype modules during the beam test, located in the CE-E and the CE-H prototypes.}
	\label{fig:ped_stability}
\end{figure}

Dedicated runs were taken to evaluate the pedestals for all memory cells of all channels. The median of the ADC-count distributions defined the pedestal values, while the standard deviation of these distributions gave an estimate of the level of noise in each channel. Files of the pedestal and noise values were created and saved for in subsequent analyses. Due to the trigger latency, the maximum amplitudes of the waveforms were reconstructed between the third and the fourth time-samples. Therefore the first time-sample of every recorded event was free of signal. This allowed the estimation of a pedestal value for each channel for each run. Figure~\ref{fig:ped_stability} shows the average pedestal value of two example modules (one in the CE-E prototype, one in the CE-H prototype) as functions of the run number. 

\begin{figure}[!ht]
	\centering
        \includegraphics[width=0.7\textwidth]{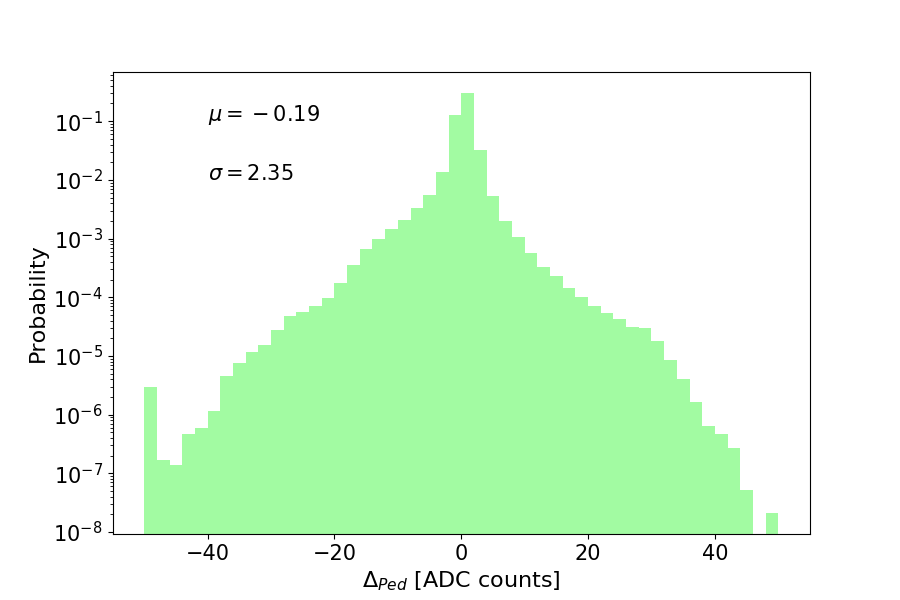}
	\caption{The distribution of differences between pedestal values evaluated from the dedicated pedestal run and those evaluated from all runs of the beam-test campaign, for all channels.}
	\label{fig:ped_diff}
\end{figure}

The pedestal stability was then evaluated by comparing these `per run' pedestal values with those obtained from the dedicated pedestal runs. Figure~\ref{fig:ped_diff} shows the distribution of the differences ($\Delta_{Ped}$) between the pedestal values obtained from the dedicated run and those obtained from all runs by using only the first time-sample for all channels of the CE prototype. The distribution peaked at zero with a standard deviation of 2.35~ADC counts (about 1/17 MIPs) and indicated a good stability of the pedestals over the full beam-test campaign.

The total noise ($\sigma_{total}$) is the quadratic sum of the intrinsic noise of the cell ($\sigma_{intrinsic}$) and any common-mode ($\sigma_{\mathrm{CM}}$) that might be present in the system.  

\begin{equation} \sigma_{total} = \sigma_{intrinsic} \oplus \sigma_{CM}  \label{eqn:noise} \end{equation}


To estimate the coherent noise, pedestal data were used to estimate the correlation coefficients, $C_{i,j}$, between channels in the same module. The coefficients $C_{i,j}$ were defined as:

\begin{equation} 
  C_{i,j} = \dfrac{ \frac{\sum^N_{e}S_{i}(e)S_{j}(e)}{N} - \overline{S_{i}}~\overline{S_{j}} }{\sigma_{i}\sigma_{j}}
  \label{eq:correlation}
\end{equation}
where N is the number of events, $S_{i}(e)$ is the high-gain ADC counts of channel $i$ after pedestal subtraction, $\overline{S_{i}}$ the average of $S_{i}$ and $\sigma_{i}$ its standard deviation. In this study, only the first time-sample was used within a module. The correlation coefficients varied between 0.3 and 0.9 and no correlation was observed between channels of different modules. The autocorrelation of the channels between time-samples was also studied to assure that the common-mode noise could be evaluated using the first time-sample alone. Equation~\ref{eq:autocorr} defines the autocorrelation coefficients $R_{t_i,t_j}$ for a channel was estimated for different time-samples.

\begin{equation}
  R_{t_i,t_j} = \dfrac{ \frac{\sum^N_{e}S_{t_i}(e)S_{t_j}(e)}{N} - \overline{S_{t_i}}~\overline{S_{t_j}} }{\sigma_{t_i}\sigma_{t_j}}
  \label{eq:autocorr}
\end{equation}
where N is the number of events, $S_{t_i}(e)$ is the amplitude of the high-gain ADC counts in time-sample $t_i$, $\overline{S_{t_i}}$ the average of $S_{t_i}$ and $\sigma_{t_i}$ its standard deviation. The average value of the autocorrelation coefficients for all channels of the CE prototype was found to be around -0.03. Thus common-mode noise could be evaluated separately for each time-sample. This result indicates the common-mode noise has a major high frequency component with a frequency $\ge$ 40 MHz.

\begin{figure}[!ht]
	\centering
	\includegraphics[width=0.6\textwidth]{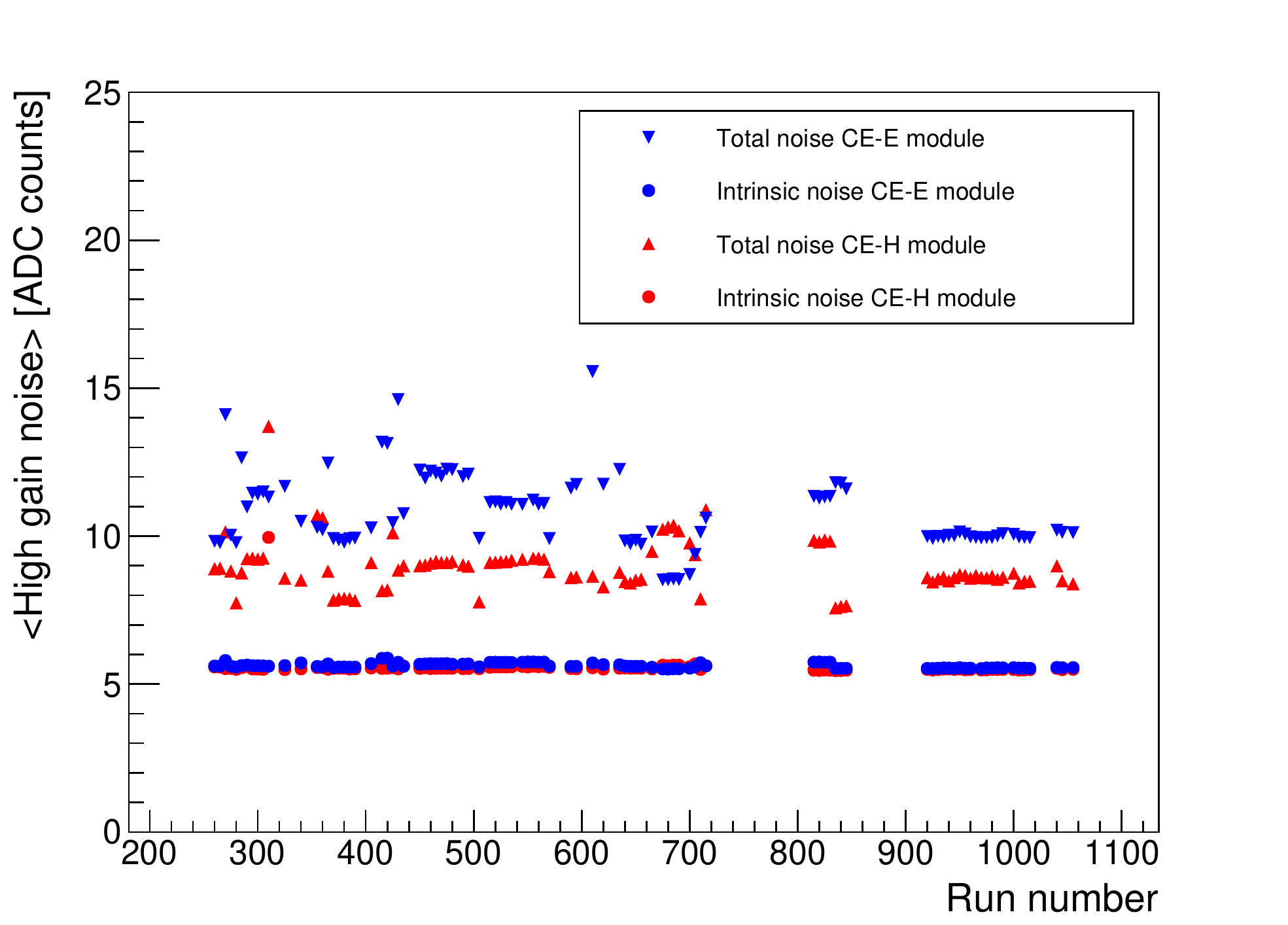}
	\caption{Stability of total and intrinsic noise during beam-test campaign in two CE prototype modules. The intrinsic noise was stable within 1-2 ADC counts throughout the full run period.}
	\label{fig:noise}
\end{figure}

Based on these studies, it was decided to evaluate and subtract a common-mode for each module and time-sample. For both the high and low-gain shaping amplifiers, the common-mode was defined as the median of the ADC counts of the full cells in a module. Only channels with an amplitude lower than 2 MIPs in the third time-sample
were considered for the common-mode evaluation. For the other cell types (half, mousebitten and calibration cells), the common-mode was scaled using the area of these special cells. After subtracting the common-mode, the residual noise was evaluated and the intrinsic noise of the cell was obtained. Figure~\ref{fig:noise} shows the total and intrinsic noise for two modules during this beam-test campaign. For both modules the intrinsic noise was around 0.12 MIP (5-6 ADC counts) and was substantially lower compared to the total noise, which was around 0.2 MIP (7-9 ADC counts) for CE-H modules and between 0.25 to 0.5 MIP (9-20 ADC counts) for CE-E modules. The intrinsic noise also showed small run-by-run variations.

\begin{figure}[!ht]
	\centering
	\includegraphics[width=0.48\textwidth]{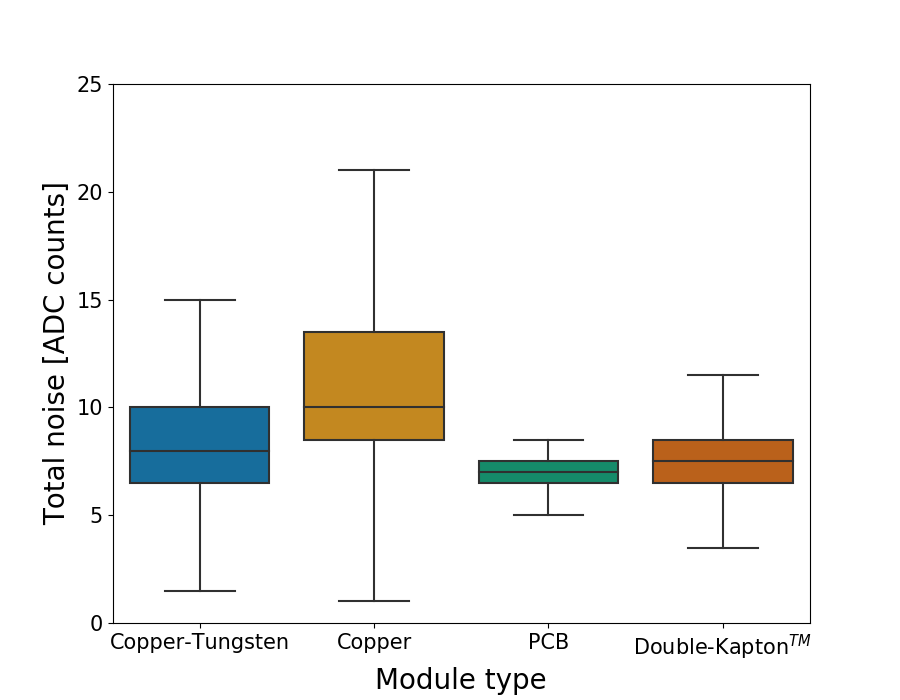}
	\includegraphics[width=0.48\textwidth]{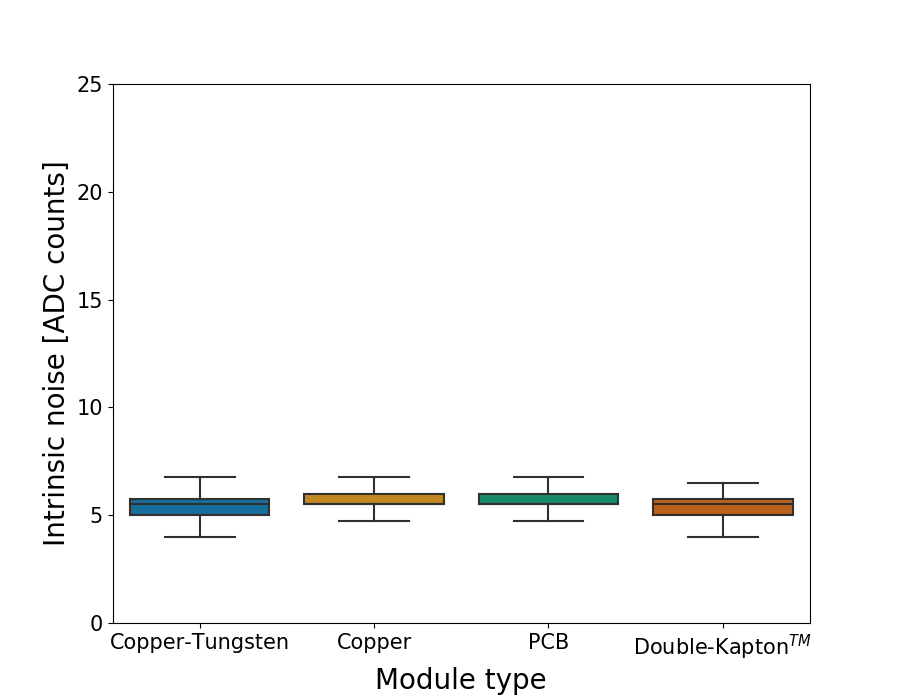}
	\caption{The high-gain total (left) and intrinsic (right) noise for modules with different baseplates and grounding schemes. The line inside the boxes indicates the median, the box indicates the interquartile range, and the whiskers indicate the 5th and the 95th percentiles.}
	\label{fig:noise_bp}
\end{figure}

As discussed in Sections~\ref{sec:modules}~and~\ref{sec:setup}, different types of prototype modules were used. In the CE-E, the modules were built with a copper-tungsten baseplate, while copper was mainly used in the CE-H. In addition, the CE-H prototype contained 1 layer with the PCB baseplate modules and 1 layer with the double-Kapton\texttrademark~modules, in which special grounding schemes were tested. Figure~\ref{fig:noise_bp} compares the total and intrinsic noise for modules from these different species. The double-Kapton\texttrademark~and PCB baseplate modules are valid options for the final CMS CE-H system. 

\begin{figure}[!ht]
	\centering
	\includegraphics[width=0.48\textwidth]{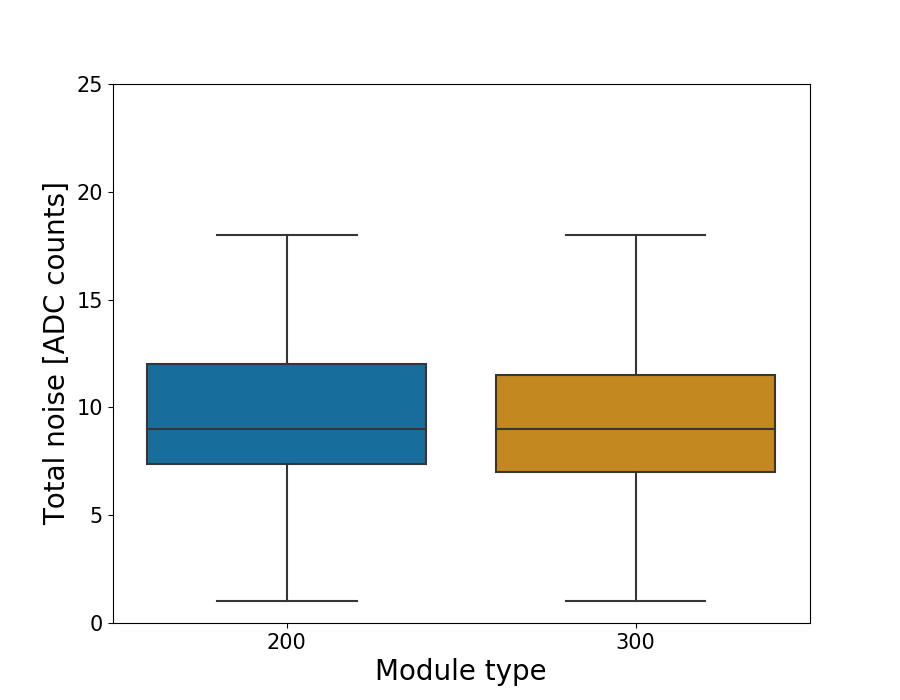}
	\includegraphics[width=0.48\textwidth]{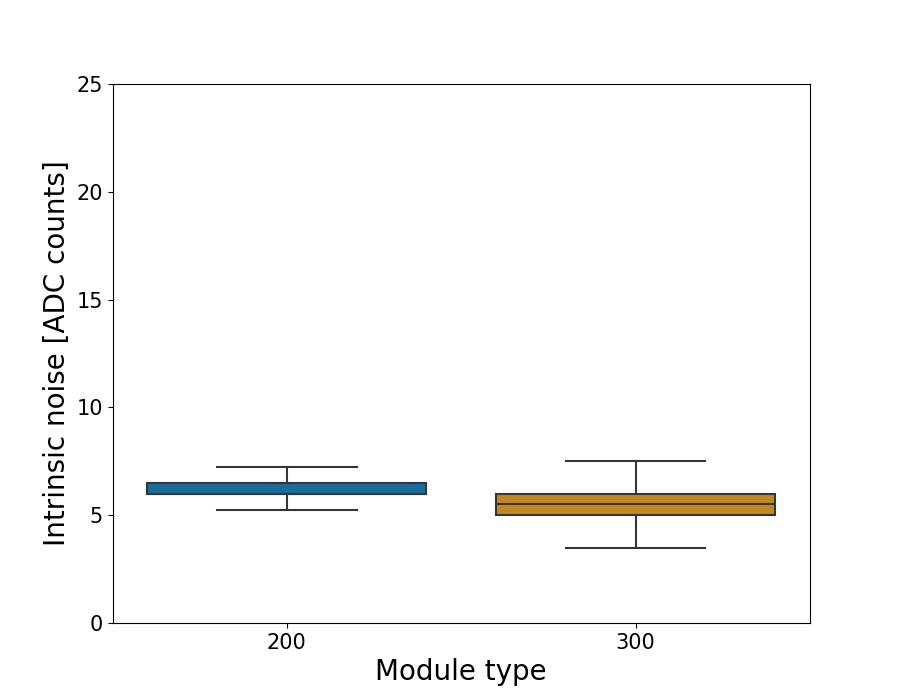}
	\caption{The high-gain total (left) and intrinsic (right) noise for modules with different sensor depletion thicknesses. The line inside the boxes indicates the median, the box indicates the interquartile range, and the whiskers indicate the 5th and the 95th percentiles.} 
	\label{fig:noise_sen}
\end{figure}

Figure~\ref{fig:noise_sen} compares the total and intrinsic noise for modules with different sensor depletion thicknesses (200~$\upmu m$ and 300~$\upmu m$). For the total noise the common-mode dominates, such that there is no appreciable difference between the two thicknesses. After the common-mode subtraction, the thicker sensor shows a slightly lower intrinsic noise due to its lower capacitance.  

\subsection{Signal reconstruction}
\label{sec:signal-reco}

\begin{figure}[!ht]
  \centering
  \includegraphics[width=0.7\textwidth]{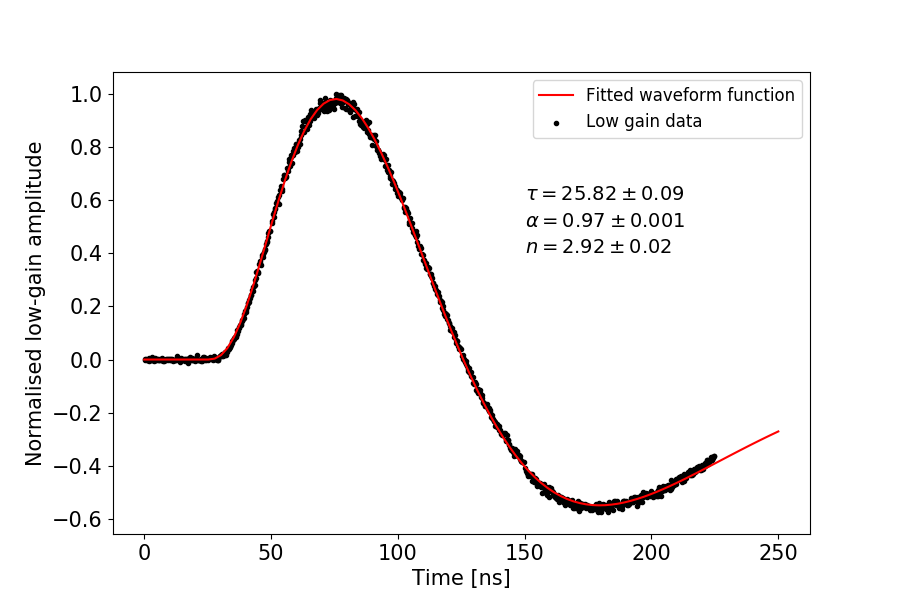}
  \caption{An example of the signal waveform from the low-gain shaper from the dedicated injection data acquired with the test stand. The waveform is fitted with the function (represented by the red line) described by Equation~\ref{eq:pulseshape}.}
  \label{fig:labpulseshapes}
\end{figure}

As explained in Section~\ref{subsec:felectronic}, the signals from the two slow shapers were sampled at 40~MHz. After the subtraction of the pedestals and the common-mode on each of these samples, the signal from the shapers could be reconstructed and a bipolar waveform was obtained. The amplitudes of these waveforms were reconstructed by fitting them as functions of time, since the beam was asynchronous with the 40 MHz sampling clock of the CE prototype. As suggested in~\cite{bib.semiconductor}, the function used to fit a bipolar waveform is given by Equation~\ref{eq:pulseshape}:

\begin{equation}
  S(t) = \left\{ \begin{array}{ll}
    A_0\bigg[\bigg(\frac{t-t_0}{\tau}\bigg)^n -\frac{1}{n+1}\bigg(\frac{t-t_0}{\tau}\bigg)^{n+1}\bigg]e^{-\alpha(t-t_0)/\tau} & \mbox{ if $t>t_0$} \\
    0 & \mbox{ otherwise}
  \end{array} \right.
  \label{eq:pulseshape}
\end{equation}
where $A_{0}$ is the amplitude of the waveform, $\tau$ is the time constant of the shapers and $n$ is a floating parameter\footnote{The function suggested in~\cite{bib.semiconductor} fixed the parameter $n$ to 2 but it was decided to use a floating parameter to agree better with the CE prototype data.}. The parameter $\alpha$ was introduced to improve the description of the undershoot region (t >125 ns in Figure~\ref{fig:labpulseshapes}) by the model. The parameters $\tau$, $n$ and $\alpha$ were calculated using data taken with the test stand introduced in Section~\ref{subsec:module-test}. Using this test stand, a known charge was injected in the channels at a given programmable time. By scanning this time in steps of 1~ns, the full waveform was reconstructed and fitted with Equation~\ref{eq:pulseshape} as shown in Figure~\ref{fig:labpulseshapes}. The calculated values of parameters $\tau$, $n$ and $\alpha$ were used as constants when fitting the waveform data from the beam test. 

The pre-selection criteria are listed in Table~\ref{tab:criteria}, in which $S(t_{max})$ is the maximum value of the high-gain ADC counts between the third and the fourth time-sample and $S(t_{max}+1)$ and $S(t_{max}+3)$ are respectively the values of the high-gain ADC counts in the next and third next time-samples.

\begin{table}[!ht]
  \centering
  \begin{tabular}{l | l}
    & Criteria                        \\
    \hline                            
    & $S(t_{max})$   > $S(t_{max}+3)$   \\
    and & $S(t_{max}+1)$ > $S(t_{max}+3)$   \\
    and & $S(t_{max})$   > 20 [ADC counts] \\
  \end{tabular}
  \caption{The pre-selection criteria applied during the zero-suppression procedure.}
  \label{tab:criteria}
\end{table}

If all of these criteria were satisfied, the high-gain and low-gain waveforms were fitted with the function of Equation~\ref{eq:pulseshape} and the high-gain ($A_{0,HG}$) and low-gain ($A_{0,LG}$) amplitudes extracted. 

\begin{figure}[!ht]
  \centering
  \includegraphics[width=0.45\textwidth]{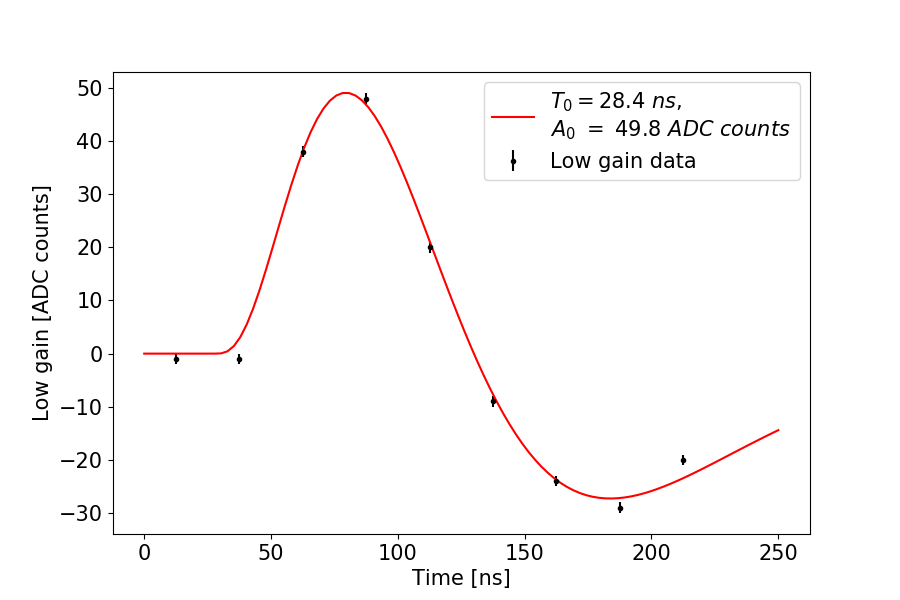}
  \includegraphics[width=0.45\textwidth]{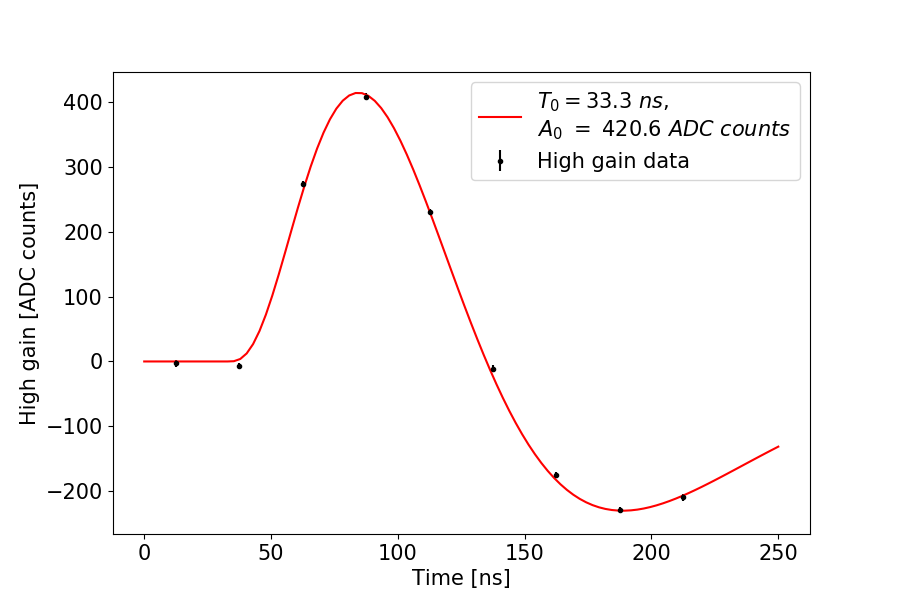}
  \caption{Two example waveforms for low gain (left) and high gain (right) due to a signal of about 10 MIPs from a 280 GeV/c electromagnetic shower. The red lines indicate the fits.}
  \label{fig:expulseshape}
\end{figure}

Figure~\ref{fig:expulseshape} shows the example waveforms for both gains in one channel of the CE prototype, due to a signal of about 10 MIPs from a 280 GeV/c electromagnetic shower. The fits are also shown.

The high-gain and low-gain amplitudes and the ToT signal were respectively transformed into equalized response measurements ($E_{HG}$, $E_{LG}$ and $E_{ToT}$) using calibration constants. The derivation of these constants is described in Section~\ref{sec:calibrations}. Finally the signal was selected following the criteria:

\begin{equation}
  E = \left\{ \begin{array}{ll}
    E_{HG} &, \mbox{ if $A_{0,HG}<HG_{sat}$} \\
    E_{LG} &, \mbox{ if $A_{0,HG}>HG_{sat}$ and $A_{0,LG}<LG_{sat}$} \\
    E_{ToT} &, \mbox{ otherwise}
  \end{array} \right.
  \label{eq:gainchoice}
\end{equation}
where $HG_{sat}$ and  $LG_{sat}$ correspond to the saturation amplitude for high and low gains. The evaluation of these saturation amplitudes is also explained in Section~\ref{sec:calibrations}.

\section{Channel-to-channel response equalization and gain linearization}
\label{sec:calibrations}

The CE prototype modules were calibrated in two stages. The first stage consisted of equalizing the response of the channels to the energy deposited by MIP-like particles at normal incidence. Simulation studies of the CE-E calorimeter showed that a precision of 3$\%$ on these equalization constants yields a constant term of 0.5$\%$ in the electromagnetic energy resolution~\cite{bib.cms-tdr}. The second stage of the procedure was gain linearization. As explained in Sections~\ref{sec:modules} and~\ref{sec:performances}, the Skiroc2-CMS ASIC offered 3 different measurements of the energy to provide the large dynamic range required by the calorimeter. Therefore, gain linearization needed to be carefully performed to ensure a linear response to signals in the CE prototype.  

\subsection{Channel-to-channel response equalization}
\label{subsec:mip}

The distribution of the energy deposited by an ionizing charged particle passing through a silicon cell approximately follows a Landau function. Its most probable value is expected to be around 57 keV and 86 keV for 200$~\upmu m$ and 300$~\upmu m$ thick sensors, respectively, for normal incidence. The relative calibration using MIP-like particles such as muons aims to correct the variations in the electronic gains and differences in the energy response, e.g. due to potential non-uniformity of the depleted thickness of the silicon sensor. Due to the overall negligible energy loss compared to the initial momentum of the MIP-like particles, the amount of deposited energy per unit distance can be regarded as being independent of the calorimeter depth. For this reason, MIP-like particles are suitable for the channel-to-channel response equalization. For this purpose, dedicated runs with muons (referred to simply as MIPs) were taken in the 2018 beam tests. 

\begin{figure}[!ht]
	\centering
	\includegraphics[width=0.75\textwidth]{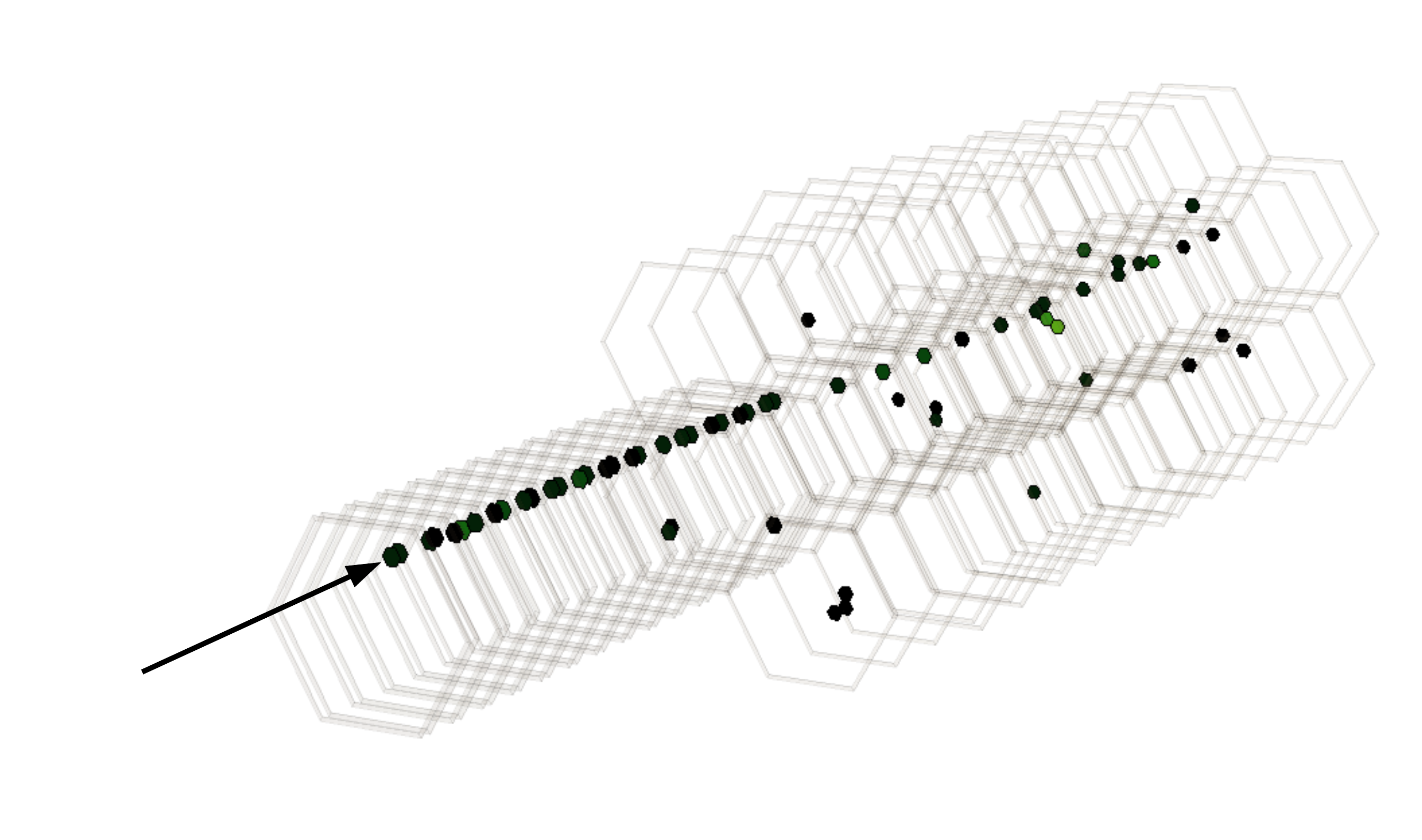}
	\caption{An event display of a 200 GeV/c muon traversing the CE-E (28~layers) and CE-H (12~layers) prototypes during the beam test of October 2018. The muon enters the detector from the left-hand side. A 0.5 MIP cut on the reconstructed energy of the hits is applied. Out of line hits correspond to noisy cells.}
	\label{fig:MuonDisplay}	
\end{figure}

At the CERN SPS, it is difficult to operate a muon beam with a momentum less than 1 GeV/c wide and enough to illuminate as many channels as possible, with an acceptable trigger rate. In addition, scanning the x-y positions of the prototype was impossible because the prototype was not installed on a moving table. A wide muon beam with a momentum of 200 GeV/c 
was used for the channel-to-channel response equalization. Although such high-energy muons are not true MIPs, they rarely initiate particle showers inside the calorimeter, and so their deposited signals are still good candidates for an energy reference. This reference was also used by the simulation of the CE prototype\footnote{The simulation of the CE prototype will be described in a future paper.}. Figure~\ref{fig:MuonDisplay} shows a typical muon event, recorded with the CE prototype during the October 2018 beam test.

Muons were initially selected using information from detectors upstream of the calorimeter prototype. The data from the DWCs of the H2 beam line \cite{bib.dwcs} incorporated into the beam tests then provided reference measurements for the beam particles' trajectories. Their extrapolated pointing precision at the calorimeter prototype amounted to better than one millimeter and allowed for a precise selection of cells traversed by an incident muon. The calorimeter itself was also used as a MIP tracking device. MIP signatures were identified and combined to reconstruct calorimeter hits consistent with one straight line trajectory per readout traversing the full calorimeter. Such a procedure was already used with the first prototype for the 2016 beam test campaign and details on the tracking algorithm can be found in~\cite{bib.hgcal-paper}.

\begin{figure}[!ht]
	\begin{subfigure}{.48\textwidth}
		\includegraphics[width=1.\textwidth]{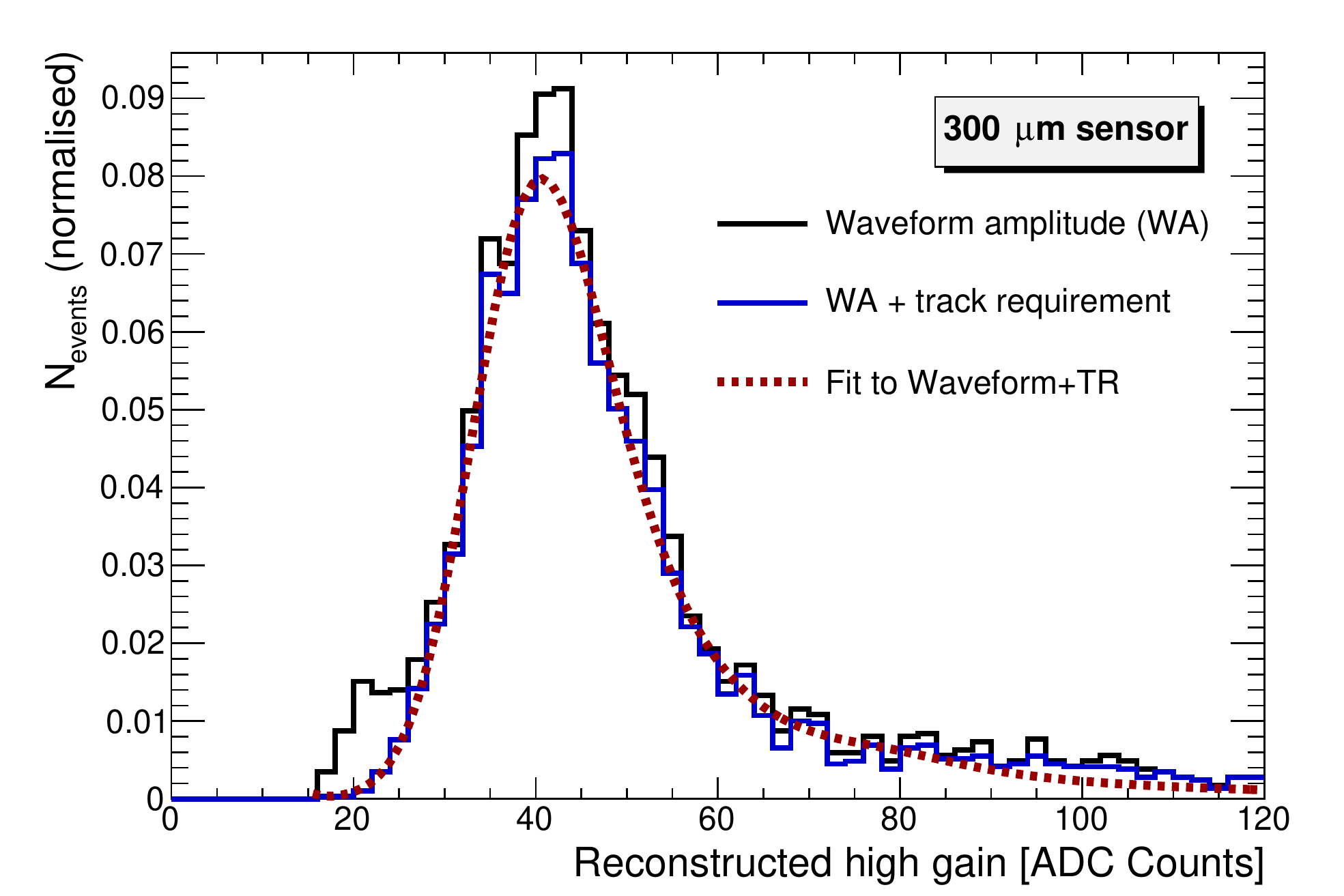}
		\label{fig:MuonEnergySpectra_A}
	\end{subfigure}
	\hfill
	\begin{subfigure}{.48\textwidth}
		\includegraphics[width=1.\textwidth]{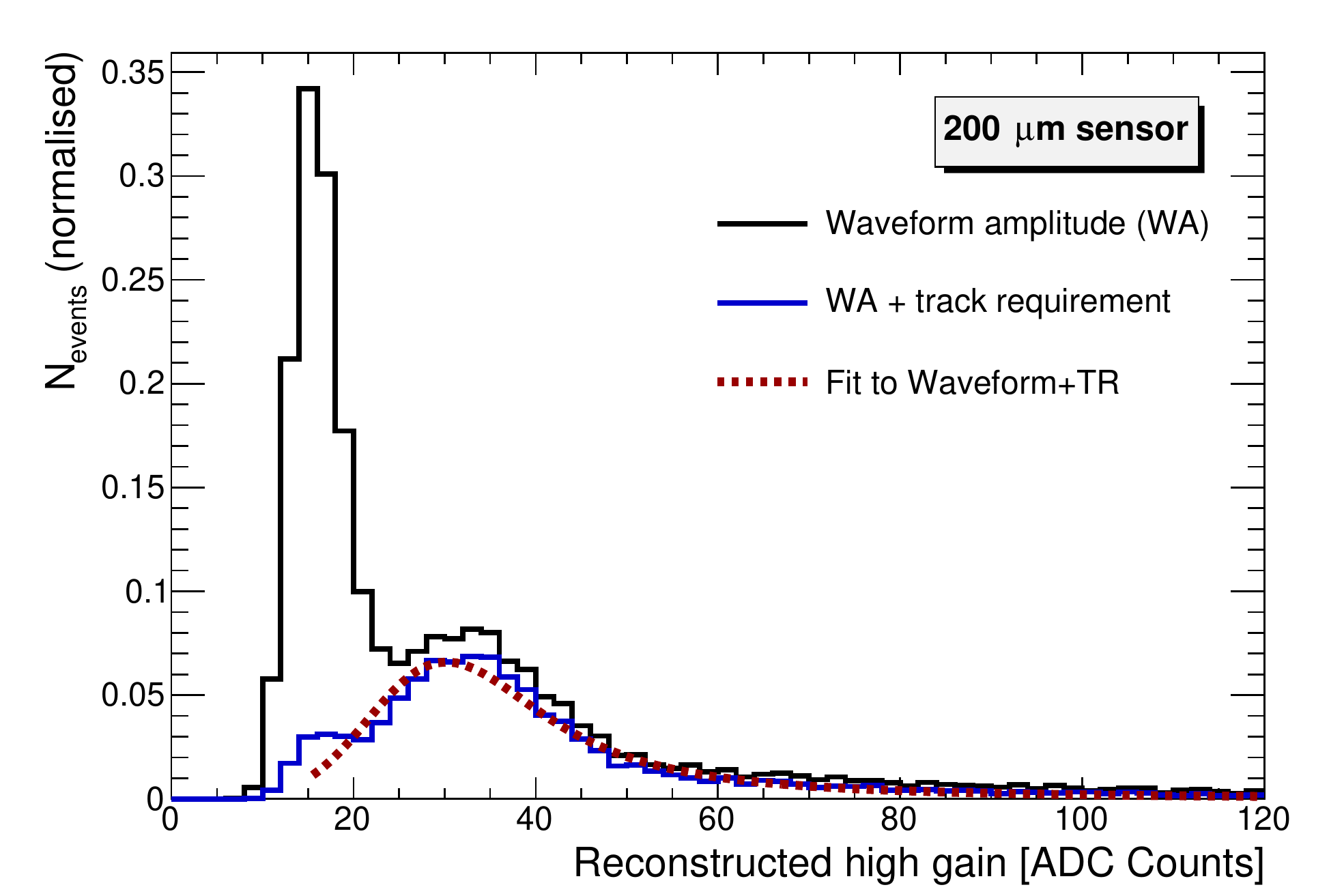}
		\label{fig:MuonEnergySpectra_C}
	\end{subfigure}	
		\hfill
	\begin{subfigure}{.48\textwidth}
		\includegraphics[width=1.\textwidth]{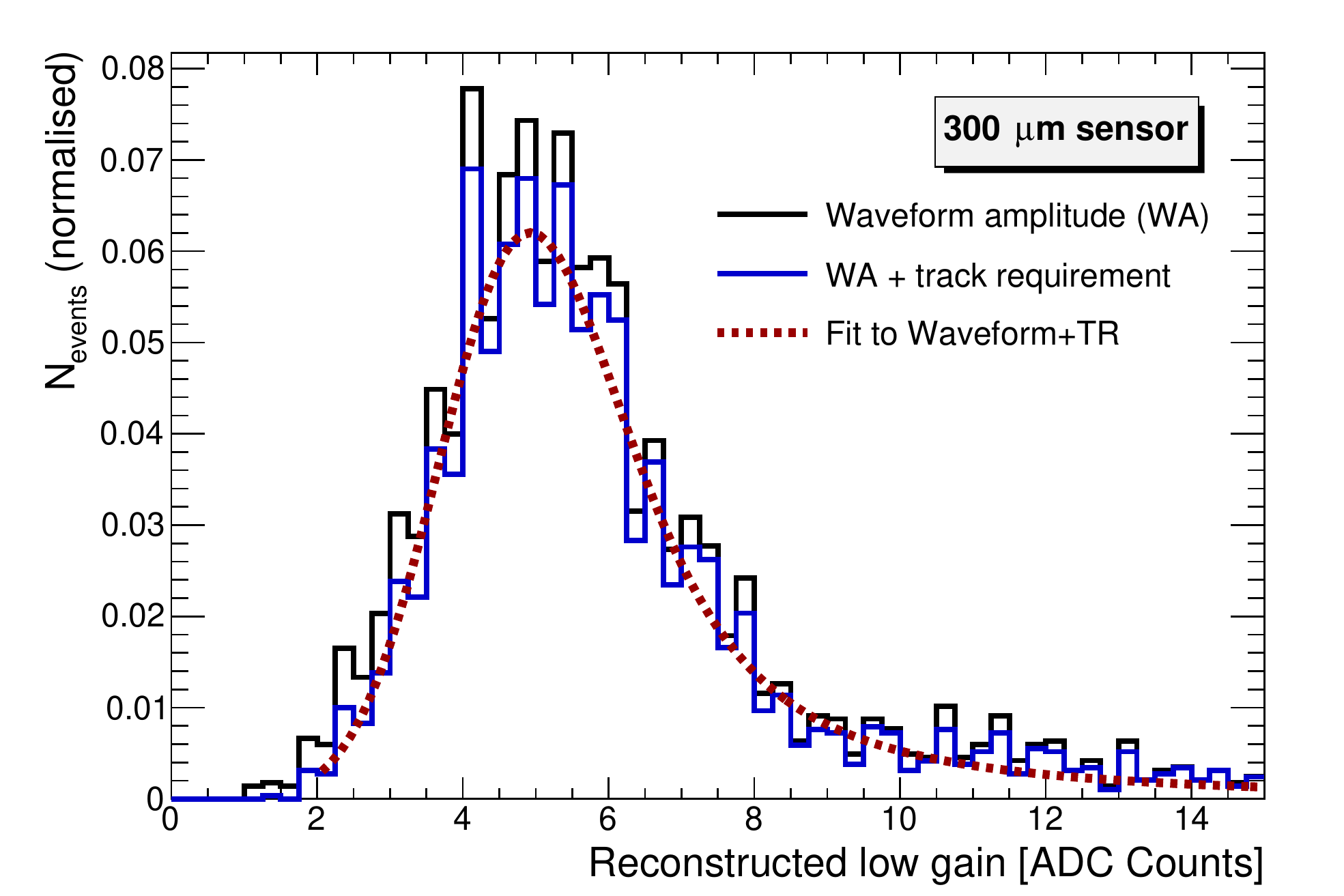}
		\label{fig:MuonEnergySpectra_D}	
	\end{subfigure}
	\hfill
	\begin{subfigure}{.48\textwidth}
		\includegraphics[width=1.\textwidth]{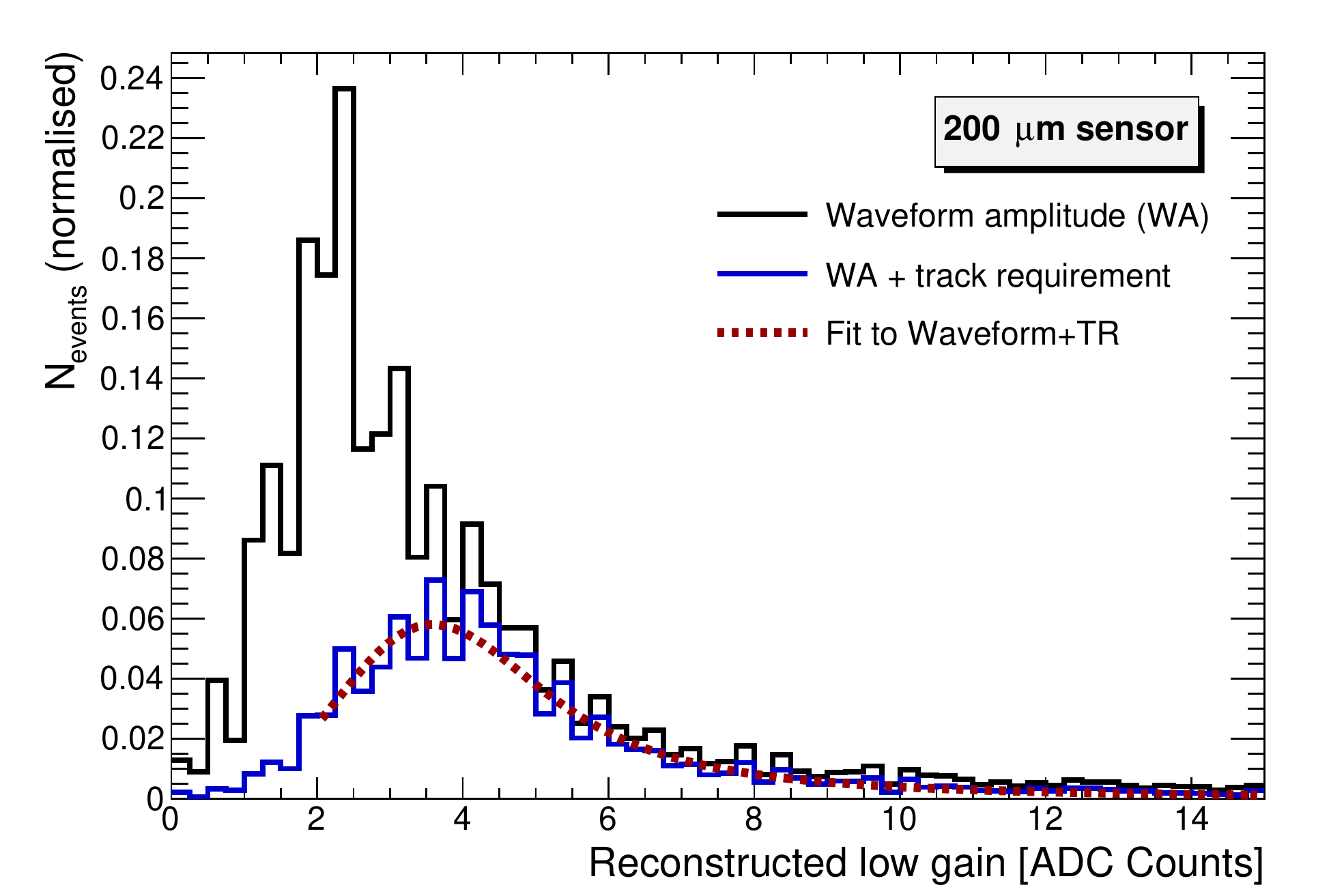}
		\label{fig:MuonEnergySpectra_F}
	\end{subfigure}		
	\caption{The equalized response spectrum of reconstructed ADC counts both in high (top left and right) and low gain (bottom left and right) for two example readout channels in a 300- (left) and 200-$~\upmu m$ (right) sensor due to incident 200 GeV/c muons. The black curves show all events recorded in the channel, while the blue curves show events additionally passing the track requirement described in the text.}
	\label{fig:MuonEnergySpectra}	
\end{figure}

Figure \ref{fig:MuonEnergySpectra} displays the reconstructed ADC counts spectra for two example channels before and after selection of physical hits.

The noise contributions of these spectra were almost eliminated when requiring a hit to be part of a straight line trajectory inside the calorimeter. Preliminary studies showed an efficiency above 98$\%$ for the hits induced by MIP-like particles~\cite{bib.thesisquast}. Equivalent spectra were obtained when applying the DWC based hit selection.

The signal spectrum induced by MIPs was also detectable in the distributions of the low-gain shaper amplitudes, where the signal-to-noise for single particles was around 3. This indicates that the CE remains sensitive to MIPs even with a lower signal-to-noise ratio.
 
Finally, these spectra were fitted with the function given by Equation~\ref{eq:MIPFitModel}.
\begin{equation}
  \begin{aligned}
    f\left(x,\upmu_L,c_L,\sigma_p,c_0,c_1,c_2\right) &= c_0\cdot G(x, 0,\sigma_p) \\
                                                 &+ c_1\cdot \left( L(\upmu_L,c_L) * G(0,\sigma_p) \right) (x)\\
                                                 &+ c_2\cdot \left( L(2\upmu_L,c_L) * G(0,\sigma_p)\right) (x)
    \label{eq:MIPFitModel}
  \end{aligned}
\end{equation}
where $G(0,\sigma_p)$ is a Gaussian function centered in 0, $L(\upmu_L,c_L)$ a Landau distribution with $\upmu_L$ and $c_L$ as location and scale parameter respectively and $c_0$, $c_1$, and $c_2$ normalization constants. The second convolution product ( $L(2\upmu_L,c_L) * G(0,\sigma_p)$ ) aimed to fit a second Landau peak, if any, and was introduced to improve the quality of the fit. The example fits to the corresponding measured spectra are included as red dotted lines in Figure~\ref{fig:MuonEnergySpectra}. The inverse of the maximum of the fitted fucntion, provided the channel-to-channel response equalization constant ($C_{MIP}$) to convert the high-gain amplitude to a response in MIP units. 

In the October 2018 beam test, due to the limited spread of the muon beam and the limited mobility of the calorimeter setup through the beam, only 31\% of all channels could be calibrated in this way. For the remaining channels, a dataset accumulated with muons of unknown energies was used. During the two weeks after the main beam test in October 2018, the CE prototype was still operated while being exposed to any particles not stopped by another experiment located further upstream. With such a setup, most of the parasitic particles expected to reach the CE were muons. The same tracking algorithm (as the one used with the previous prototype as described in~\cite{bib.hgcal-paper}), using the calorimeter as a tracking device, was chosen. Channel-to-channel response equalization constants for an additional 54\% of the channels could be derived from this dataset accumulated during the parasitic beam time.

\begin{figure}[!ht]
	\centering
	\includegraphics[width=.9\textwidth]{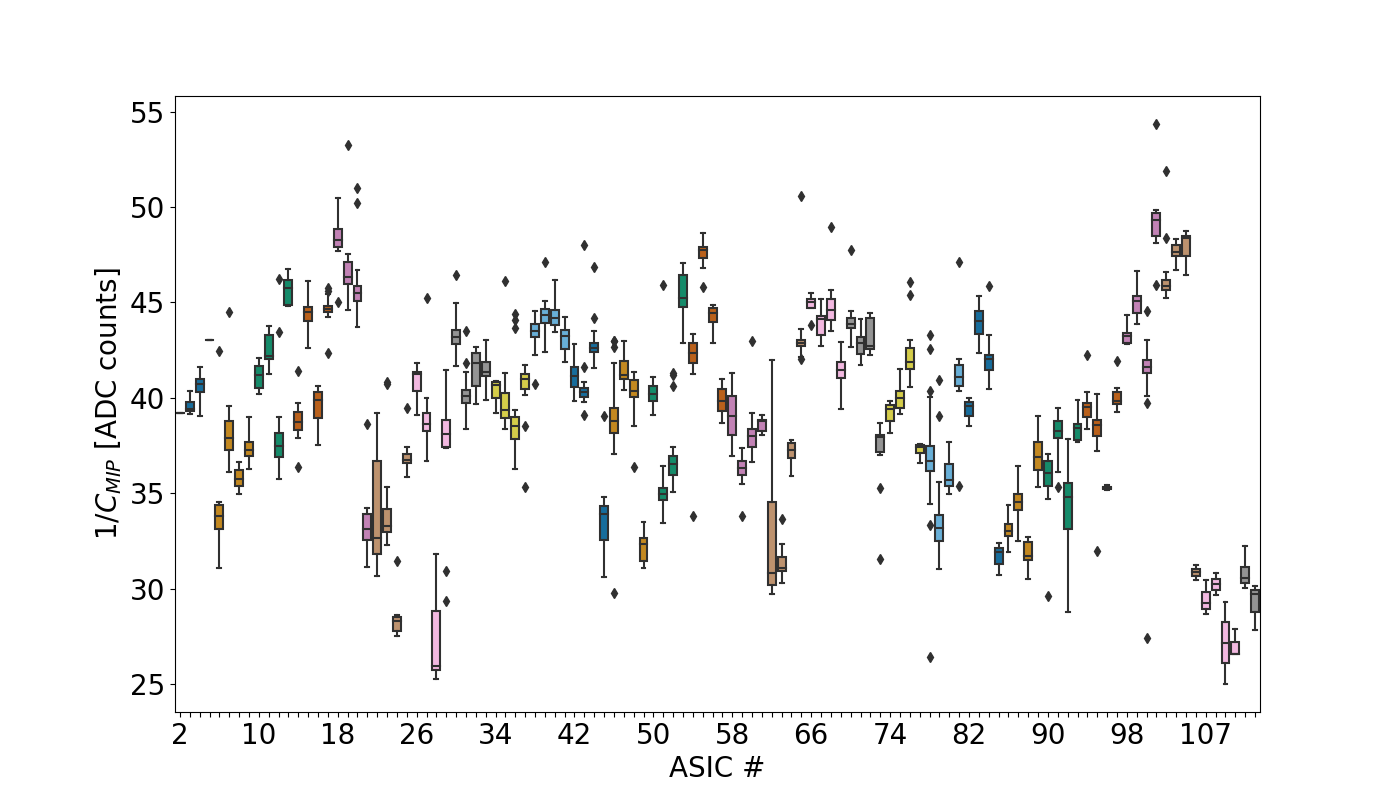}
	\caption{The per-ASIC distribution of high-gain ADC counts per MIP for the October 2018 beam tests of the CE-E prototype. The line inside the boxes indicates the median, the box indicates the interquartile range, the whiskers indicate the 5th and the 95th percentiles and the single points indicate the outliers. Each color corresponds to different modules. The two last modules of the CE-E prototype were built with 200$~\upmu m$ thick silicon sensors.}
	\label{fig:JuneVsOctoberMIPs}	
\end{figure}

Figure~\ref{fig:JuneVsOctoberMIPs} summarizes the distribution of high-gain ADC counts per MIP for all 28 modules in the CE-E prototype. Their intra-module variation amounted to $\approx10\%$ and was dominated by the difference in the mean electronic response of each ASIC. The per-ASIC dispersion of this calibration constant was on average about 4$\%$.

\begin{figure}[!ht]
	\centering
	\begin{subfigure}{.49\textwidth}
		\includegraphics[width=1.\textwidth]{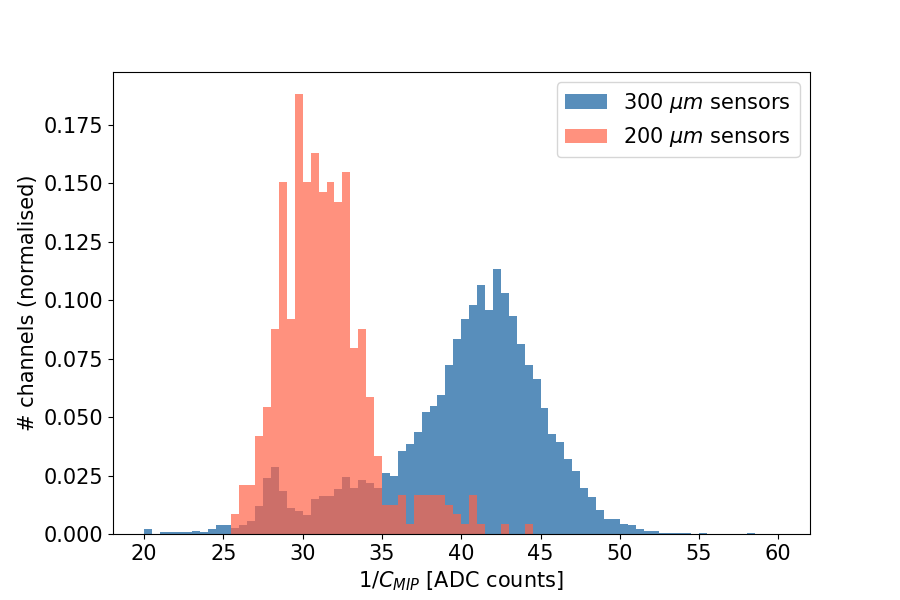}
		\label{fig:MIPs300Vs200}			
	\end{subfigure}
	\hfill
	\begin{subfigure}{.49\textwidth}
		\includegraphics[width=1.\textwidth]{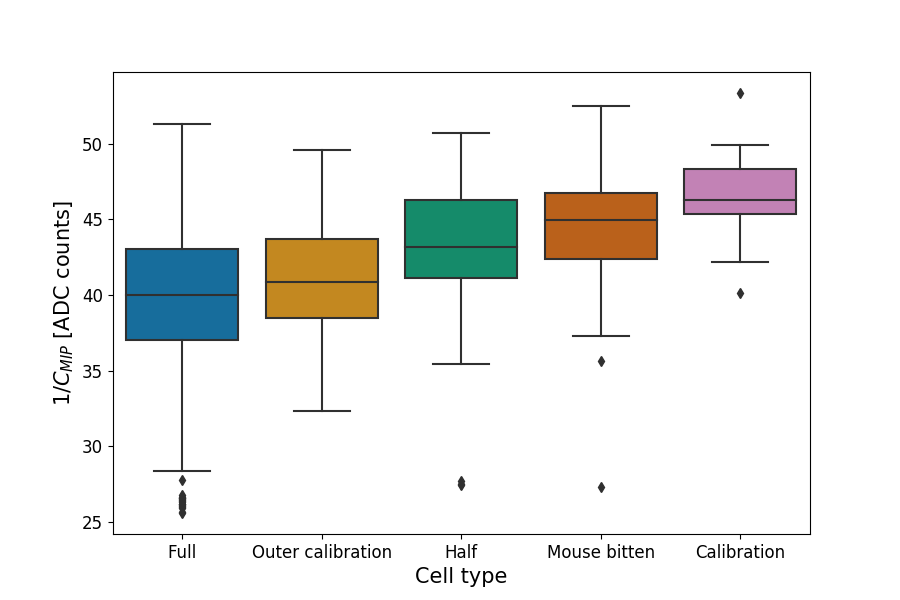}		
		\label{fig:MIPsCellTypes}	
	\end{subfigure}
	\caption{The distribution of the high-gain ADC counts per MIP constant of all full hexagon cells with 200 GeV muons (left). High-gain ADC counts per MIP constant as a function of the cell type (right). The line inside the boxes indicates the median, the box indicates the interquartile range, the whiskers indicate the 5th and the 95th percentiles and the single points indicate the outliers.}
         \label{fig:MIPsCellTypesandsensorthickness}	
\end{figure}

Further analysis confirms the expected lower response to MIPs for 200$~\upmu m$ than for 300$~\upmu m$ thick sensors, shown in Figure~\ref{fig:MIPsCellTypesandsensorthickness} (left). Figure~\ref{fig:MIPsCellTypesandsensorthickness} (right) shows the channel-to-channel reponse equalization constant as a function of the cell type. It confirms that smaller silicon pads leads to larger signals as introduced in Section~\ref{subsec:sensor}. This effect was already observed with the previous CE prototype~\cite{bib.hgcal-paper} and explained by the lower capacitance of the calibration cells leading to a faster rise time when shaped in the SKIROC2 ASIC. 

\begin{figure}[!ht]
        \centering
	\includegraphics[width=.6\textwidth]{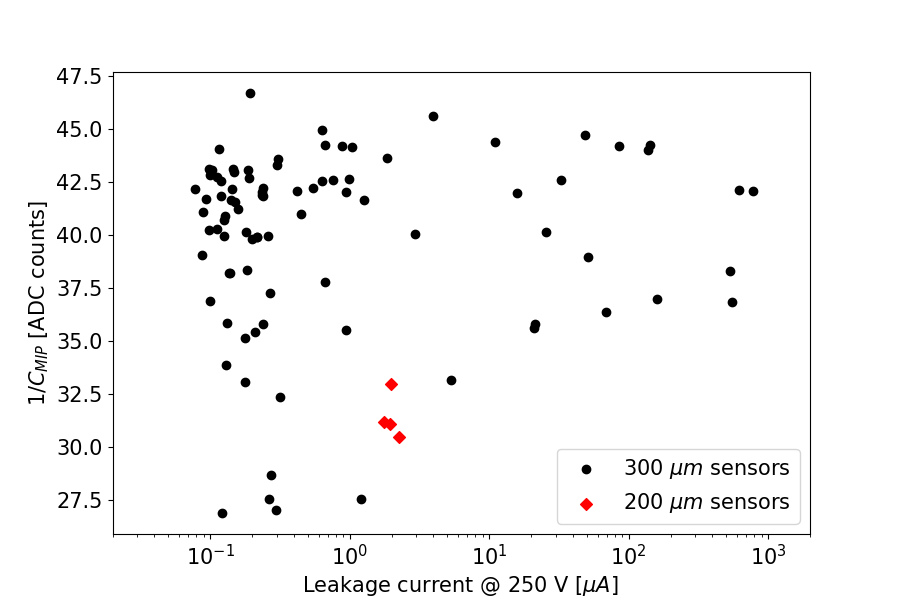}
	\caption{The average channel-to-channel response equalization constant of a module, as a function of the leakage current with a bias voltage of 250V.}
	\label{fig:mipvsleakage}
\end{figure}

In Section~\ref{sec:modules}, high leakage current in a few modules was mentioned. This high leakage current could lead to under-depleted silicon sensors and hence lower response to deposited energy from particles. This had been observed during a previous beam test with an early version of the CE prototype where modules were operated with a bias voltage of 200V. 
Figure~\ref{fig:mipvsleakage} shows the average channel-to-channel response equalization constant of a module, as a function of the module leakage current with a bias voltage of 250V.

The leakage current seems to have a limited impact on the channel responses. This could be explained by operating silicon sensors with a bias voltage of 250V while the full depletion bias voltage was expected around 200V for 300$~\upmu m$ sensors. Therefore, even though few modules could not be operated in an optimal way, consequences on their charge collection efficiency and then on energy reconstruction and resolution were limited.

\begin{figure}[!ht]
  \centering
  \includegraphics[width=.49\textwidth]{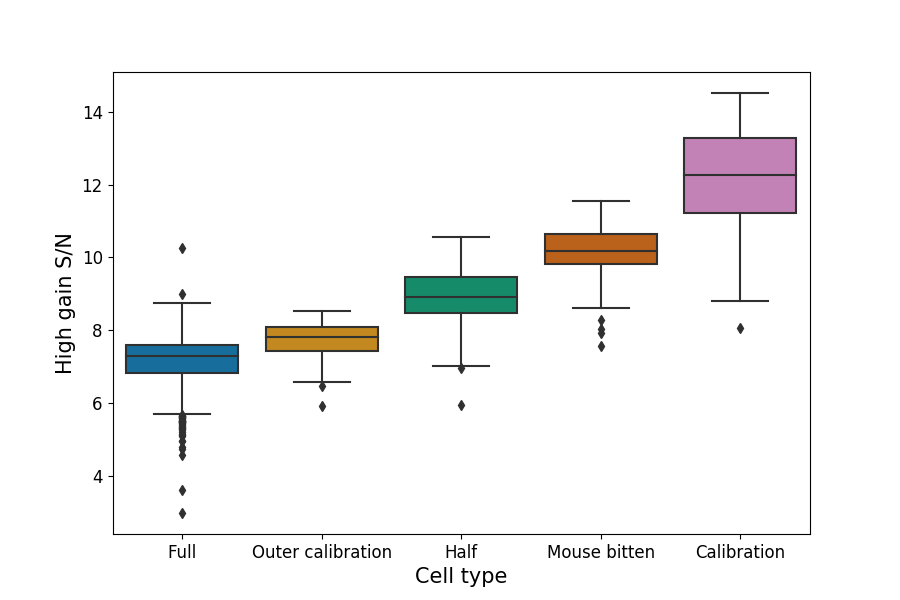}
  \includegraphics[width=.49\textwidth]{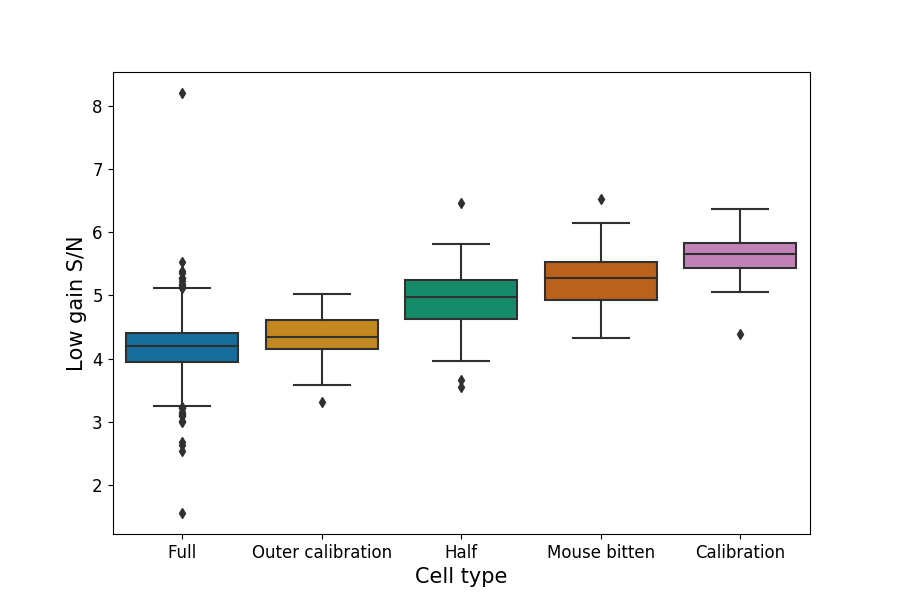}
  \caption{Signal over noise ratio for MIP in high (left) and low (right) gains as a function of the cell type for modules with 300$~\upmu m$ sensor. The line inside the boxes indicates the median, the box indicates the interquartile range, the whiskers indicate the 5th and the 95th percentiles and the single points indicate the outliers.}
  \label{fig:SNR_celltype}
\end{figure}

\begin{figure}[!ht]
  \centering
  \includegraphics[width=.49\textwidth]{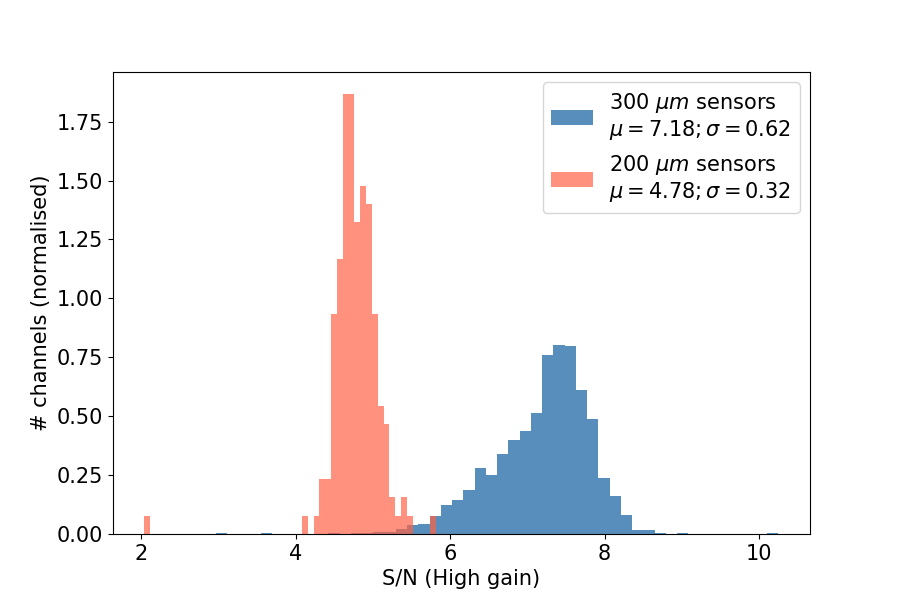}
  \includegraphics[width=.49\textwidth]{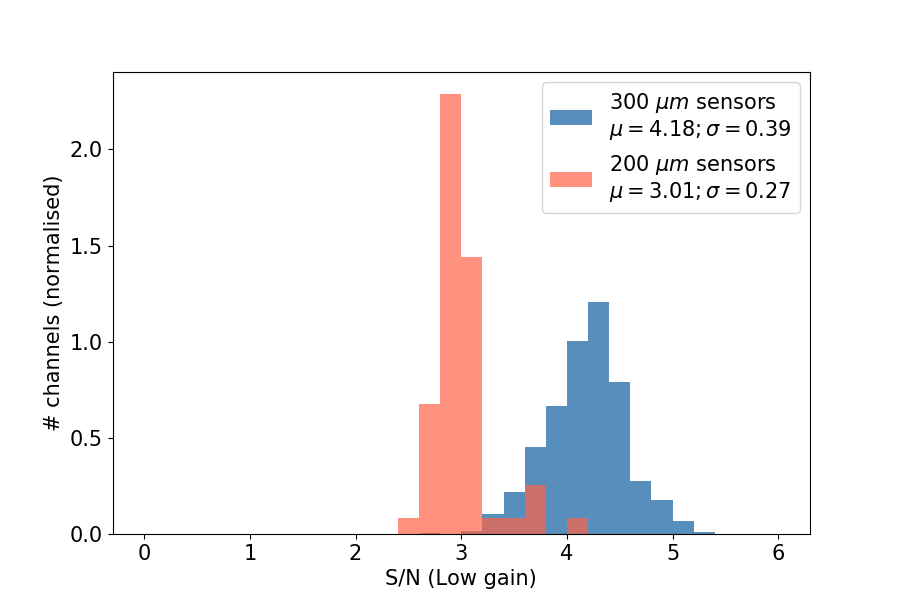}
  \caption{Distributions of S/N for the high gain (left) and for the low gain (right) for the two sensor-depletion thicknesses used during the CE beam tests.}
  \label{fig:son-distrib}
\end{figure}

Figure~\ref{fig:SNR_celltype} shows the MIP signal to intrinsic noise ratio (S/N) as a function of the cell type for both high- (left) and low-gain (right) MIP signal. As expected, smaller cells have a better  MIP signal to intrinsic noise ratio.
Figure~\ref{fig:son-distrib} shows distributions of S/N for the two gains for the two sensor-depletion thicknesses used during the CE beam tests. From the characterization studies of the SKIROC2-CMS ASIC \cite{bib.Skiroc2cms}, the high-gain intrinsic noise was expected to be about 13$\%$ of a MIP. The measured signal over noise ratio of the CE prototype modules are in good agreement with the expectations.

In summary, the data taken with muon and parasitic beams provided the channel-to-channel response equalization constant for roughly 85$\%$ of all silicon cells. For the other channels, the average channel-to-channel response equalization constant from the channels from the same SKIROC2-CMS ASIC was used. If the ASIC did not have any calibrated channels, then the equalization constant was evaluated from all the channels of the CE prototype. The resulting variation of these constants as well as the S/N ratio were consistent with the targeted calorimeter endcap design values ($\approx$3$\%$ of dispersion per ASIC)~\cite{bib.cms-tdr}.

\subsection{Gain linearization}
\label{subsec:gainintercalib}
The ASIC gain linearization was a very important aspect of the calibration chain of the CE prototype, since it had to ensure a linear response over a large dynamic range. The SKIROC2-CMS ASIC was designed to use the high- and low-gain shapers for collected charges up to about 200 and 600 fC, respectively, and the ToT readout for larger charges. The threshold of 200(600) fC corresponds to the energy deposit of about 60(200) MIPs. Two methods for the gain linearization were designed. The first directly used the beam-test data, while the second was based on charge injection into the FE ASIC. This second method is close to the gain linearization method foreseen for the final CE calorimeter.

\subsubsection{Gain linearization using beam-test data} 
\label{subsubsec:datadriven-calib}
\begin{figure}[!ht]
  \centering  \includegraphics[width=1.0\textwidth]{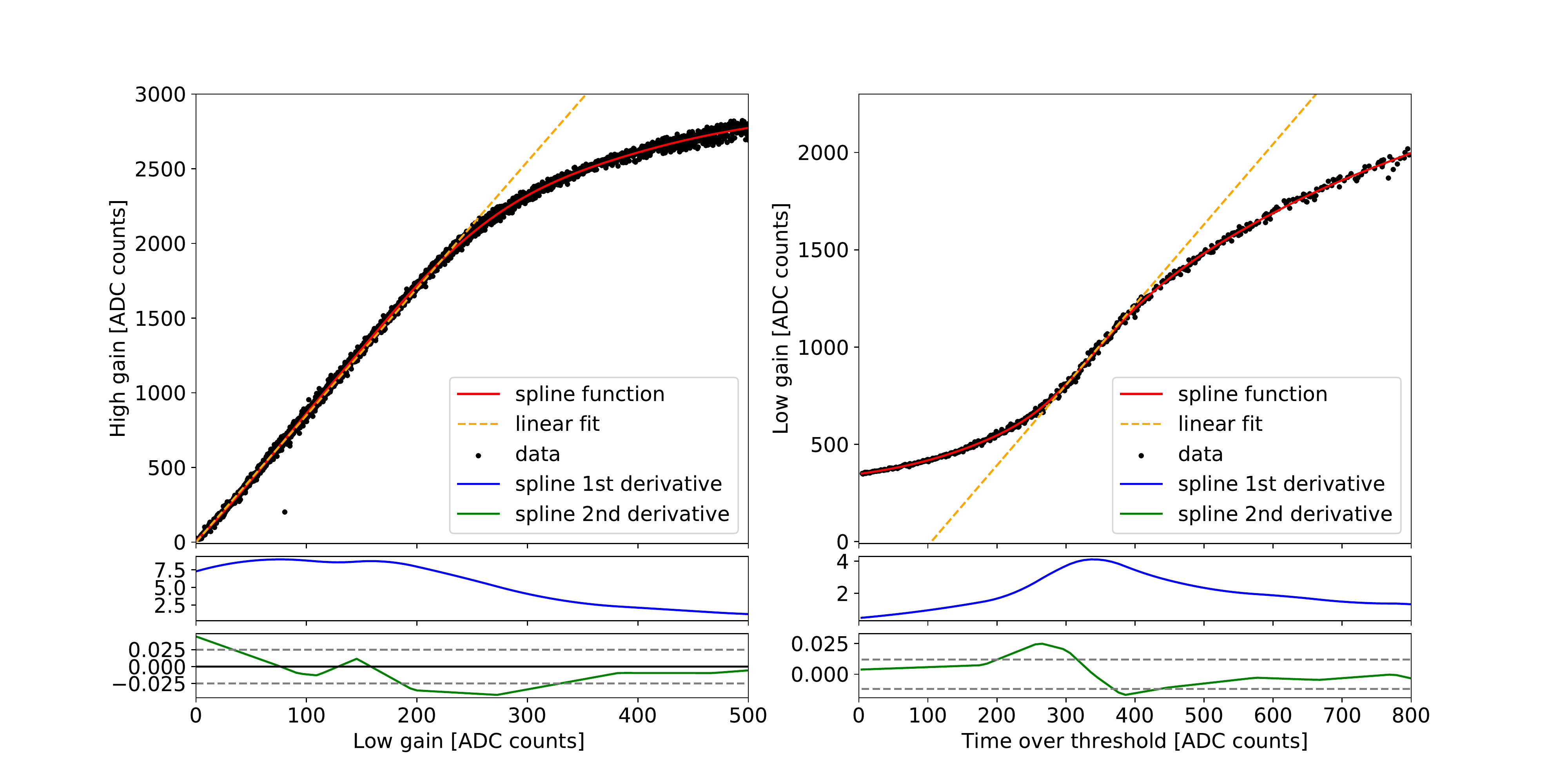}
  \caption{The high-gain amplitude as a function of the low-gain amplitude (left) in a CE-E prototype channel from 280 GeV electromagnetic shower data and the low-gain amplitude as a function the ToT (right) from the same channel. On the latter, the intercept between the dashed orange line and the x-axis defines the ToT offset.}
  \label{fig:beamtestadccalib}	
\end{figure}

The first gain linearization method is similar to the one used for the previous CE prototype described in~\cite{bib.hgcal-paper}. A linear correlation between high- and low-gain response was measured for events with energies falling within the range accessible by both shapers.  Similarly, a separate linear correlation was measured between the low-gain and ToT response.  The slopes of the linear fits were then used to transform the ToT and the low-gain amplitude to an equivalent high-gain amplitude, which was then transformed into a energy equivalent by using the channel-to-channel response equalization constants. This gain linearization method relied on fitting data in an energy range in which the correlation between high- and low-gain amplitudes was linear and the correlation between low-gain amplitude and ToT was linear. Figure~\ref{fig:beamtestadccalib} shows the high-gain amplitude as a function of the low-gain amplitude (left) and the low-gain amplitude as a function of the ToT (right) in a CE-E prototype channel from 280 GeV electromagnetic shower data.

The range of linear correlation between the high- and low-gain amplitudes was not known {\it{a priori}}. A spline interpolation was performed on the data, and the first and second derivatives of the interpolation were calculated and used to derive the linear region. This region was then fitted with a linear function. The high-gain value for which the difference between the high-gain value and the linear function exceeded 3$\%$ defined the amplitude of saturation for the high gain. The spline interpolation, the derivatives, and the linear fit are also shown in Figure~\ref{fig:beamtestadccalib}. The same method was also applied for the low-gain amplitude and the ToT. In addition to the low-gain-amplitude-to-ToT coefficient and the low-gain saturation amplitude, a ToT offset due to insensitivity at very low charges was also reconstructed and stored.

Finally,  when using this method, the calibrated response of a signal in a silicon cell is given by Equation~\ref{eq:calibenergy}:
\begin{equation}
  E = \left\{ \begin{array}{ll}
    E_{HG}  = C_{MIP}\cdot A_{0,HG} &, \mbox{ if $A_{0,HG}<HG_{sat}$} \\
    E_{LG}  = C_{MIP}\cdot C_{HL}\cdot A_{0,LG} &, \mbox{ if $A_{0,HG}>HG_{sat}$ and $A_{0,LG}<LG_{sat}$} \\
    E_{ToT} = C_{MIP}\cdot C_{HL}\cdot C_{LT}\cdot \left(ToT-ToT_{offset}\right)  &, \mbox{ otherwise}
  \end{array} \right.
  \label{eq:calibenergy}
\end{equation}
where $A_{0,HG}$ and $A_{0,LG}$ are the amplitude of the high-gain and low-gain waveforms respectively, $C_{MIP}$ is the channel-to-channel response equalization constant, and $C_{HL}$ and $C_{LT}$ are the gain linearization constants.

Similar to the channel-to-channel response equalization, the gains of a large fraction of channels of the CE prototype could not be linearized using this method. Events with energies between 0 and about 300-400 MIPs equivalent were needed to obtain the ToT-to-low-gain constant for a channel. On average for the 280 GeV/c electromagnetic shower events, the number of channels with a response larger than 300 MIPs was about 20, which corresponded to 0.5$\%$ of the number of channels in the CE-E prototype. After combining all the beam-test data from electromagnetic shower events for the CE-E prototype and from the hadronic shower events for the CE-H prototype, the fraction of channels having the full set of constants was about 17$\%$ for the full CE prototype, representing 44$\%$ and 5.5$\%$ of the CE-E and CE-H prototypes, respectively. For the other channels, the average of the constants from the channels belonging to the same ASIC were used. For the channels of an ASIC where no constant could be derived, global approximations were used. 

\begin{figure}[!ht]
  \centering
  \includegraphics[width=0.45\textwidth]{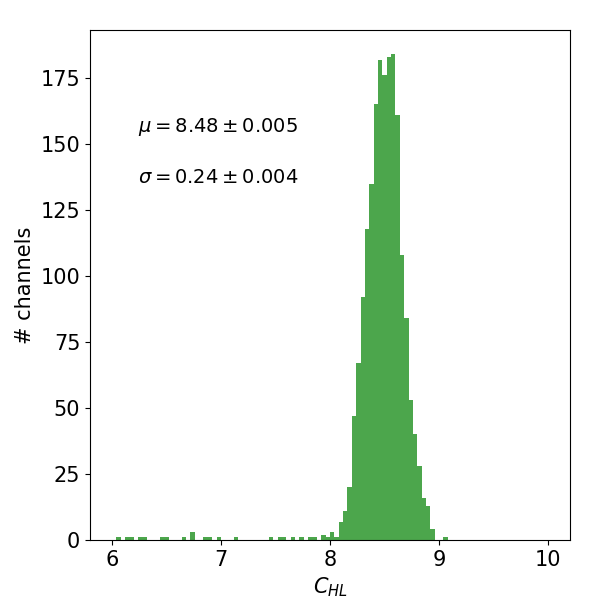}
  \includegraphics[width=0.45\textwidth]{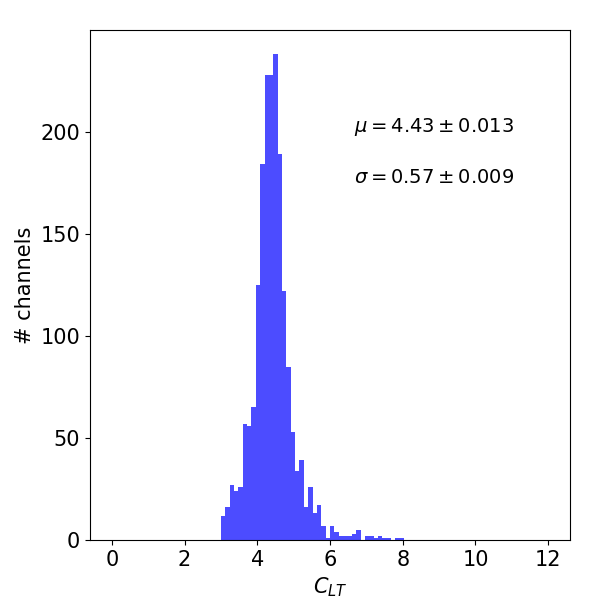}
  \includegraphics[width=0.45\textwidth]{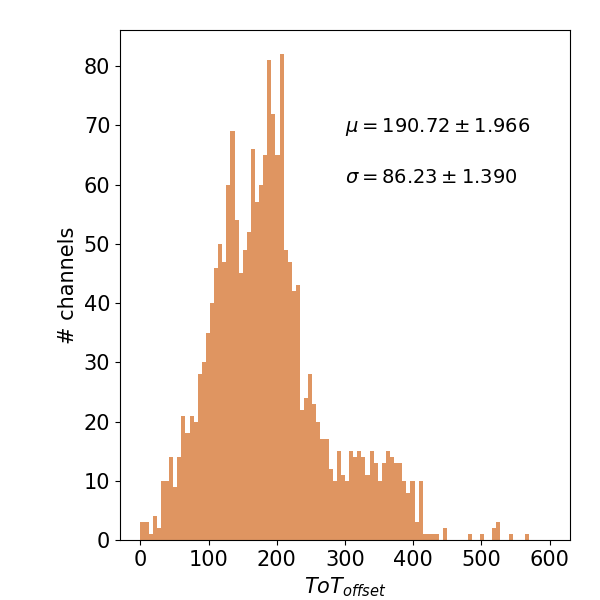}
  \caption{The distributions of the gain linearization parameters $C_{HL}$ (top left), $C_{LT}$ (top right) and $ToT_{offset}$ (bottom) derived from beam-test data.}
  \label{fig:data-driven-intercalib}
\end{figure}
Figure~\ref{fig:data-driven-intercalib} shows the distributions of the gain linearization constants, showing only channels for which the full set of constants could be derived. The spread of the variable $C_{HL}$ is relatively small (2$\%$). Hence using the average, when needed, would not have a major impact on the reconstructed signals. However, the spreads for the ToT to low-gain conversion constants, particularly $ToT_{offset}$, were significant (>~10$\%$). This could affect the reconstrucetd signals in the channels where gain linearization constants could not be derived. Fortunately, the lack of data for the gain linearization of these channels also meant that these channels were not located in the core of the showers. Therefore a mis-estimation of the ToT for such channels had a limited impact on the reconstructed shower energy and resolution in the beam-test data. 

\subsubsection{Gain linearization using charge injection}
In the future CMS CE, it will not be possible to use the method described in Section~\ref{subsubsec:datadriven-calib}. Indeed the ASIC foreseen for the future CMS CE will not have an overlap in charge sensitivity between the ADC and the linear region of the ToT. Therefore, a second gain linearization using controlled charge injections was studied. Moreover, using injection techniques allowed the gain linearization to be performed for all the channels of the CE prototype. As introduced in Section~\ref{subsubsec:module-test-electronic}, the test stands used to test the hexaboards and modules were equipped with a 12-bit DAC, which allowed the injection of a controlled charge into the channels of the SKIROC2-CMS ASICs. During the testing procedure, an injection scan was performed for all channels of the CE prototype modules. The scans were performed in a temperature-controlled box (28$^{\circ}C$) in order to have the same temperature condition as during the beam test. With this procedure, gain linearization constants could be derived for about 98$\%$ of the channels of the CE prototype modules. About half of the remaining 2$\%$ corresponded to the masked channels as described in Section~\ref{subsubsec:module-test-electronic}. For the others, various causes (dead channels, noisy channels, fail in fitting procedure) prevented the computation of the gain linearization constants.

\begin{figure}[!ht]
  \centering
  \includegraphics[width=0.6\textwidth]{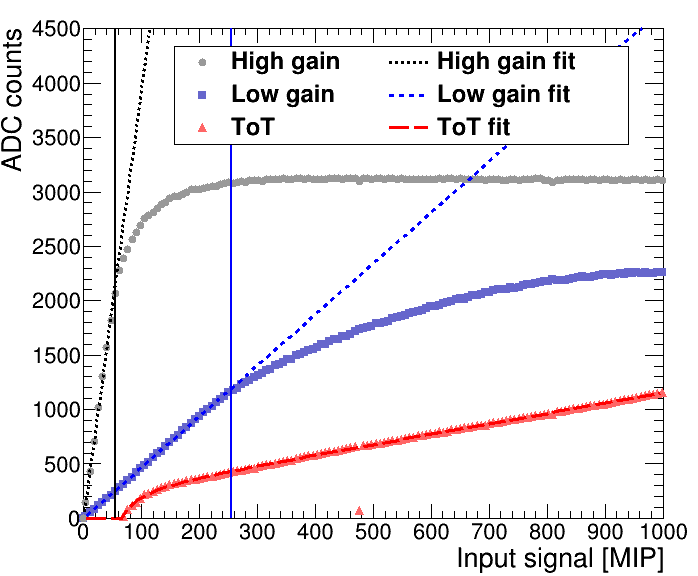}
  \caption{The high-gain and low-gain waveform amplitudes and ToT output as a function of the input charge expressed in MIP units. The dashed lines correspond to the fitted functions used to derive the gain linearization constants. The vertical solid lines mark the maximum input signals where the high-gain and low-gain shaper outputs were linear.}
  \label{fig:hglgtot_vs_mip}
\end{figure}

The same analysis procedure as described in Section~\ref{sec:signal-reco} was performed on the injection data, and the amplitudes of the two gains and the ToT were obtained as functions of the input charge. The aim was to provide the constants for transforming the electronic outputs (waveform amplitudes, ToT) into calibrated responses. Therefore, the first step of the procedure was to transform the input charge in DAC units to an energy equivalent in MIP units. This was achieved by fitting the high-gain waveform amplitude as a function of the input charge with a linear function over a limited range. The slope of the linear function was extracted, and then using the channel-to-channel response equalization constant, the input signal ($S_I$) could be expressed in MIP units using Equation~\ref{eq:calib-input}:

\begin{equation}
  \label{eq:calib-input}
  S_{I} = C_{MIP} \cdot k_{HG,I} \cdot Q_{I}
\end{equation}
where $C_{MIP}$ is the channel-to-channel response equalization constant, $Q_{I}$ is the input charge in DAC units and $k_{HG,I}$ is the slope of the correlation between high-gain amplitude to input charge. Then the amplitude of the waveforms and the ToT could be expressed as functions of the input signal in MIP units as shown in Figure~\ref{fig:hglgtot_vs_mip}.

The waveform amplitudes as functions of the input signal were fitted with linear functions and the inverse of the slopes $C_{MIP,HG}$ and $C_{MIP,LG}$ were stored. The gain saturation ($HG_{sat}$ and $LG_{sat}$) points were derived from the minimum injected signal for which the high-gain and low-gain waveform amplitudes deviated from their linear fit by 3$\%$. The saturation values of high and low gains are indicated by the vertical solid lines in Figure~\ref{fig:hglgtot_vs_mip}. The ratio $C_{HL}=\dfrac{C_{MIP,LG}}{C_{MIP,HG}}$ is equivalent to the parameter $C_{HL}$ obtained from the beam-test data as described in Section~\ref{subsubsec:datadriven-calib}.

\begin{figure}[!ht]
  \centering
  \includegraphics[width=0.45\textwidth]{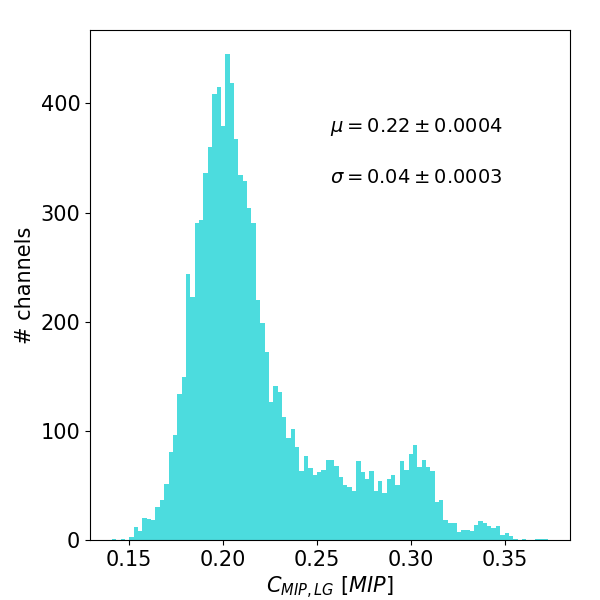}
  \includegraphics[width=0.45\textwidth]{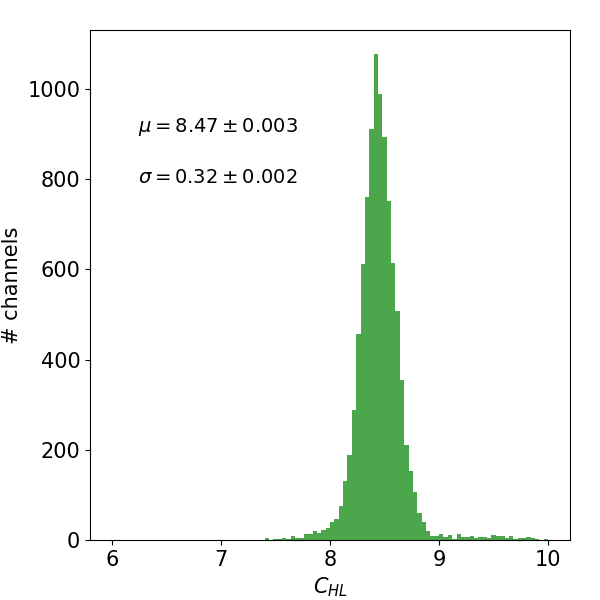}
  \caption{The distributions of the gain linearization parameter $C_{MIP,LG}$ (left) and distribution of the ratio $C_{HL}$ derived from the injection data.}
  \label{fig:injection-calib-hg-lg}
\end{figure}

Figure~\ref{fig:injection-calib-hg-lg} shows the distribution of the parameters $C_{MIP,LG}$ and $C_{HL}$ derived from the injection data. The distribution of the $C_{HL}$ is very close to the one obtained from the beam-test data. The ratio of the $C_{HL}$ parameter, obtained with the previous method, with the one obtained from injection data was studied using only the channels with full set of constants by the data driven method. The average value of this ratio was about 0.99 and the standard deviation about 0.08.

The full ToT function could be derived by fitting the ToT data as a function of the input signal with Equation~\ref{eq:tot-fit}:
\begin{equation}
  \label{eq:tot-fit}
  ToT = \left\{ \begin{array}{ll}
    0 &, \mbox{ if $S_{I}<S_{Thr}$} \\
    \frac{S_{I}}{C_{MIP,ToT}} + ToT_{offset} + \dfrac{N}{(S_{I}-S_{Thr})^\alpha} &, \mbox{ otherwise}
  \end{array} \right.
\end{equation}
where $S_{Thr}$ is the minimum input signal to trigger the ToT and $C_{MIP,ToT}$, $ToT_{offset}$, $N$ and $\alpha$ are free parameters of the fit. 

\begin{figure}[!ht]
  \centering
  \includegraphics[width=0.45\textwidth]{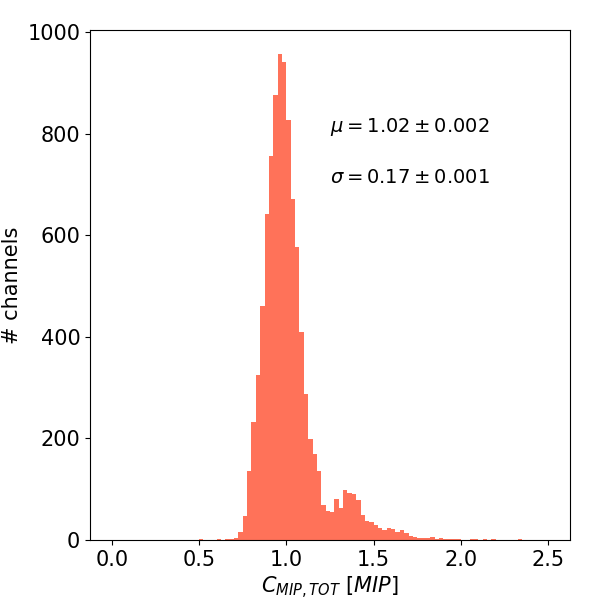}
  \includegraphics[width=0.45\textwidth]{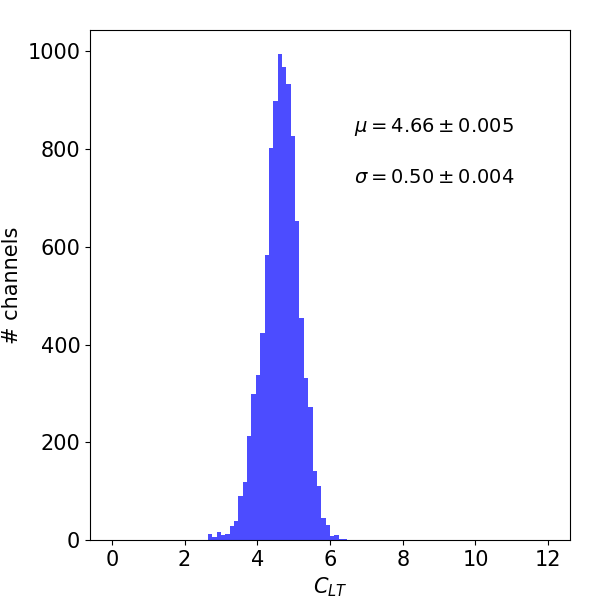}
  \includegraphics[width=0.45\textwidth]{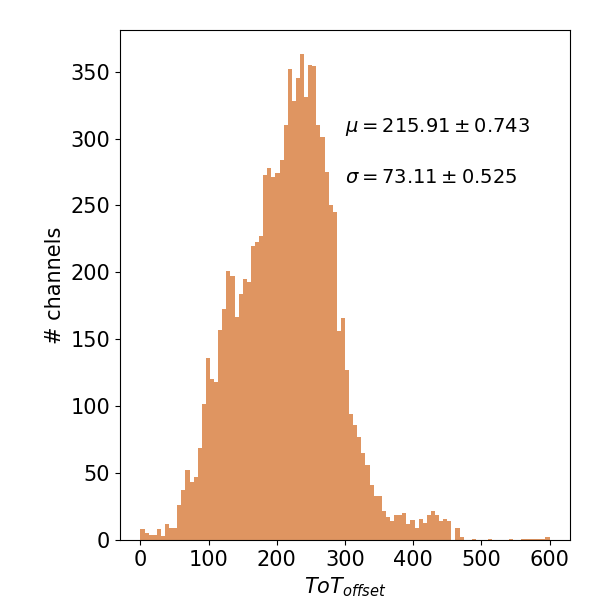}
  \includegraphics[width=0.45\textwidth]{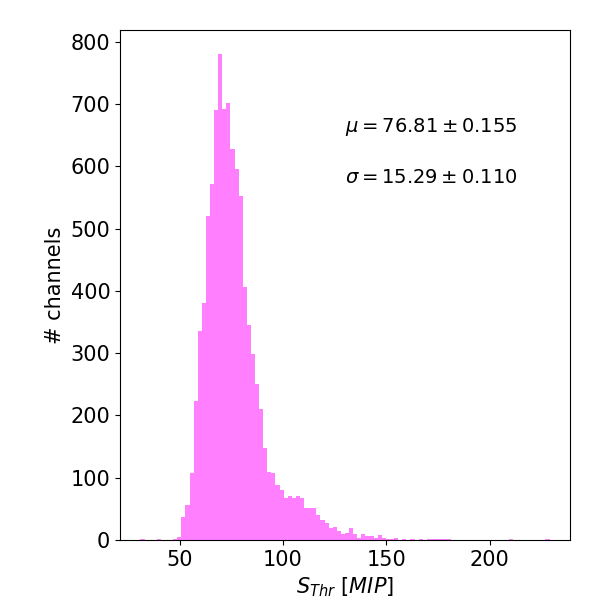}
  \caption{The distributions of the gain linearization parameters $C_{MIP,ToT}$ (top left), $C_{LT}$ (top right), $ToT_{offset}$ (bottom left) and $S_{Thr}$ (bottom right) derived from the injection data.}
  \label{fig:injection-intercalib}
\end{figure}

The ratio $C_{LT}=\dfrac{C_{MIP,ToT}}{C_{MIP,LG}}$ is equivalent to the parameter $C_{LT}$ obtained from the beam-test data as described in Section~\ref{subsubsec:datadriven-calib}. Figure~\ref{fig:injection-intercalib} shows the distributions of the parameters $C_{MIP,ToT}$, $C_{LT}$, $ToT_{offset}$ and $S_{Thr}$. The distribution of the $ToT_{offset}$ parameter is significantly different than the one obtained from the beam-test data as shown in Section~\ref{subsubsec:datadriven-calib}. The mean of the distribution of the $ToT_{offset}$ parameter is shifted by about 13$\%$ when compared to the mean of the $ToT_{offset}$ parameter obtained from the method using beam-test data (shown in Figure~\ref{fig:data-driven-intercalib}). The main reason is the lack of statistics in channels with the beam-test data. The distribution of the $ToT_{offset}$ from the injection data by using only channels with enough statistics from the beam-test data, has a mean value of $197 \pm 2$ and a standard deviation of $85 \pm 1.4$ which are in better agreement with the $ToT_{offset}$ distribution of Figure~\ref{fig:data-driven-intercalib}.

With the full ToT function, the non-linear part of the ToT could also be used. A binary search algorithm was used to compute the calibrated signal from the ToT since the ToT function is monotonic. The requested tolerance was set to 0.01 and the maximum number of iterations was 100. Finally, the calibrated signal in a silicon cell, using the gain linearization constants derived from the injection data, is given by Equation~\ref{eq:reco-energy-injection}:
\begin{equation}
  E = \left\{ \begin{array}{ll}
    E_{HG}  = C_{MIP,HG}\cdot A_{0,HG} &, \mbox{ if $A_{0,HG}<HG_{sat}$} \\
    E_{LG}  = C_{MIP,LG}\cdot A_{0,LG} &, \mbox{ if $A_{0,HG}>HG_{sat}$ and $A_{0,LG}<LG_{sat}$} \\
    E_{ToT} &, \mbox{ otherwise}
  \end{array} \right.
  \label{eq:reco-energy-injection}
\end{equation}
where $E_{ToT}$ was calibrated signal from the ToT computed with the bisection method.

\section{Conclusion}
\label{sec:conclusion}
A new high granularity calorimeter prototype has been built in 2017-2018 and tested with beam of particles at CERN and DESY. Around 100 prototype modules, based on 6-inch hexagonal silicon sensors with cell areas of 1.1~$cm^{2}$, have been constructed using the SKIROC2- CMS readout ASIC for the final beam test in October 2018 at CERN-SPS.  This ASIC contains the ToT and ToA functionality, which will be important features of the final ASIC of the CMS CE. The prototype,  including more than 12,000 readout channels, comprised sensors of two different full depletion thicknesses (300~$\upmu$m and 200~$\upmu$m). 

Different grounding schemes, pedestal and common-mode estimation methods have been studied, validating the options proposed for the final CMS CE system. The pedestal values were stable over the full beam test campaign. The modules showed good S/N performance when tested by muons. The total noise level, dominated by common-mode noise, was comparable for the two sensor depletion thicknesses.  After common-mode noise removal, the thicker sensors showed a slightly lower intrinsic noise due to their lower capacitance.

The prototype modules were calibrated in two stages: equalizing the response of the channels to the signal energy deposited by MIP-like particles and linearizing the gains for the different energy measurements offered by the prototype ASIC to provide the large dynamic range.  Both the calibration procedure and two different types of gain linearization methods were established.

\acknowledgments
We thank the technical and administrative staffs at CERN and at other CMS institutes for their contributions to the success of the CMS effort.  We acknowledge the enduring support provided by the following funding agencies: BMBWF and FWF (Austria); CERN; CAS, MoST, and NSFC (China); MSES and CSF (Croatia); CEA and CNRS/IN2P3 (France); SRNSF (Georgia);  BMBF, DFG, and HGF (Germany); GSRT (Greece); DAE and DST (India); MES (Latvia); MOE and UM (Malaysia); MOS (Montenegro); PAEC (Pakistan); FCT (Portugal); JINR (Dubna); MON, RosAtom, RAS, RFBR, and NRC KI (Russia); MST (Taipei); ThEPCenter, IPST, STAR, and NSTDA (Thailand); TUBITAK and TENMAK (Turkey); STFC (United Kingdom); DOE (USA).



\end{document}